\documentclass[pra,twocolumn,showpacs,amsmath,amssymb,floatfix]{revtex4}

\usepackage{graphicx}
\usepackage{color}

\begin{document}

\title{Pairing in a two-dimensional Fermi gas with population imbalance}

\author{M. J. Wolak$^1$,  B.~Gr\'emaud$^{1,2,3}$,  R. T. Scalettar$^4$,
 G. G. Batrouni$^{5,1,6}$}

\affiliation{ $^1$ Centre for Quantum Technologies, National
University of Singapore; 2 Science Drive 3 Singapore 117542}
\affiliation{ $^2$ Laboratoire Kastler Brossel, UPMC-Paris 6, ENS,
CNRS; 4 Place Jussieu,F-75005 Paris, France} 
\affiliation{ $^3$ Department of Physics, National University of
Singapore, 2 Science Drive 3, Singapore 117542, Singapore}
\affiliation{ $^4$ Physics Department, University of California, Davis, California
95616, USA}
\affiliation{ $^5$ INLN, Universit\'e de Nice--Sophia Antipolis, CNRS;
1361 route des Lucioles, 06560 Valbonne, France}
\affiliation{ $^6$ Institut Universitaire de France}

\begin{abstract}
  Pairing in a population imbalanced Fermi system in a two-dimensional
  optical lattice is studied using Determinant Quantum Monte Carlo
  (DQMC) simulations and mean-field calculations. The
  approximation-free numerical results show a wide range of stability
  of the Fulde-Ferrell-Larkin-Ovshinnikov (FFLO) phase. Contrary to
  claims of fragility with increased dimensionality we find that this
  phase is stable across wide range of values for the polarization,
  temperature and interaction strength. Both homogeneous and
  harmonically trapped systems display pairing with finite center of
  mass momentum, with clear signatures either in momentum space or
  real space, which could be observed in cold atomic gases loaded in
  an optical lattice. We also use the harmonic level
    basis in the confined system and find that pairs can form between particles
    occupying different levels which can be seen as the analog of the
    finite center of mass momentum pairing in the translationally
    invariant case. Finally, we perform mean field calculations for
  the uniform and confined systems and show the results to be in good
  agreement with QMC. This leads to a simple picture of the different
  pairing mechanisms, depending on the filling and confining
  potential.
\end{abstract}


\maketitle

\section{Introduction}

The question of pairing in polarized fermionic systems came to the
fore shortly after superconductivity in unpolarized systems was
explained by BCS~\cite{BCS} as being due to the formation of Cooper
pairs with zero center of mass momentum. Fulde and
Ferrell~\cite{fulde} and independently Larkin and
Ovchinnikov~\cite{larkin} proposed similar but not identical
mechanisms whereby the fermions form pairs with nonzero center of mass
momentum. We will refer to such a phase as the FFLO phase. On the
other hand, Sarma~\cite{sarma} proposed a mechanism where, in spite of
the mismatch in the Fermi momenta due to the spin population
imbalance, the pairing occurs with zero center of mass
momentum. Verifying these predictions experimentally proved difficult
in condensed matter systems~\cite{radovan}. However, thanks to rapid experimental
progress in the domain of ultra-cold atoms, it is now possible to
study such population imbalanced systems. Fermionic atoms are made to
occupy two hyperfine states thus emulating a system with ``up'' and
``down'' spins. An advantage of these systems is that the population
imbalance (the polarization) and the interaction strengths are highly
tunable. Such experiments have been performed in
three-dimensional~\cite{zwierlein06, partridge06} and
one-dimensional~\cite{hulet} systems.

It is by now widely accepted that at $T=0$ the FFLO phase is robust
over a wide range of parameters in one-dimensional systems with
imbalanced fermion populations. This was shown in various numerical
studies using, for example, QMC \cite{batrouni08,casula} and DMRG
\cite{feiguin07,luscher,rizzi,tezuka07}. In a previous work we also
showed that the FFLO phase is stable over a wide range of parameters
in the temperature-polarization (TP) phase diagram \cite{us2010}. This
exotic pairing occurs both in homogeneous and confined systems, and has
been shown to survive up to relatively high temperatures ($T/T_F\approx0.1$)
which are achievable in current experiments.

The question of the stability of this phase in higher dimensions
remains a subject of debate. It is believed that ``nesting'' of the
Fermi surfaces stabilizes FFLO pairing.  For example in one dimension
one wave-vector connects all points on the Fermi surfaces of each
species, which would enable all particles from the Fermi surfaces to
participate in the formation of pairs with finite-momentum. The effect
of ``nesting'' is considerably weaker in higher dimensions. In a
two-dimensional lattice system, the shape of a Fermi surface depends
on the filling. At half filling the Fermi surface becomes a square and
touches the edge of the first Brillouin zone (Van Hove
singularity). Around this filling, matching of the Fermi surfaces
becomes more efficient, in other words the ``nesting'' is enhanced as
compared to the situation when both Fermi surfaces are circular (low
filling). This reasoning leads us to expect that FFLO pairing should
be more prevalent around half filling than at lower fillings. This
lattice enhanced stability of FFLO was studied using mean field (MF)
methods in \cite{Koponen2008} and \cite{Trivedi}. In the latter the
authors point also at Hartree corrections and domain wall formation as
additional reasons for enhancement.

Numerous theoretical studies of the system in higher dimensions do not
offer a clear conclusion on the stability of the FFLO mechanism. In a
variational MF study of a three dimensional system in the
continuum with and without a trapping potential it is observed that
FFLO is a fragile state which can be realized only in a tiny sliver of
the interaction-polarization phase diagram \cite{Sheehy}. Furthermore,
this study showed that in a trap, FFLO can exist only in a thin shell
of the atomic cloud. Another study of a three dimensional Fermi gas at
unitarity~\cite{Bulgac} shows that this phase is competitive over a
large region in the phase diagram. However, the trap would need to be
adjusted to allow FFLO to occupy a large enough spatial region to be
observed. On the other hand, in a Bogoliubov-de Gennes study
\cite{machida06} of a trapped system, the calculated phase diagram
indicates that the ground state of the system is always FFLO for any
imbalance up to some critical value.

The unsettled status of this phase in higher dimensionality may be
clarified with exact numerical simulations. However, simulations of
the Hubbard model in three dimensions are not feasible for large
systems at low enough temperatures due to the severity of the
``fermion sign problem''. On the other hand, exact QMC simulations in
two dimensions are feasible but so far none have been done. In
addition, two dimensional systems are intermediate between one
dimension where MF is almost certain to fail and three dimensions
where MF is more reliable. Consequently, there has been a concerted,
yet inconclusive, effort to understand FFLO physics theoretically in
two-dimensional systems. Homogeneous and trapped two-dimensional
polarized Fermi gases have been studied with MF calculations which
exclude the possibility of FFLO pairing (for \textit{e.g.} \cite{He}
and \cite{Schaeybroeck}). An interaction-polarization phase diagram is
shown in \cite{Rombouts} where FFLO pairing is seen to occupy a wide
region.  Koponen \emph{et al.} \cite{Koponen2008} obtain MF phase
diagrams in the polarization versus filling plane for one-, two- and
three-dimensional systems. In the two-dimensional system there is a
very strong feature around the van Hove singularity of the majority
component and the FFLO phase is present over a wide range of
parameters around this value. They also show temperature-polarization
phase diagrams of one dimensional system which were shown not to agree
with exact QMC results \cite{us2010}. The temperature-polarization
phase diagram in three dimensions is shown as well but not the two
dimensional case. Studies of quasi two-dimensional systems have been
done using MF and they predict a first-order transition to FFLO at
finite temperature \cite{DeSilva}.  Another mean field study of
two-dimensional two-orbital Hubbard model with p-orbitals and highly
unidirectional hopping shows enhancement of the FFLO region in the
phase diagram due to the one-dimensional character of the Fermi
surface \cite{Cai}. A DMRG study of population imbalanced Fermi gas on
two-leg ladders has found FFLO pair correlations \cite{Feiguin09}.

In this paper we present a Determinant QMC (DQMC) \cite{DQMC} study of
the two dimensional Hubbard model with imbalanced populations of up
and down spins. In section II we present the model and discuss our
results for the uniform system in section III. Our main result here is
the demonstration of the robustness of the FFLO phase and the
determination of the phase diagram in the temperature-polarization
plane at low filling. We also compare the behavior of the system at
low and half fillings. In section IV we examine the system in a
harmonic trap. Our conclusions are in section V.

\section{Model and Methods}
The system of interest is governed by the two-dimensional fermionic
Hubbard Hamiltonian
\begin{eqnarray}
\label{Hamiltonian}
H&=&-t\sum_{<i,j>\, \sigma} (c_{i\,\sigma}^{\dagger}
c_{j\,\sigma}^{\phantom\dagger} + c_{j\,\sigma}^\dagger c_{i\,
\sigma}^{\phantom\dagger}) - \sum_i ( \mu_1\hat n_{i\,1} +
\mu_2 \hat n_{i\,2}) \nonumber \\ 
\nonumber
&&+U \sum_{i} \left(\hat n_{i\,1}-\frac{1}{2}\right)\left(\hat
n_{i\,2}-\frac{1}{2}\right)\\ 
&&+V_T \sum_{j} \left(j-j_c\right)^2 \left(\hat n_{j \,1} +
\hat n_{j \, 2}\right) 
\end{eqnarray}
Where $c_{i\,\sigma}^\dagger$ ($c_{i\, \sigma}^{\phantom\dagger}$)
create (annihilate) a fermion of spin $\sigma=1,\,2$ on lattice site
$i$ and $\hat n_{i\,\sigma}=c_{i\,\sigma}^\dagger c_{i\, \sigma}^{\phantom\dagger}$ is the corresponding number operator. The near neighbor, $<i,j>$, hopping parameter is $t$ which we
take equal to unity to set the energy scale. We consider only on-site
interaction with an attractive coupling constant $U<0$. The number of
particles in each population is governed by its chemical potential
($\mu_{\sigma}$). The harmonic trap is introduced via the $V_T$ term
in the Hamiltonian where $j_c$ is the position of the center of the trap (also middle
of the lattice).  All simulations are performed with periodic boundary
conditions. In the confined case we ensured that the
  density vanishes at the edge of the lattice.

The main quantities of interest in this study are the single particle
Green functions $G_{\sigma}$ and the pair Green function, $G_{pair}$,
\begin{eqnarray}
\label{gfct}
G_\sigma(l) &=& \langle c_{i+l\,\sigma}^\dagger c_{i
\,\sigma}^{\phantom\dagger} \rangle,\\
\label{pairgfct}
G_{\rm pair}(l) &=& \langle
\Delta^{\dagger}_{i+l}\,\Delta_{i}^{\phantom\dagger} \rangle,\\ 
\label{pairoperator}
\Delta_i &=& c_{i\, 2} \,c_{i \, 1},
\end{eqnarray}
where $\Delta_{i}^{\phantom\dagger}$ creates a pair on site $i$.  The
Fourier transform of $G_\sigma$ gives the momentum distribution of the
spins-$\sigma$ species while the transform of $G_{\rm pair}$ yields
the pair momentum distribution. In the trapped case, the density
profiles of the two species are also studied.

We studied this system numerically using the DQMC \cite{DQMC}
algorithm. In this approach, the Hubbard-Stratonovich (HS)
transformation is employed to decouple the quartic interaction term
into two quadratic terms coupling the number operator of each species,
$n_\sigma(i)$, to the HS field which effectively acts as a site and
imaginary time dependent chemical potential. The fermion operators can
now be traced out leading to a partition function in the form of a
product of two determinants, one for each spin, summed over all
configurations of the HS field. For $U<0$ and equal populations,
$\mu_1=\mu_2$, the determinants are identical; their product is always
positive. But in the imbalanced case, $\mu_1\neq\mu_2$, the two
determinants are no longer equal and their product can, and does,
become negative leading to the known ``fermion sign problem''. This is
the main obstacle to the simulation of this system. We found that at
low total filling the sign problem is manageable even at large
polarizations and low temperatures. This was not the case closer to
half filling. Typical simulations of the harmonically confined system
at low temperature took about two weeks on a 3 GHz processor.

In the presence of the trapping potential, we have also studied the
system using a mean-field approach. Starting from the full Fermi-Hubbard
Hamiltonian~\eqref{Hamiltonian}, one can derive the mean-field
Hamiltonian:
\begin{equation}
 \begin{aligned}
  H_{MF}&= \psi^{\dagger}M\psi+\frac{1}{U}\sum_{i}\Delta_i^*\Delta_i-\sum_i\mu_{i\,2}\\
M=&\left(\begin{array}{cc}
 h_{ij\,1} & -\Delta_i \\
 -\Delta_i^* & -h_{ji\,2}	
\end{array}\right),
 \end{aligned}
\end{equation}
where $\Delta_i^*=U\langle
c^{\dagger}_{i\,1}c^{\dagger}_{i\,2}\rangle$ are on-site pairing
amplitudes;
$\Psi^{\dagger}=\left(\cdots,c_{i\,1}^{\dagger},\cdots,c_{i\,2},\cdots\right)$
is the Nambu spinor.  The matrix $h$ depicts the one particle
Hamiltonian, namely hopping terms between nearest neighbors
$h_{ij\sigma}=-t\quad i\ne j$ and chemical potential terms
$h_{jj\sigma}=-\mu_{j\sigma}=-\mu_{\sigma}+V_T\left(j-j_c\right)^2
\hat n_{j \,\sigma}$.  To account properly for spatial
inhomogeneities, the BCS order parameter at each site, $\Delta_i$, is
an independent variable~\cite{Iskin_08,Fujihara_10,Gremaud_12}, whose
value is determined, for a given temperature, by a global minimization
of the free energy, $F=-\frac{1}{\beta}\ln{(\mathcal{Z})}$ associated
with the mean-field Hamiltonian:
\begin{equation}
 F=-\frac{1}{\beta}\sum_k\ln{\left(1+e^{-\beta\lambda_k}\right)}+\frac{1}{U}\sum_{i}\Delta_i^*\Delta_i-\sum_i\mu_{i\,2},
\end{equation}
where the $\lambda_k$ are the $2N$ eigenvalues of the Nambu matrix
$M$; $N$ is the number of sites.  The minimization of the free energy
is performed using a mixed quasi-Newton and conjugate gradient method;
additional checks were performed to ensure that the global minimum has
been reached.

\section{Homogeneous system}

\begin{figure}[!htb]
\begin{center}
\includegraphics[width=0.35\textwidth,clip]{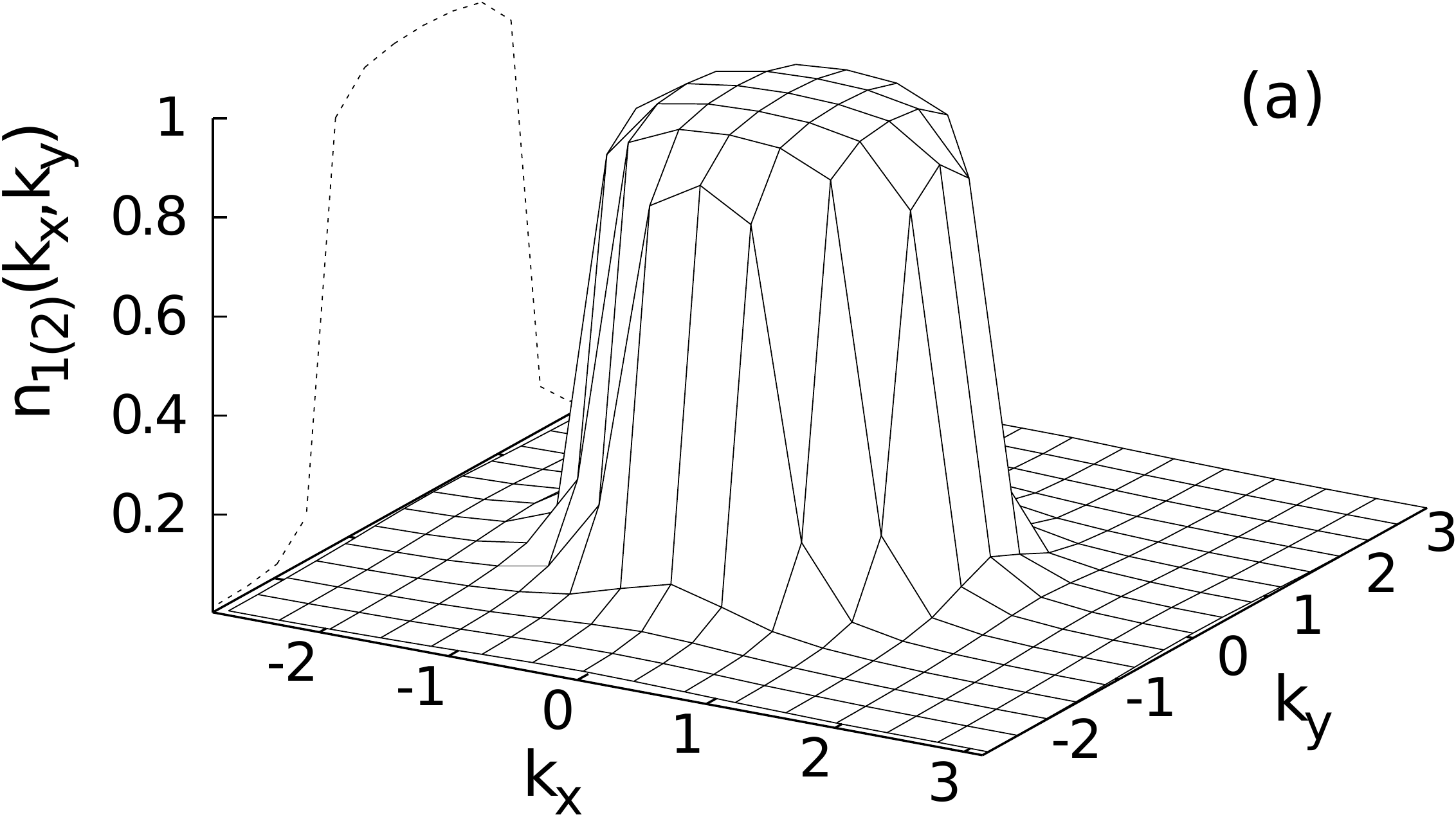}\\
\includegraphics[width=0.35\textwidth,clip]{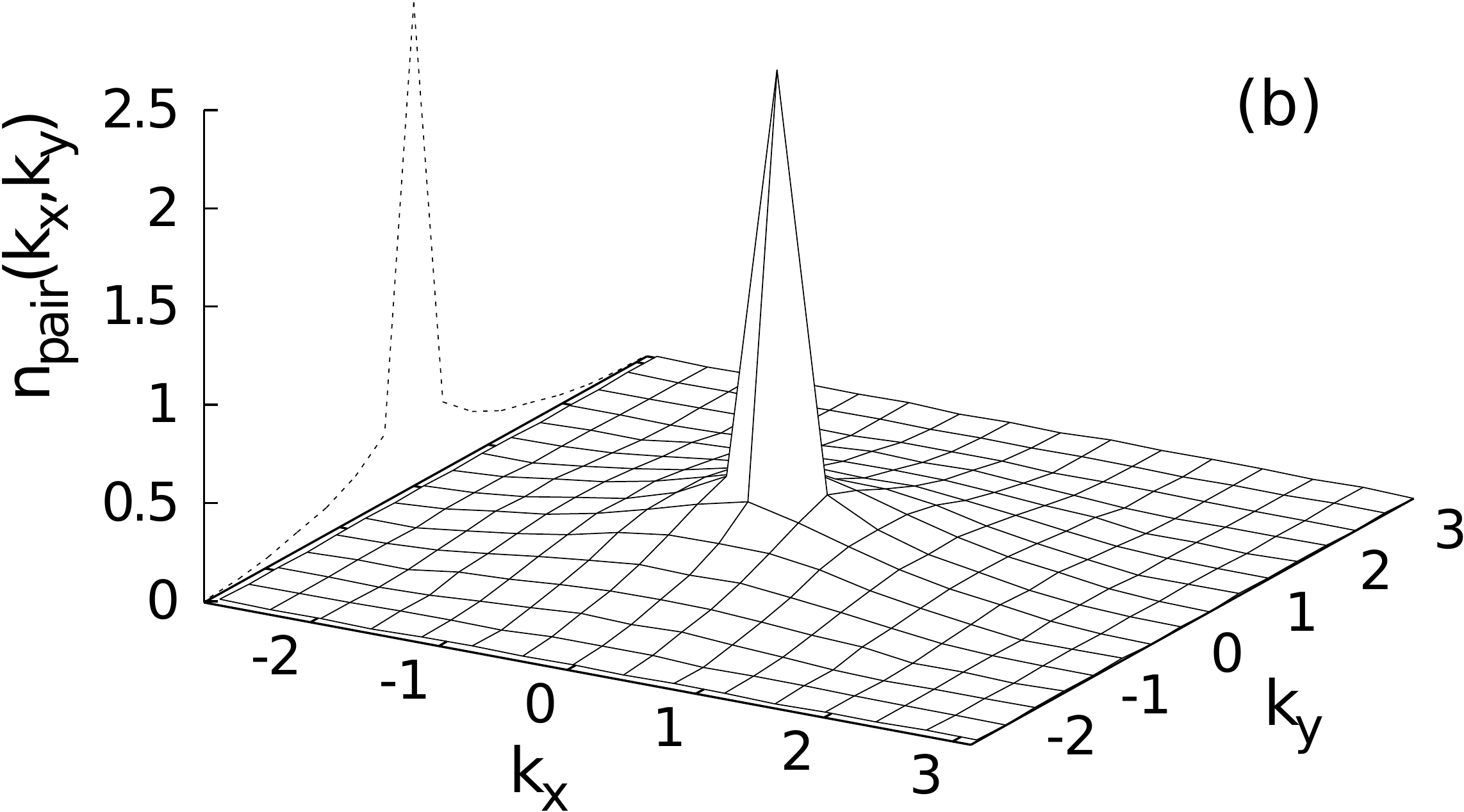}
\end{center}
\caption{\label{fig:balancedmomentumNoTrap} (a) Single particle momentum
distribution, $n_1(k_x,k_y)$ (the same as $n_2(k_x,k_y)$). (b) Pair
momentum distribution, $n_{pair}(k_x,k_y)$
exhibiting a sharp peak at zero momentum. The total
density is $\rho_1+\rho_2=0.3$ ($\rho_1=\rho_2$), $\beta=30$, $U=-3.5t$
and the system size is $16\times 16$. }
\end{figure}

To set the stage, we start with the homogeneous two dimensional
Hubbard model with balanced populations at total density
$\rho=\rho_1+\rho_2=0.3$. With balanced populations, the pairs form
with zero center of mass momentum and a sharp peak in the pair
momentum distribution is expected at ${\vec k}=0$. Figure
\ref{fig:balancedmomentumNoTrap} shows the momentum distributions for
a system with $U=-3.5t$, $\rho_1+\rho_2=0.3$ and $\beta=30$ in a
$16\times 16$ optical lattice. The single particle momentum
distribution, identical for the two spins, is shown in Fig.
\ref{fig:balancedmomentumNoTrap}(a) while (b) shows the pair momentum
distribution. As expected for weak to moderate values of $|U|$, the
single particle distribution has the usual Fermi form and the pair
momentum distribution exhibits a very sharp peak at ${\vec k}=0$
indicating pairing with zero center of mass momentum.

We now examine the polarized system. To this end, the chemical
potentials $\mu_1$ and $\mu_2$ are made unequal so that $\rho_1\neq
\rho_2$ but $\rho=\rho_1+\rho_2$ remains constant. This requires
tuning the chemical potentials appropriately. The polarization, $P$,
is defined by
 \begin{equation}
P=\frac{N_1-N_2}{N_1+N_2},
\label{polarization}
\end{equation}
where $N_1$ and $N_2$ are the total populations of the two species.

\begin{figure}[!htb]
\begin{center}
\includegraphics[width=0.35\textwidth,clip]{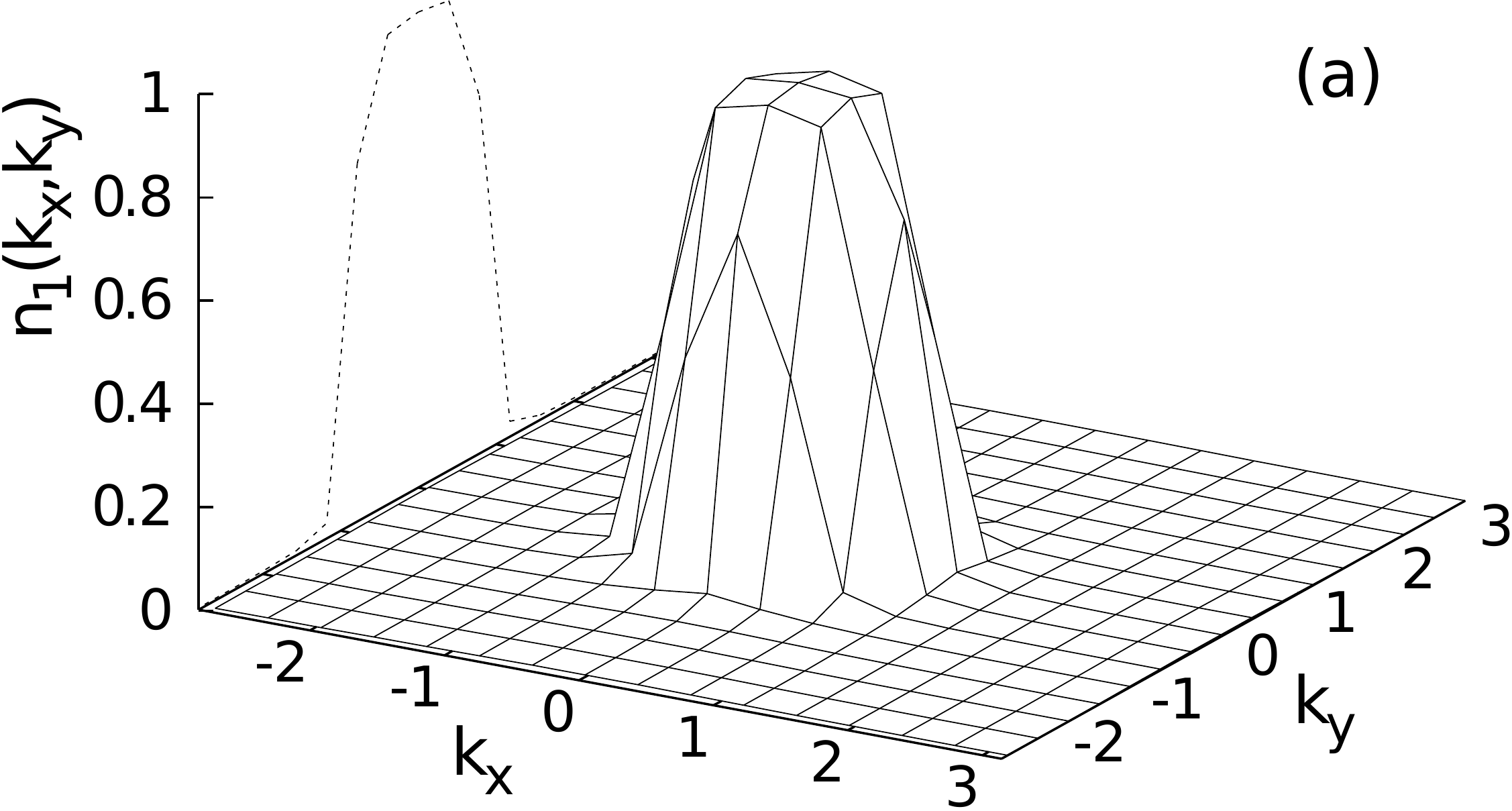}
\includegraphics[width=0.35\textwidth,clip]{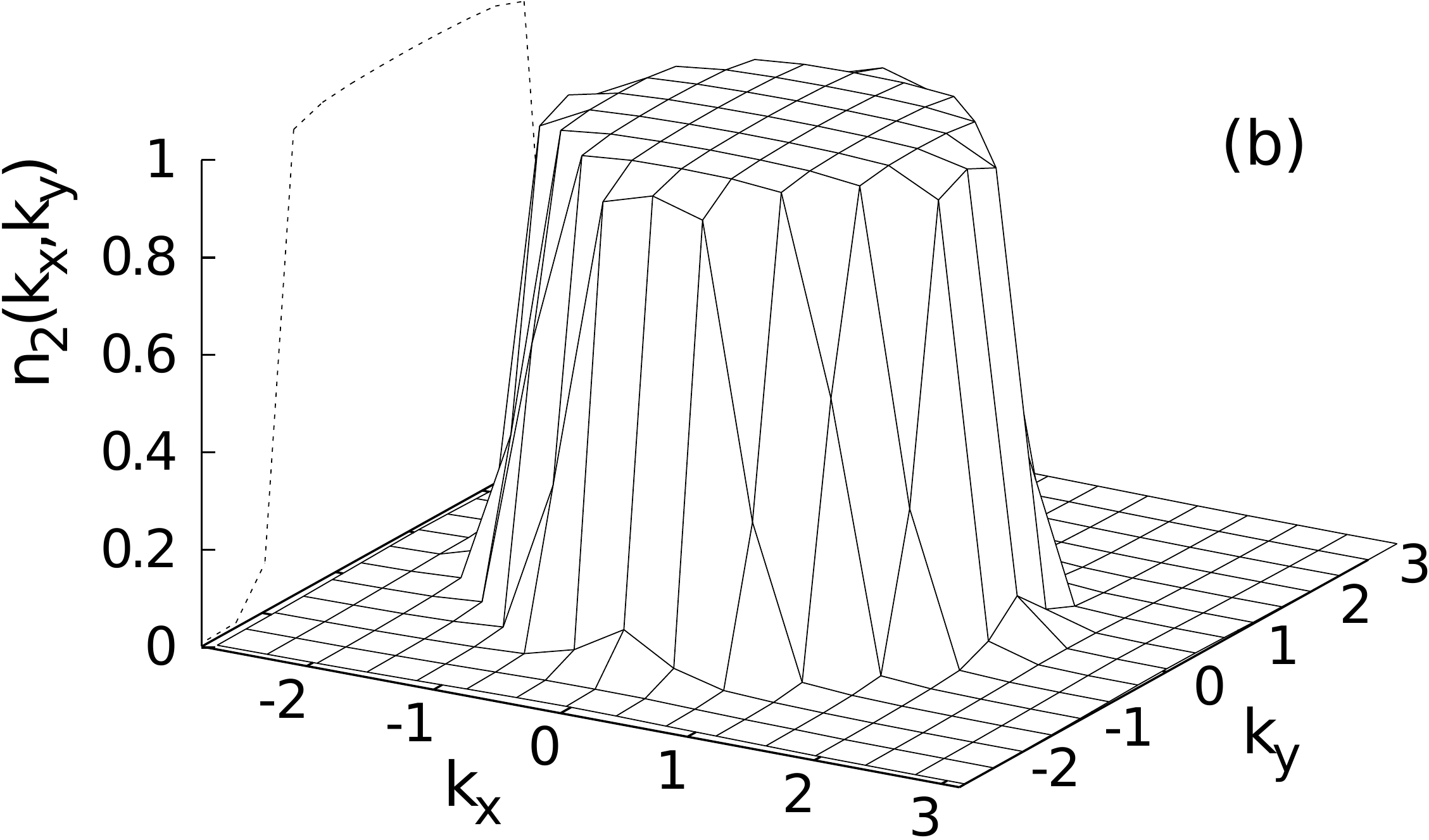}
\includegraphics[width=0.35\textwidth,clip]{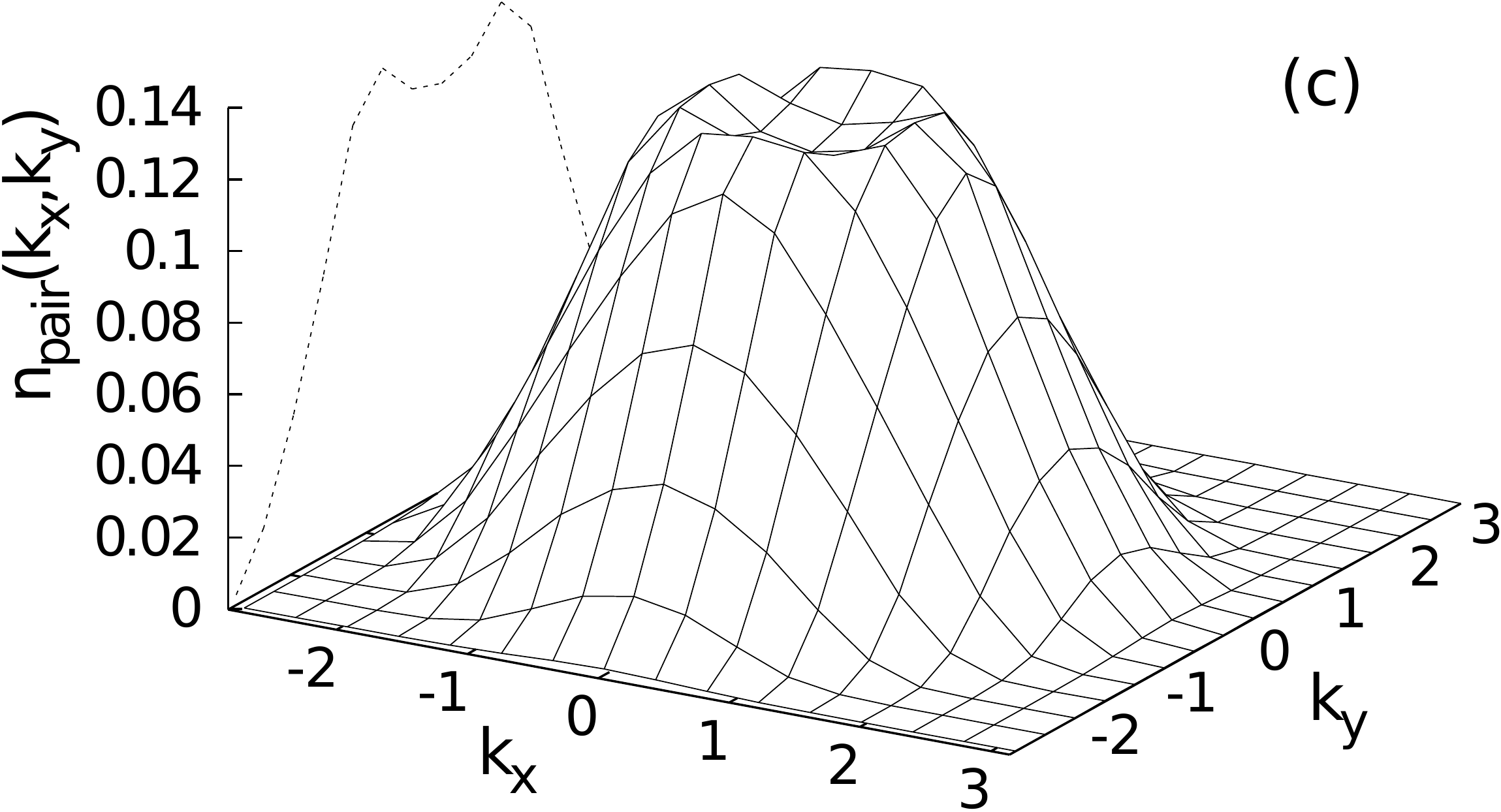}
\includegraphics[width=0.35\textwidth,clip]{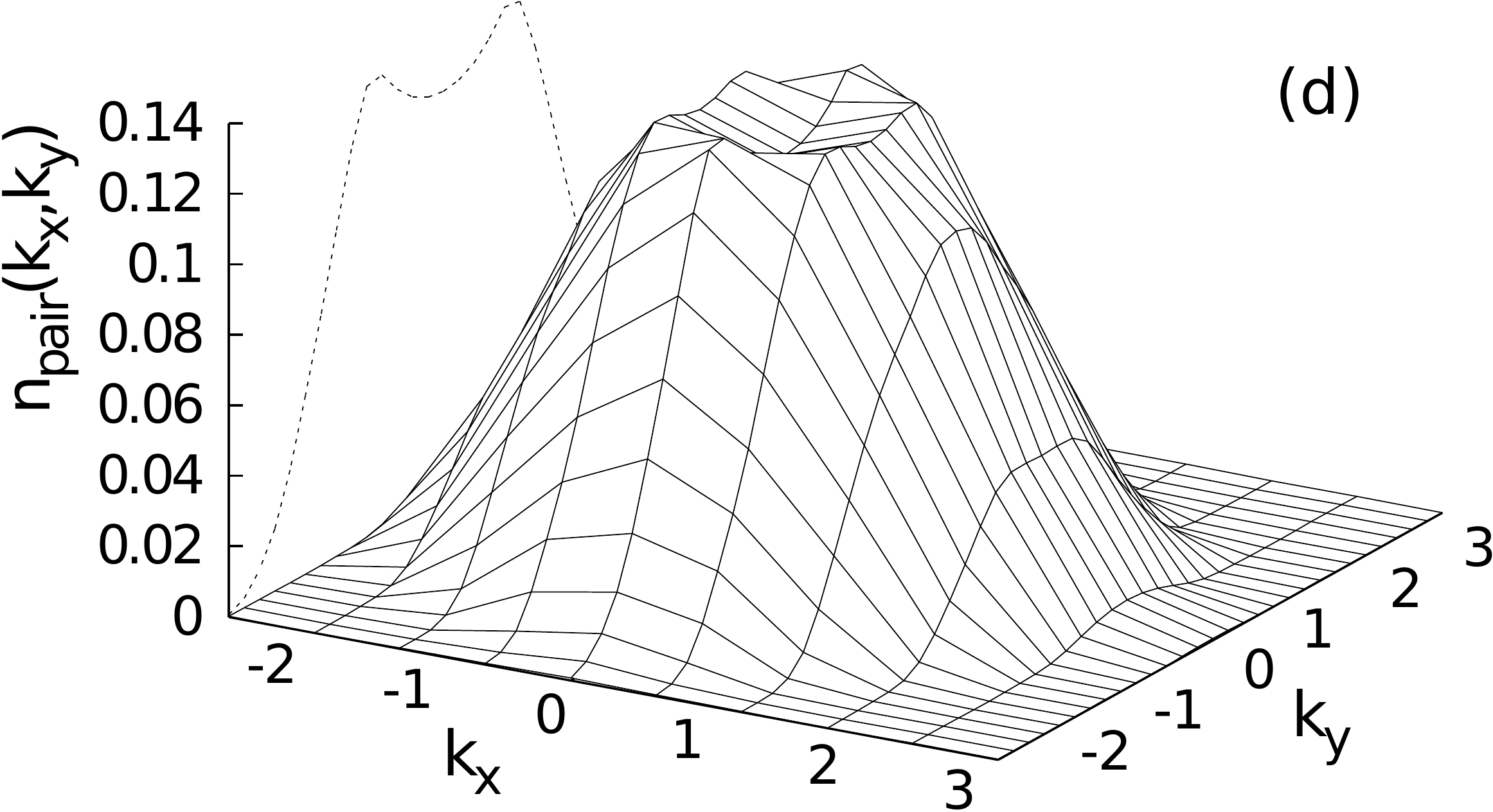}
\end{center}
\caption{\label{fig:imbalancedmomentumNoTrap} Momentum distributions
  of (a) minority and (b) majority populations. (c) shows the pair
  momentum distribution. The parameters are $\rho=\rho_1+\rho_2=0.3$,
  $P=0.6$, $\beta=10$, $U=-3.5t$ in an optical lattice of size
  $16\times 16$. (d) The pair momentum distribution for the same
  system but for a lattice of size $10\times 30$.}
\end{figure}

Figure \ref{fig:imbalancedmomentumNoTrap} shows the momentum
distributions for a system with $U=-3.5t$, $P=0.6$, $\rho=0.3$ and
$\beta=10$ in an optical lattice of size $16 \times 16$ for (a), (b)
and (c) and $10\times 30$ for (d). Panels (a) and (b) show the
minority and majority single particle momentum distributions,
$n_1(k_x,k_y)$ and $n_2(k_x,k_y)$, respectively. They exhibit usual
Fermi-like distributions. However, the pair momentum distribution,
$n_{\rm pair}(k_x,k_y)$, is strikingly different from the balanced
case: It has a volcano-like shape with the maximum of the distribution
at the rim of the crater of radius $|{\vec k}|=|{\vec k}_{F2}-{\vec
  k}_{F1}|$. ${\vec k}_{F1}$ and ${\vec k}_{F2}$ are the minority and
majority Fermi momenta respectively.

In two dimensions, the Fermi surface geometry changes with the
filling. The behavior exhibited in
Fig.~\ref{fig:imbalancedmomentumNoTrap} is for low filling where the
Fermi distributions of both species have cylindrical shape and the
pairs are formed with equal probability in all radial directions. In
this density regime, the signature for the FFLO phase is the presence
of a circular ridge in the pair momentum distribution as seen in
Fig.~\ref{fig:imbalancedmomentumNoTrap}(c). Studying the system in the
low density regime is interesting because it also approximates the
continuum conditions.

To study possible finite size effects, we performed our simulations
for systems of various sizes. In particular,
Fig. \ref{fig:imbalancedmomentumNoTrap}(d) shows the pair momentum
distribution for the same parameters as (a,b,c) but with a system of
size $10\times 30$. It is seen that the peak in the pair momentum
distribution is at the same values of $|{\vec k}|=|{\vec k}_{F1}-{\vec
  k}_{F2}|$ as the $16\times 16$ system.

\begin{figure}[!htb]
\begin{center}
 \includegraphics[width=0.4\textwidth,angle=0,clip]{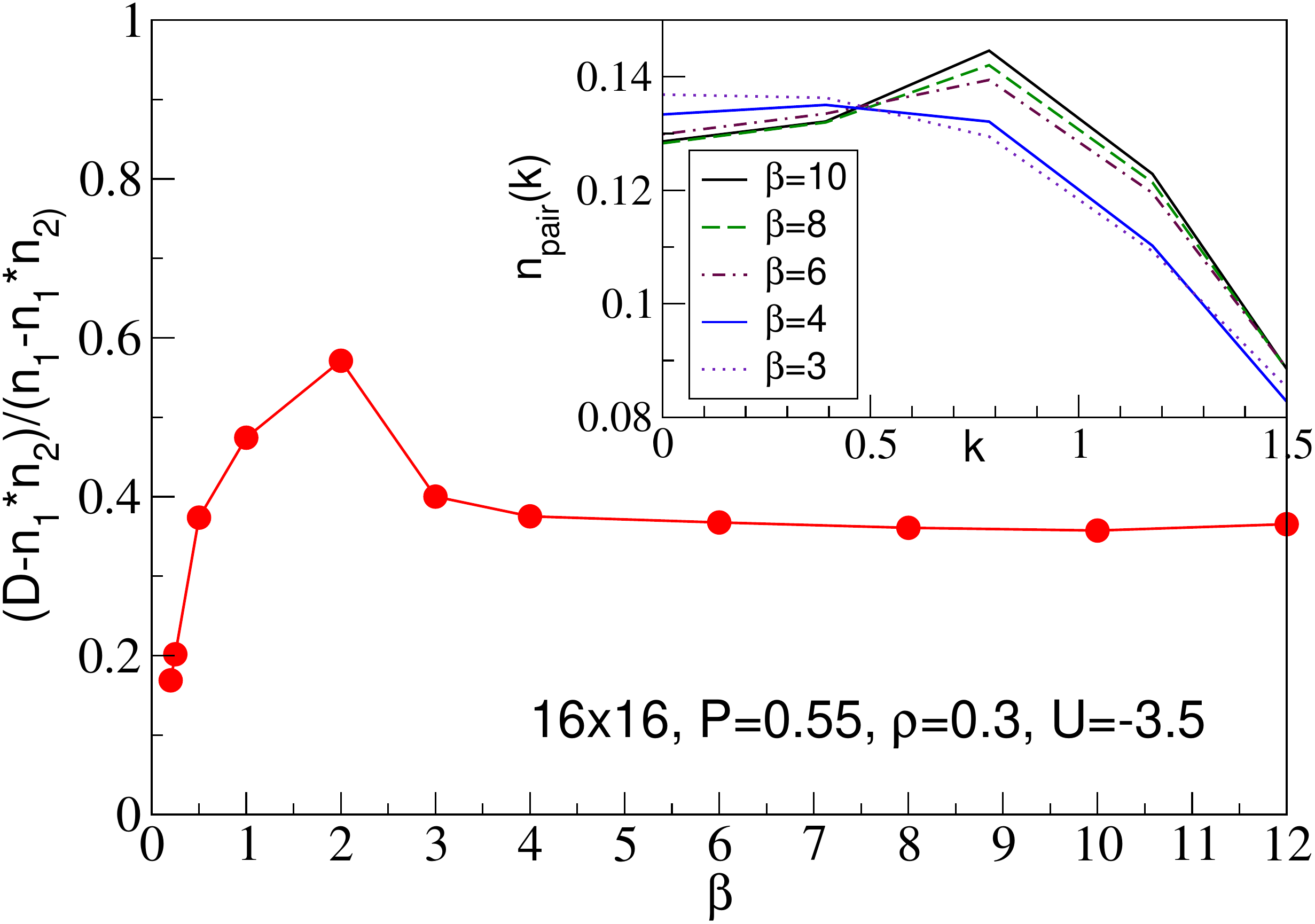}
\end{center}
\caption{\label{fig:finiteTdoubleocc}(Color Online). The normalized double occupancy
as a function of inverse temperature, $\beta$ for $\rho=0.3$, $P=0.55$
and $U=-3.5t$. The lattice size is $16\times 16$.  Inset: Behavior of
the  pair momentum distribution as the temperature is increased
($\beta$ is decreased). }
\end{figure}

We now examine the effect of temperature on the FFLO phase. In
particular, we map out the phase diagram in the
temperature-polarization plane. Thermal effects are very important in
experiments due to the difficulty in cooling fermionic atoms. The
inset in Fig.~\ref{fig:finiteTdoubleocc} shows two-dimensional cuts in
the three-dimensional pair momentum distribution for a $16\times 16$
system with $U=-3.5t$, $\rho=0.3$ and $P=0.55$. We see that as the
temperature is increased ($\beta$ decreased) the FFLO peak at nonzero
momentum decreases and, in fact, shifts towards zero momentum. Our
criterion for the appearance of the FFLO phase is when the peak of the
pair momentum distribution is no longer at zero momentum. The question
is then what replaces the FFLO phase: have the pairs been broken by
thermal fluctuations or has the system been homogenized, resulting in
a uniform mixture of pairs and excess unpaired particles of the
majority population? The double occupancy, $D=\langle n_1({\vec r})
n_2({\vec r})\rangle$, offers a measure of how tightly bound the pairs
are: In the absence of pairing, $D=\rho_1 \rho_2$ while when the
pairing is complete, $D=\rho_1$ where $\rho_1$ is the minority
population. These limits suggest the use of a normalized form,
$(D-\rho_1\rho_2)/(\rho_1-\rho_1\rho_2)$, which is now bounded by $0$
and $1$. Note that $\rho_1=N_1/L^2$ while $\langle n_1({\vec
  r})\rangle$ is the average number of type $1$ particles at ${\vec
  r}$.  In the absence of pairing the two quantities coincide. We see
in Fig.~\ref{fig:finiteTdoubleocc} that for $\beta>3$ the normalized
double occupancy is essentially constant signaling the continued
presence of pairs. This means that when the FFLO peak first disappears
at $4 < \beta < 3$, the pairs are still formed. We conclude therefore
that the system leaves the FFLO phase to enter a polarized paired phase
(PPP) phase.

When the thermal energy, $T=1/\beta$ is of the order of the pair
binding energy, $|U|$, the pairs are expected to break. We see in
Fig.~\ref{fig:finiteTdoubleocc} that the double occupancy decreases
precipitously only for $\beta<1$ which is consistent with the value of
$1/|U|=1/3.5$ in our simulation. Similar behavior was found for the
one-dimensional system \cite{us2010}.

Note in Fig.~\ref{fig:finiteTdoubleocc} that the double occupancy
increases just before it drops signaling the breaking of the
pairs. This increase can be understood physically as follows. As the
temperature is increased, the Fermi distribution near the Fermi
momentum gets rounded but for $|{\vec k}|< |{\vec k}_{F}|$ the
distribution remains saturated. This means that pairing can happen
only near the Fermi surface while inside the Fermi sea the particles
are still blocked by the Pauli exclusion principle. Eventually, as $T$
continues to increase, the occupation of momentum states inside the
Fermi sea drops, rather suddenly as shown by our simulations, which
makes available for pairing a larger number of particles causing the
double occupancy to rise.

\begin{figure}[!htb]
\begin{center}
 \includegraphics[width=0.4\textwidth,clip]{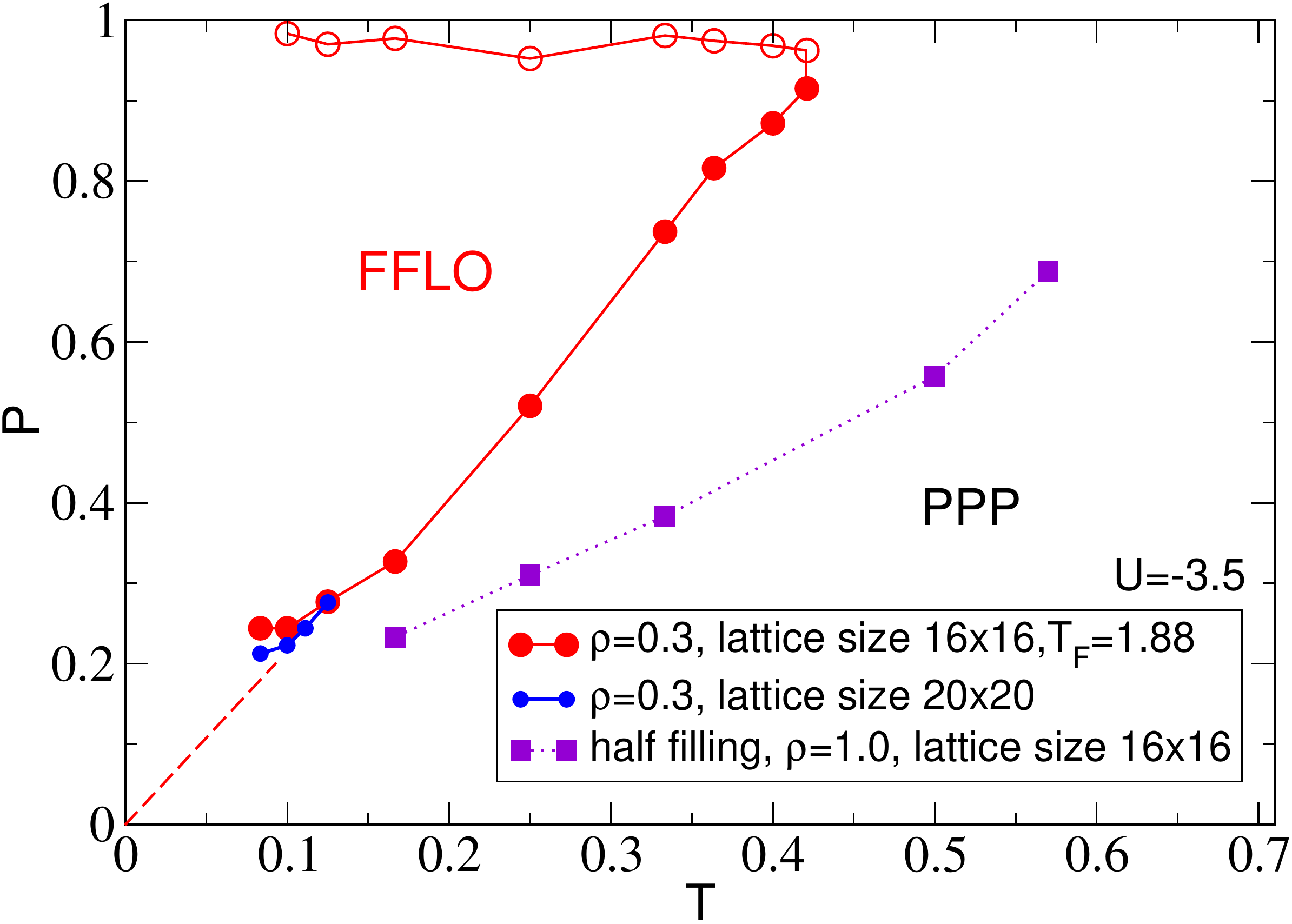}
\end{center}
\caption{\label{fig:phasediagram}(Color Online). Finite
temperature phase
    diagram of the system at $\rho=0.3$ }
\end{figure}

The phase diagram is mapped by fixing the polarization, $P$, and
increasing $T$ until the peak in the pair momentum distribution shifts
to zero momentum (inset Fig.~\ref{fig:finiteTdoubleocc}). The phase
diagram for $\rho=\rho_1+\rho_2=0.3$ (circles) and $\rho=1$ (squares)
is shown in Fig.~\ref{fig:phasediagram}. The solid circles show the
boundary of the FFLO phase; the open circles indicate the largest $P$
at which we were able to study the system. Up to these high
polarizations the system remained in the FFLO phase. The FFLO phase
boundary at low $P$ appears to extrapolate to $P\neq0$ as $T\to 0$ for
the $16\times 16$ system. However, this is an effect of the coarseness
of the lattice grid. As $P$ decreases, the peak in the pair momentum
distribution falls between $0$ and $2\pi/L$ and gives the impression
of peaking at zero momentum. The solid triangles show the phase
boundary for a $20\times 20$ system; we see that effect is corrected
for a while, but then even larger systems are needed. This is not
possible because as $T$ decreases the sign problem becomes too
severe. We believe that as soon as the system is polarized it goes
into the FFLO phase if $T$ is low enough. The long dashed line
connecting this FFLO boundary to the origin schematizes this. Outside
the FFLO phase the system is in the polarized paired phase (PPP)
since the pairs are still formed and break only at higher $T$ than
shown in the figure. The squares in Fig.~\ref{fig:phasediagram} show
the phase boundary at these temperatures for the case of $\rho=1$
(discussed below).

It is important to emphasize here that, in our discussion, the FFLO
state is characterized by the behavior of the pair momentum
distribution: If the peak is at non-zero momentum the system is in the
FFLO phase. The question naturally arises as to whether the FFLO pairs
have phase coherence and are, consequently, superfluid. In the
balanced case, the phase diagram in the temperature versus filling
plane was determined for $U=-4$ in Ref.\cite{rtsbalanced}. By studying
the pairing susceptibility as a function of $T$ as in
Ref.\cite{rtsbalanced}, we find that in the balanced case of our
system with $U=-3.5$, the critical temperature is $T_c\approx 0.1$ in
good agreement with the $U=-4$ results \cite{rtsbalanced}. However,
studying the same pairing susceptibility in the polarized case showed
no sign of s-wave superfluidity in the temperature range attainable by
QMC. Our numerical results suppport approximate analytic results which
indicate that polarization may suppress superfluidity in the FFLO
phase \cite{tempere_09}. It is, therefore, currently not clear if
when $T$ is reduced even further, the FFLO phase will become
superfluid. We note, however, that the current focus of most experimental measures of FFLO is the same non-zero momentum peak on which our simulations concentrate.

The phase diagram, Fig.~\ref{fig:phasediagram}, resembles the one
found in one dimension \cite{us2010} and shows that FFLO is very
robust.  The Fermi temperature is calculated as usual by considering a
balanced ideal system and gives for $\rho=0.3$ a value
$T_F=1.88t$. The FFLO phase at high $P$ survives up to $T=0.2T_F$
while in one dimension \cite{us2010} at $\rho=0.25$, FFLO survives up
to $T=0.8T_F$ at high $P$. So, while FFLO is still robust in two
dimensions, it is more easily destroyed by finite $T$. This is
important to keep in mind in experiments.


In a two-dimensional lattice, the Fermi surface geometry evolves with
the filling from closed, rotationally symmetric surfaces for low
filling to a square at half filling to open surfaces for higher
filling. Consequently, pairing at finite momentum occurs with
different symmetries depending on the filling. The pairs form with
equal probability in all radial directions in the case of low filling
while they form in preferred directions when the Fermi surfaces are
anisotropic.

As discussed in the introduction, there are claims that around the Van
Hove singularity the FFLO pairing could be enhanced due to increased
nesting. Indeed we observe that FFLO is stable over a wider range of
temperatures and polarizations for $\rho=1$. The squares in
Fig.~\ref{fig:phasediagram} show the FFLO-PPP boundary in the half
filled case. It is seen that the FFLO phase persists to higher $T$
than the low density case. However, when compared to $T_F=6.28t$, FFLO
is destroyed for $T\approx 0.08T_F$ as compared with $T\approx 0.2T_F$
for the half filled case in one dimension.

\begin{figure}[!htb]
\includegraphics[width=0.235\textwidth,clip,angle=0]{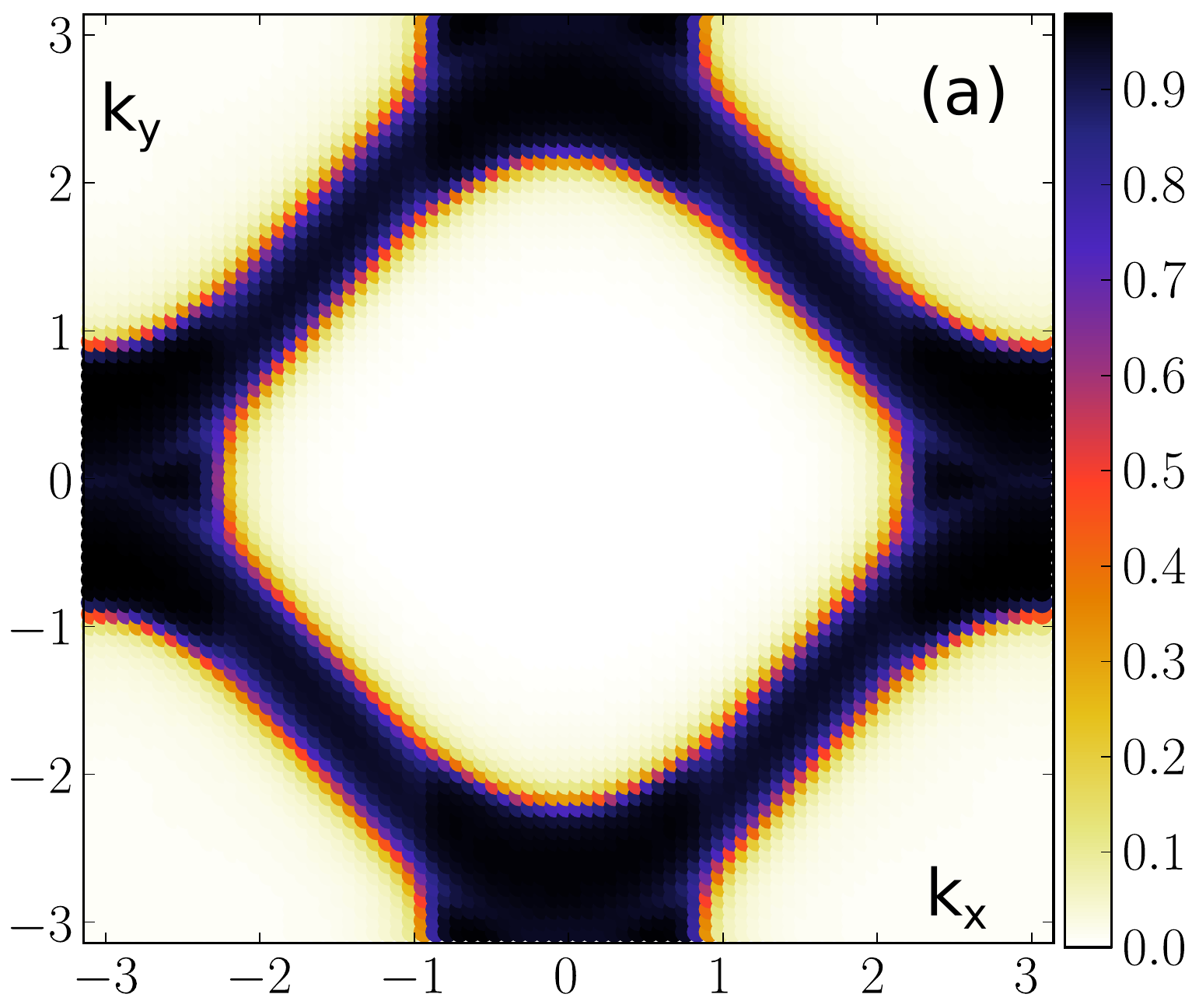}
\includegraphics[width=0.235\textwidth,clip,angle=0]{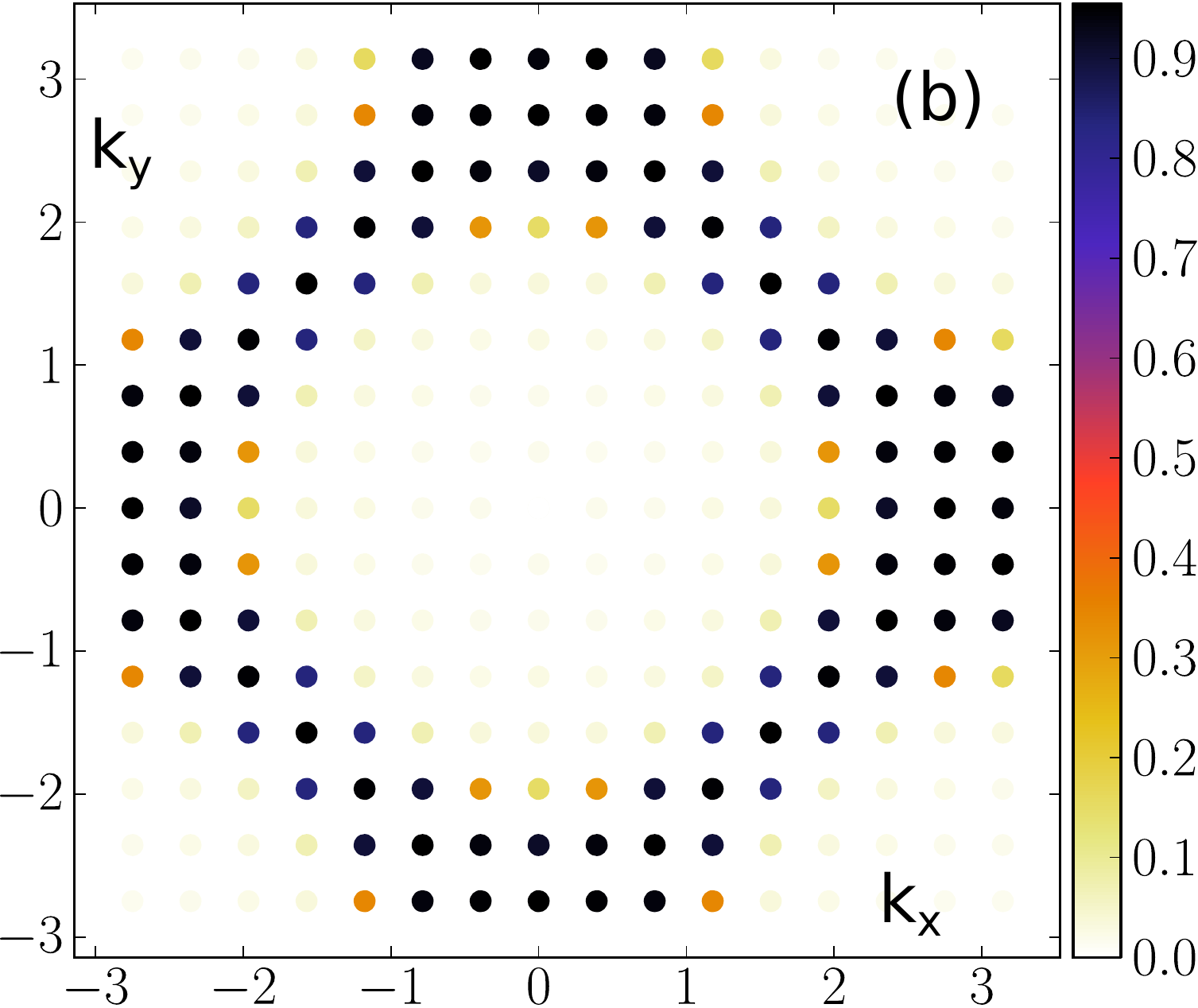}
\includegraphics[width=0.235\textwidth,clip,angle=0]{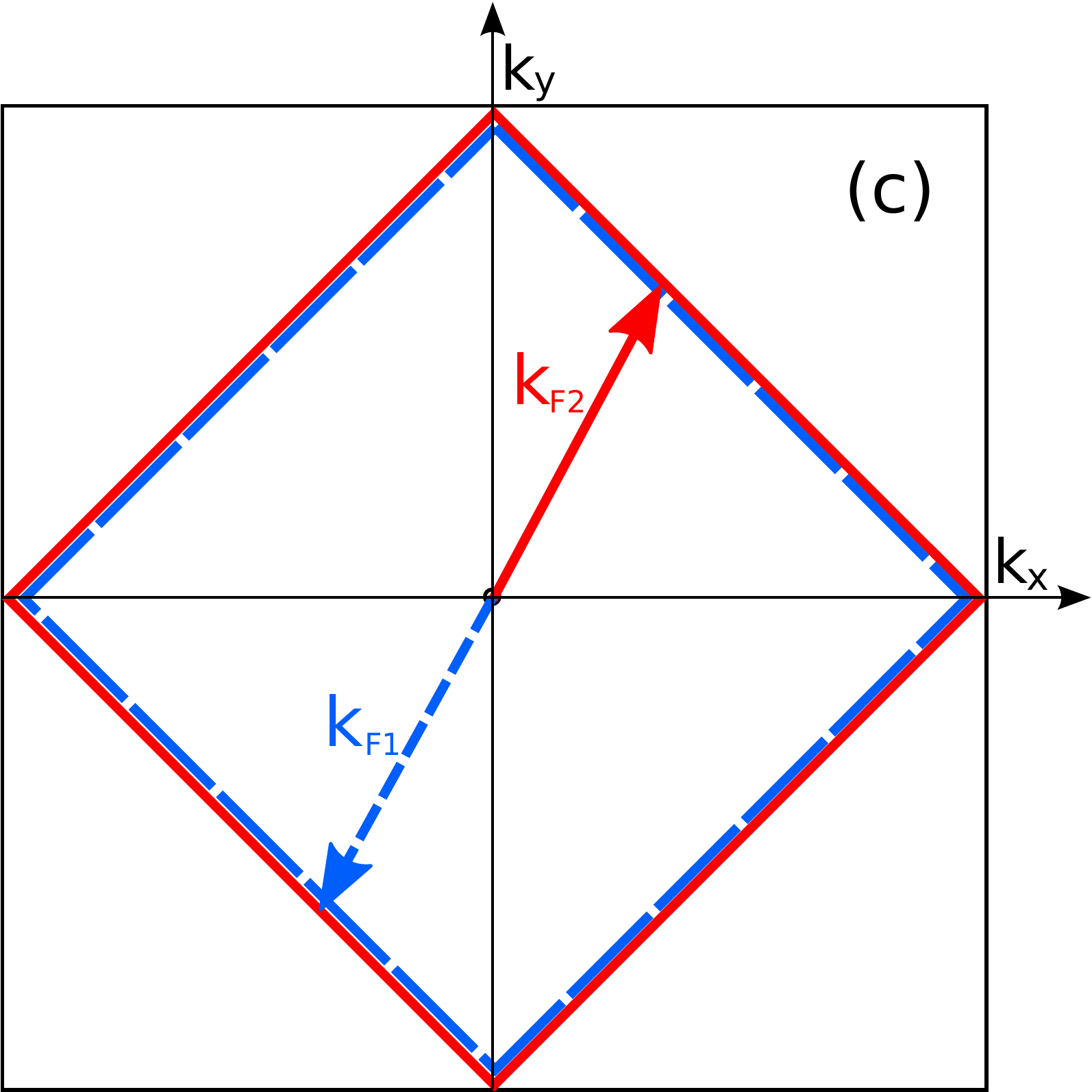}
\includegraphics[width=0.235\textwidth,clip,angle=0]{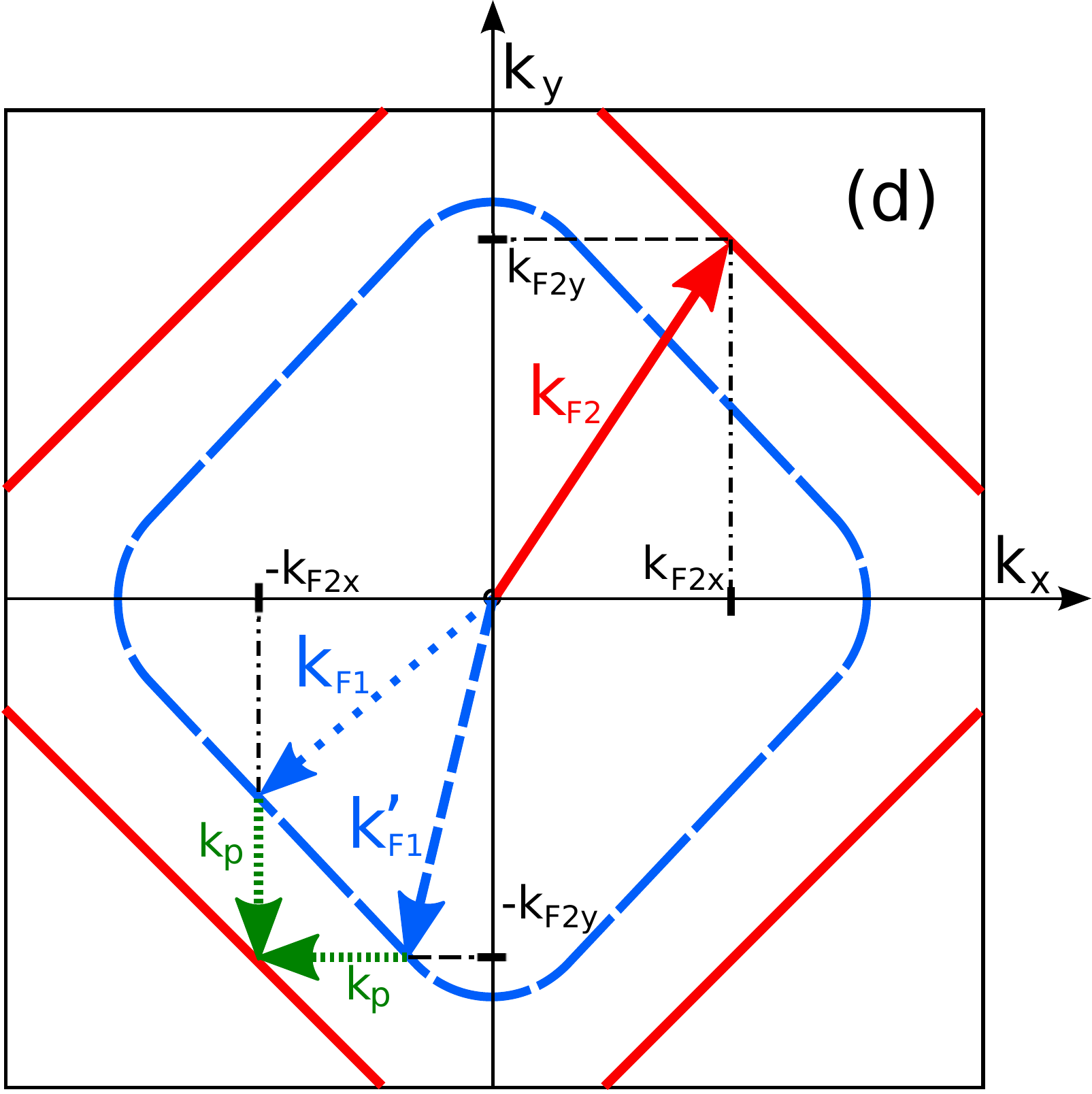}
\caption{(Color online)\label{fig:pairingschematic} Top row:
  difference in the momentum distributions of majority and minority,
  showing parallel Fermi surfaces from the Mean-Field method (a) and
  from QMC (b).  Bottom row: pairing schematic for balanced (c) and
  imbalanced (d).  In the situation when the populations of fermionic
  species are imbalanced (diagram on the right) a particle from the
  majority species forms a pair with a particle from the minority
  whose Fermi momentum either matches the $k_x$ or $k_y$ coordinate of
  the majority particle Fermi vector. The pair formed has a finite
  momentum equal to the distance of the two Fermi surfaces either
  along $k_x$ or $k_y$.  }
\end{figure}
\begin{figure}[!htb]
\begin{center}
\includegraphics[width=0.4\textwidth,clip,angle=0]{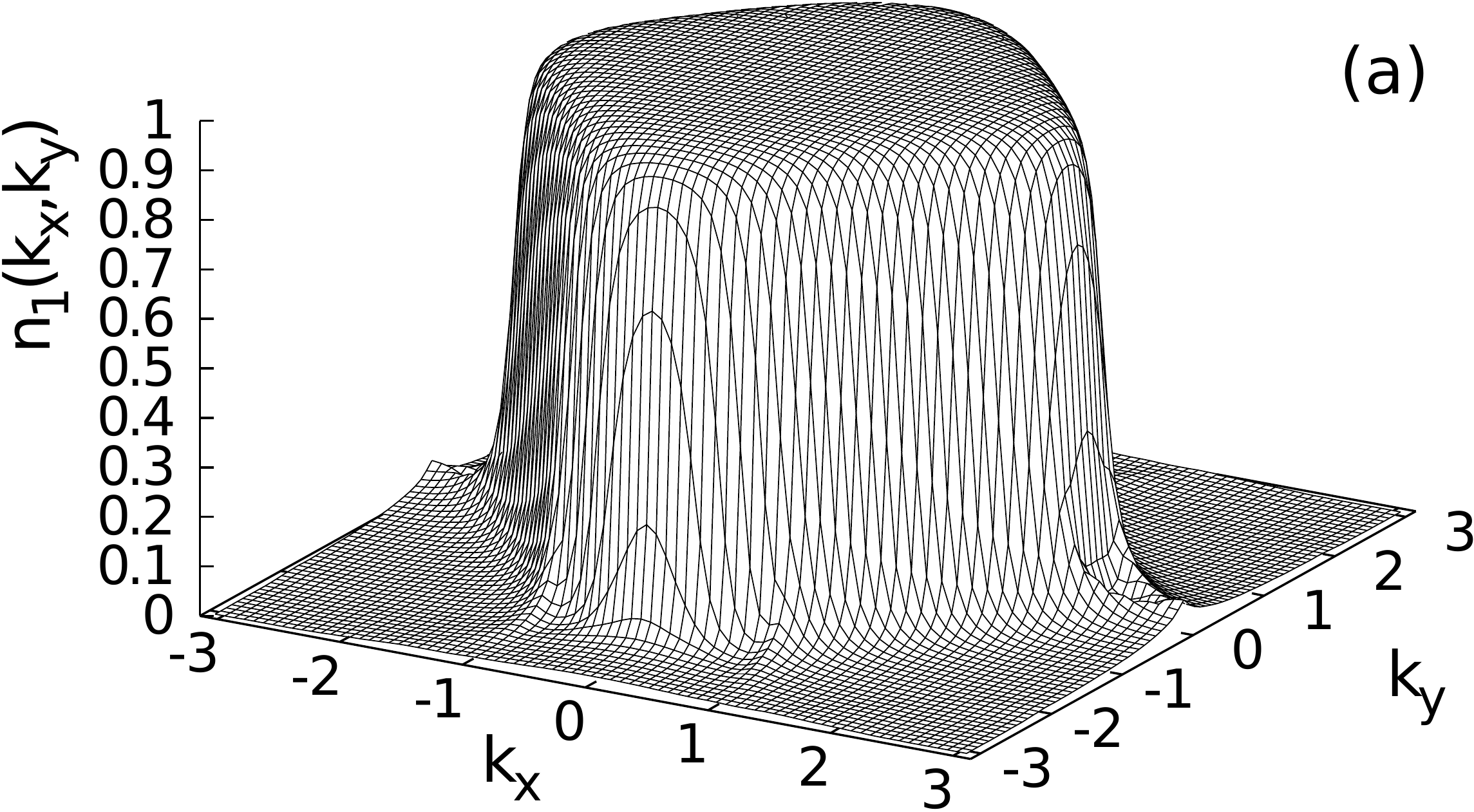}
\includegraphics[width=0.4\textwidth,clip,angle=0]{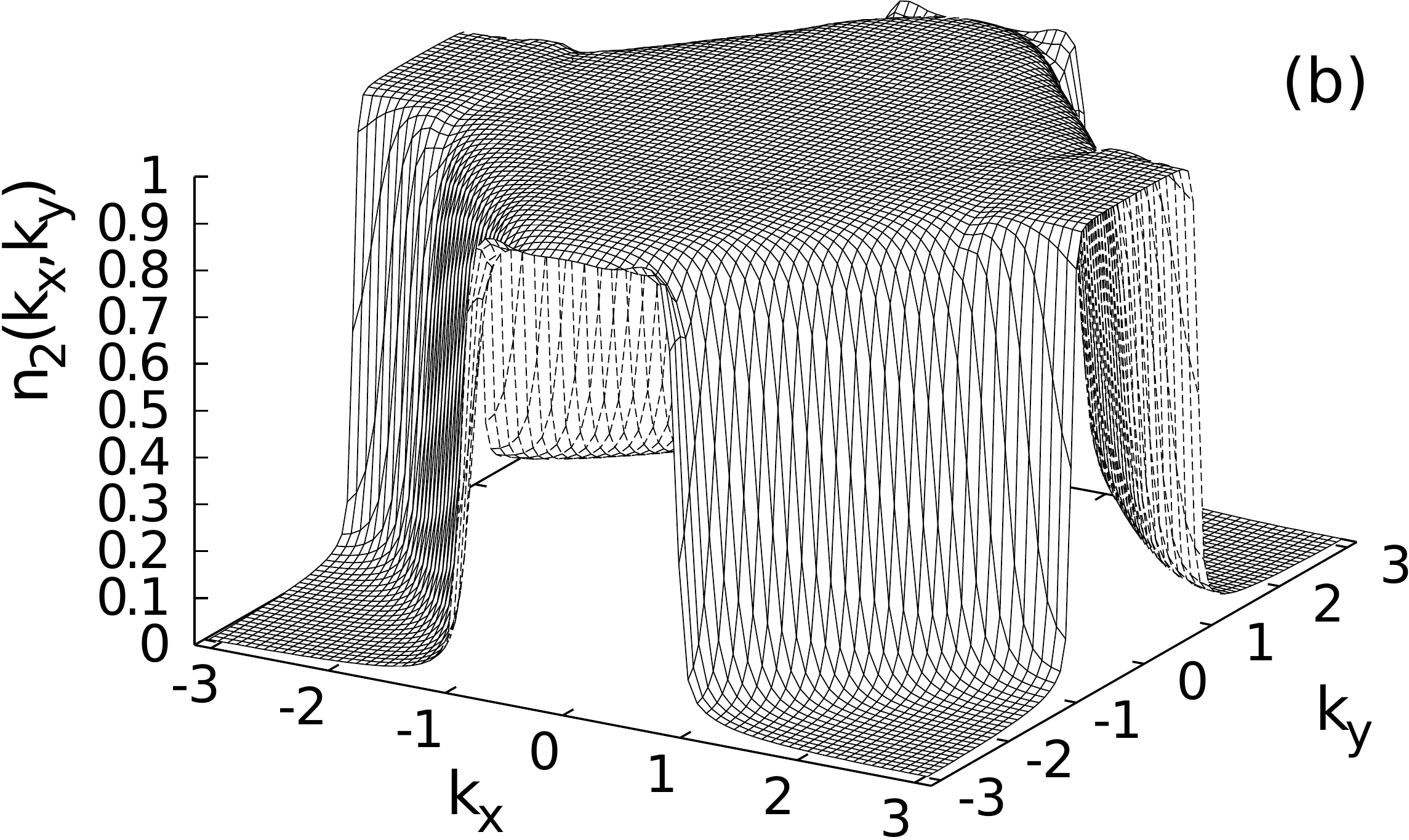}
\includegraphics[width=0.4\textwidth,clip,angle=0]{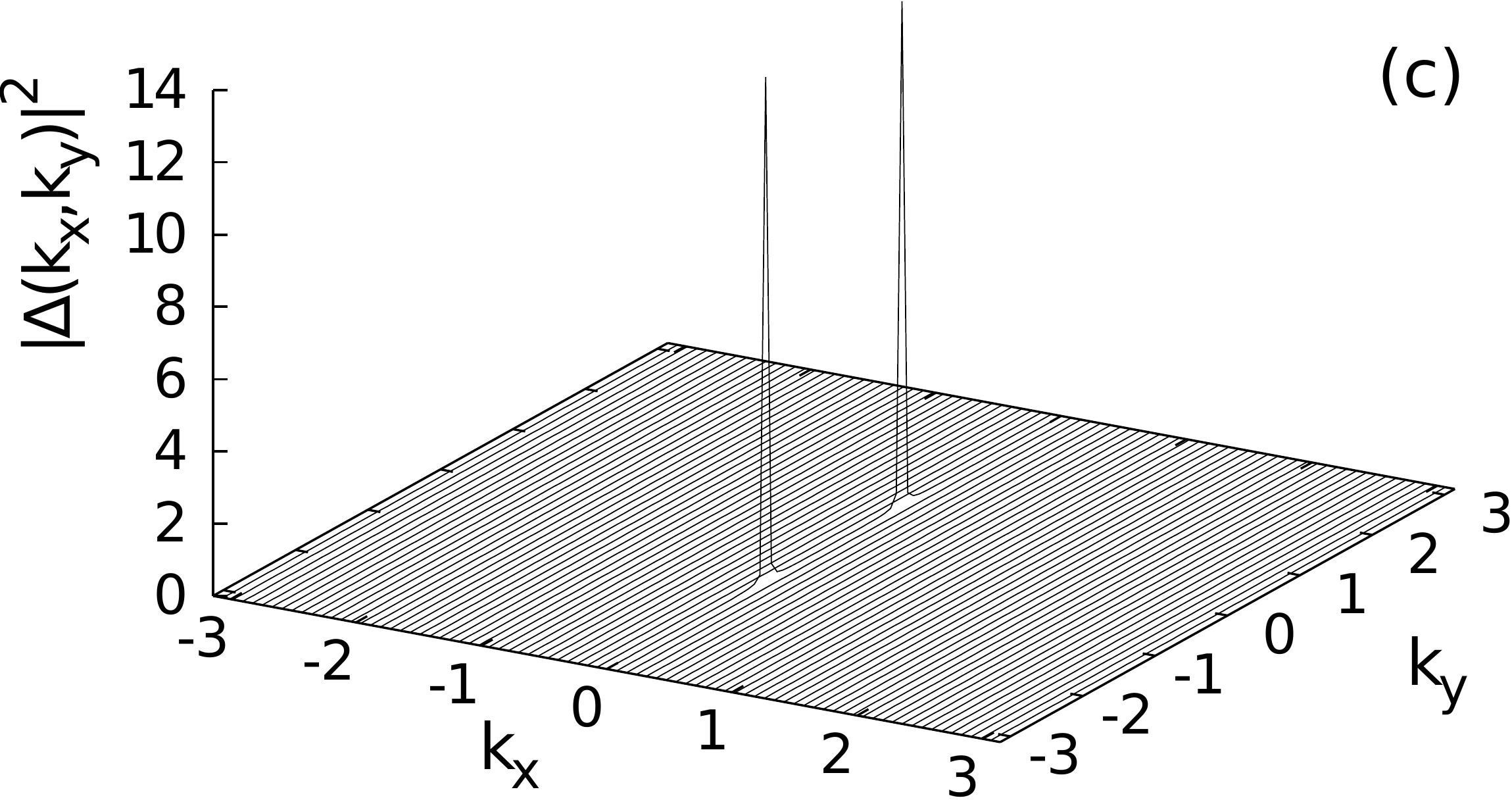}
\end{center}
\caption{\label{fig:MFHF} Momentum distributions of (a) minority, (b)
  majority and and (c) order parameter in k-space calculated using the
  Mean-Field method.  $\rho=\rho_1+\rho_2=1$.  Here $P=0.32$,
  $\beta=25$, $U=-3.5t$ and the lattice size is $79\times 79$. The
  pairing peaks are symmetric along $k_x$ or $k_y$ depending on the
  realization. }
\end{figure}

\begin{figure}[!htb]
\begin{center}
\includegraphics[width=0.35\textwidth,clip]{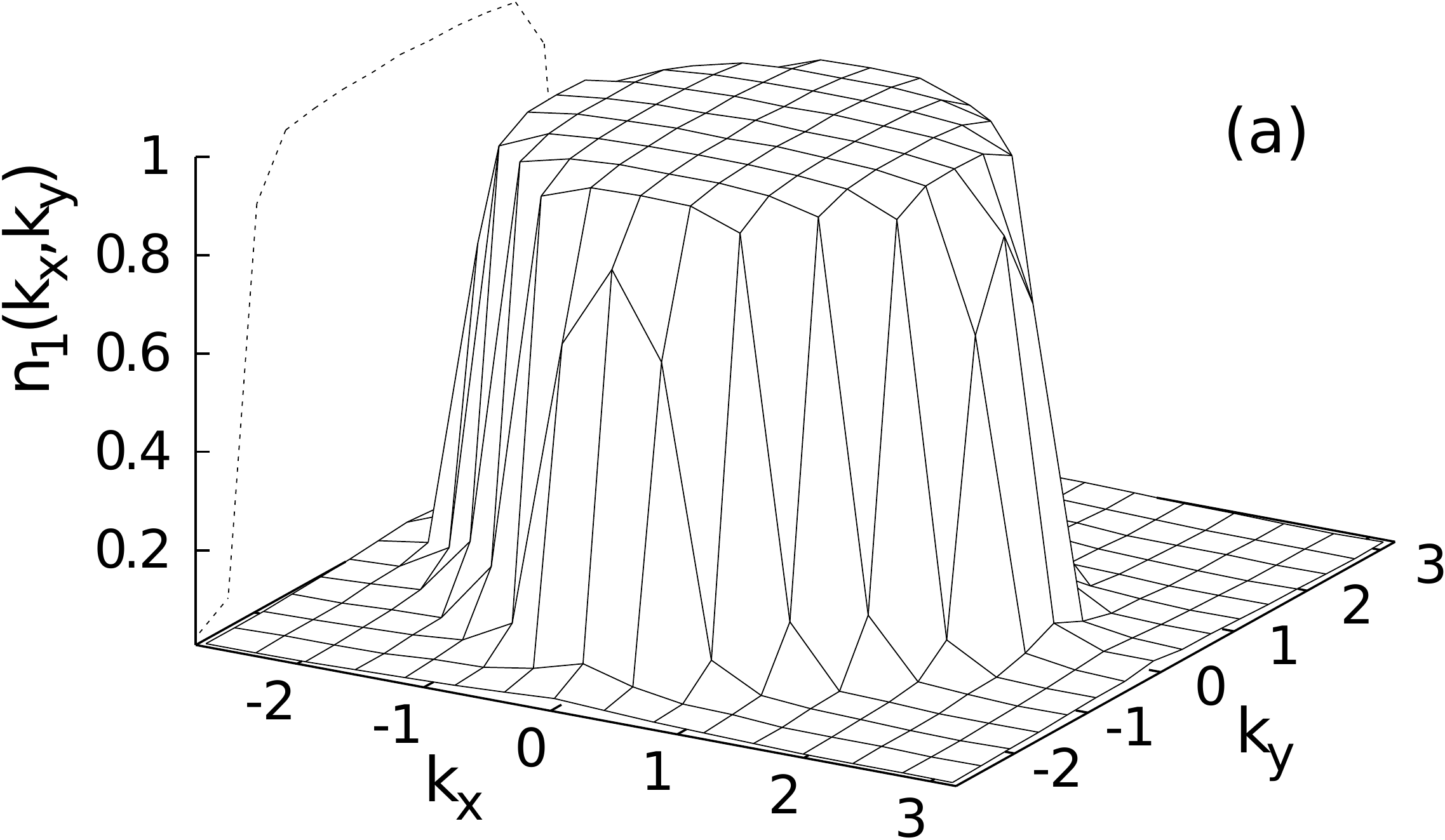}
\includegraphics[width=0.35\textwidth,clip]{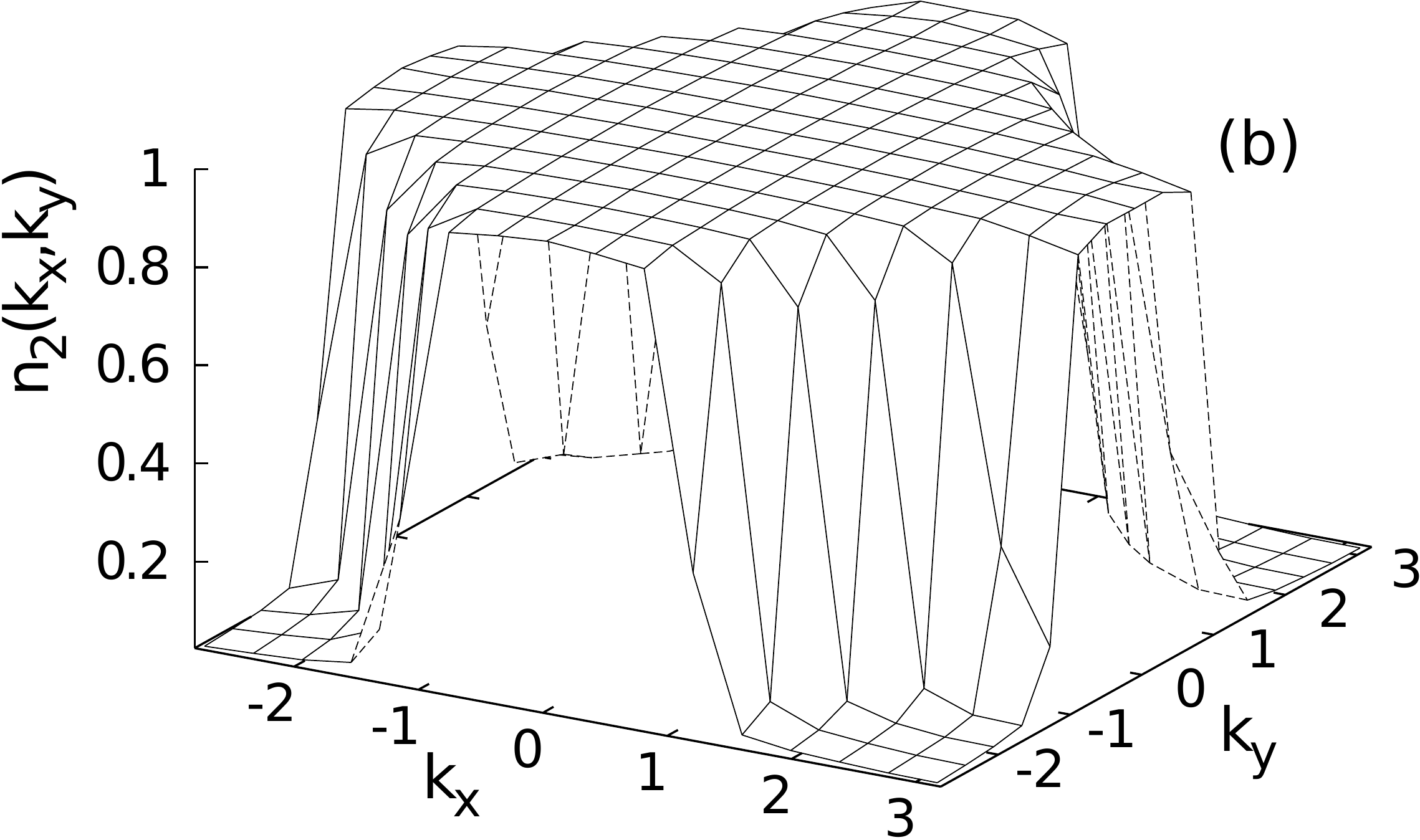}
\includegraphics[width=0.35\textwidth,clip]{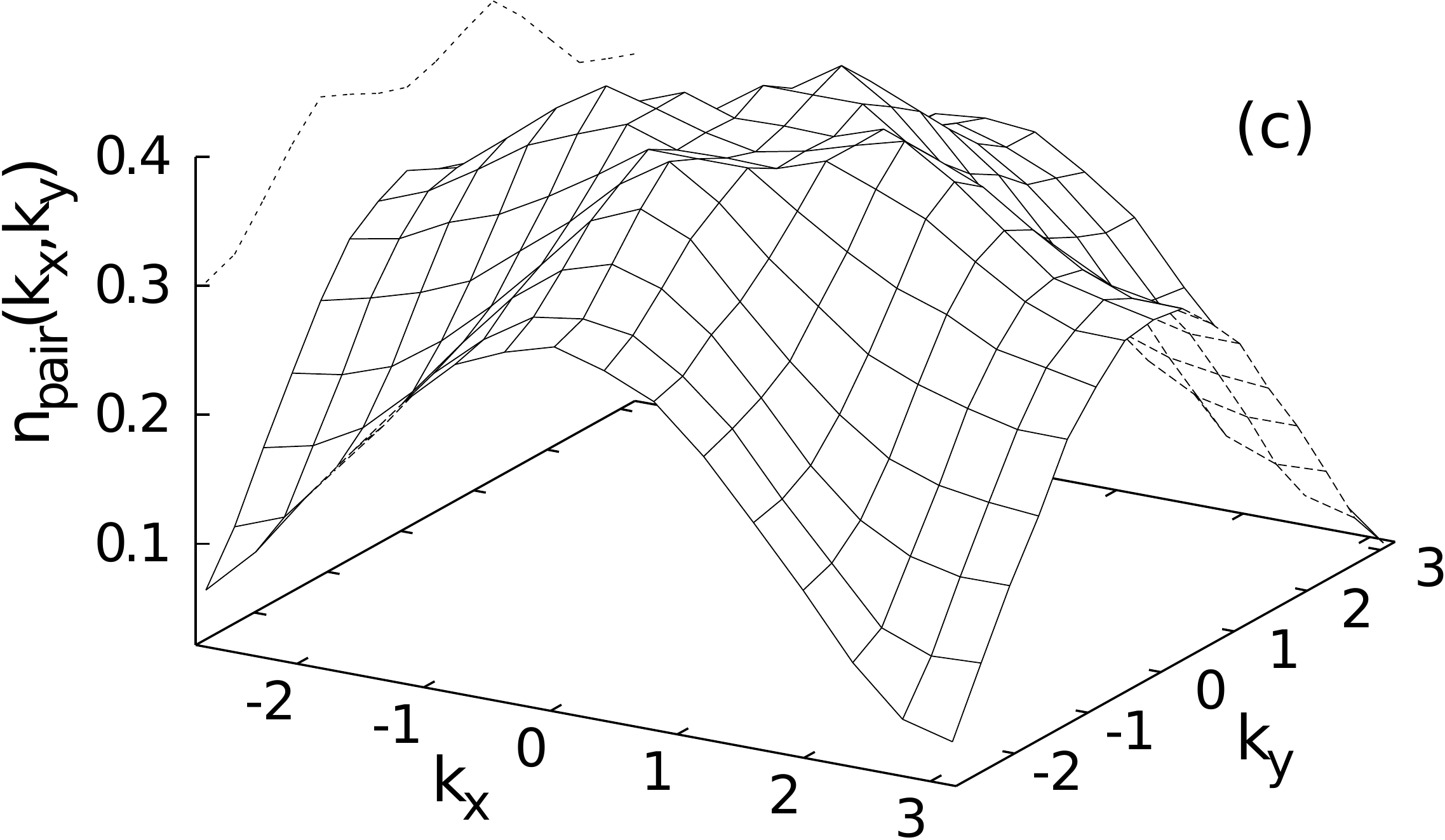}
\end{center}
\caption{\label{fig:QMCHF} Momentum distributions of (a) minority, (b)
  majority and (c) pairs at $\rho=\rho_1+\rho_2=1$, obtained from QMC.
  Here $P=0.38$, $\beta=10$, $U=-3.5t$ and the lattice size is
  $16\times 16$. The pair momentum distribution depicts four peaks
  along the $k_x$ and $k_y$ axis. }
\end{figure}

When the populations are imbalanced around half filling of the
lattice, one can readily see the effect of the interaction on the
Fermi surfaces in Fig.~\ref{fig:pairingschematic}, depicting the
difference between the Fermi distributions of the species calculated
using both mean field and QMC methods: they almost look like nested
squares parallel to each other in most of the momentum states, whereas
the non-interacting ones would look more rounded and not as parallel.
Similar Fermi surface geometry in the context of LO states in 3D have
been shown in \cite{Trivedi}.

The reason why the system exhibits such Fermi surfaces can be
understood as follows.  If we look at the region of $k_x>0$ we can
parametrize the linear part of the majority Fermi line as
$k^{+}_{f,2}(k_x)=-k_x+\alpha_2$ for positive values and
$k^{-}_{f,2}(k_x)=k_x-\alpha_2$ for negative values and doing the same
for the minority we have $k^{+}_{f,1}(k_x)=-k_x+\alpha_1$ and
$k^{-}_{f,1}(k_x)=k_x-\alpha_1$.  (see
Fig.~\ref{fig:pairingschematic}).  Pairing happens here for a given
$k_x$ between the upper part of the majority branch and the lower part
of the minority branch ($k^{+}_{f2}(k_x)$ pairs with
$k^{-}_{f1}(-k_x)$).  The momentum of the pair along y is the sum of
these momenta and is equal to
$q_y=k^{+}_{f2}(k_x)+k^{-}_{f1}(-k_x)=\alpha_2-\alpha_1$.  Therefore,
thanks to the parallel Fermi lines, the pairing momentum is
independent of $k_x$, leading to a strong enhancement of the pairing
efficiency. The same construction can be done in the $k_x$ direction,
matching the y coordinate of the momentum vectors and the pairs will
be moving along $x$ with $\pm q_x$.  In other words, for each $k_x$,
we have, along $k_y$, the usual imbalanced 1D situation, \textit{i.e.}
two rectangular Fermi distributions, with different Fermi
momenta. Again, the crucial point is both the majority and minority
effective 1D Fermi momentum values change the same way with $k_x$: the
two Fermi surfaces remain always at the same distance from each other.
This pairing mechanism is illustrated in
Fig.~\ref{fig:pairingschematic}d. The excess fermions
correspond to the part of the majority Fermi surface which can't be
paired this way, \textit{i.e.} the four regions around
$(k_x=0,ky=\pm\pi)$ and $(k_x=\pm\pi,ky=0)$.  Note that in the
balanced case, this corresponds to the usual BCS pairing on a
lattice: a particle of one species from the Fermi surface can form a
pair with a particle from the other species with the Fermi vector of
equal length but opposite direction (as shown in
Fig.~\ref{fig:pairingschematic}c).  The resulting pair has,
as expected, a zero center-of-mass momentum. The pairing along $k_x$ and $k_y$ might not seem the
most intuitive scenario, since one can imagine the pairs forming with
momentum along the diagonal with smaller $|\vec{k}_p|$. Since this
pairing was not observed in any of our simulations, this probably
means that, in a mean field approach, it only corresponds to a local
minimum of the free energy. However, since the shape of Fermi surfaces
is affected by the nature of the pairing, one can not directly compare
both situations from the present results and a more detailled study is
needed, which is beyond the scope of this paper. On the contrary, the
mean-field simulations show sharp peaks either along $k_x$ or $k_y$
depending on the realization (see Fig.~\ref{fig:MFHF}) and in the
Quantum-Monte Carlo simulations, since we average over all
realizations, we see that the pair momentum distribution exhibits four
peaks: two along $k_x$ and two along $k_y$ (see Fig.~\ref{fig:QMCHF}).
It is important to notice a very good agreement between the results
obtained by MF and QMC methods. Finally, we have also observed, as
expected, that the value of the position of the peaks, \textit{i.e.}
the center of mass of the pairs, increases with large population
imbalance.


\section{Harmonically confined system}

One is used to describing free fermions on a lattice using intuition
built on the free electron model.  Each particle occupies a state with
particular momentum $\vec{k}$ and at T=0 the filled state with the
highest $\vec{k}$ is called the Fermi level. BCS pairing mechanism is
understood as pairing between fermions from the Fermi surface with
opposite spins and opposite momenta. In this description the FFLO
pairing model predicts forming a pair of fermions from different spin
species with a finite momentum, where the momentum of the pair is the
difference of the Fermi momenta of each involved fermion.  When we
turn to study a harmonically confined system at low filling, for which only few harmonic levels are actually
filled, the translationally invariant momentum space description is no longer the obvious one. An ideal gas confined in a
harmonic trap is known to be fully characterized by the basis formed
by harmonic oscillator wave functions. In addition, for low fillings
of the lattice, only the bottom of the band structure will be
filled. Then the kinetic part of the Hamiltonian is well described by
the free particle one with an effective mass $m^*$ given by
$m^*=1/2a^2t$, where $a$ is the lattice spacing and $t$ the tunneling
amplitude. In the present case, setting the units $t=1$ and $a=1$, the
effective mass is therefore $m^*=1/2$.

In this section of the paper we explore the description of the
interacting system in the harmonic basis. This transformation is the
analog of the Fourier transformation used to go from real space to
momentum space in the case of the free system. We will show that both
BCS and FFLO models can be translated into the harmonic level basis as
pairing of particles between harmonic levels and look into the
limitations of this description. Since we are studying a
two-dimensional system we use the eigenstates of the two-dimensional
harmonic oscillator (see for example \cite{berge}). Due to rotational
symmetry, the $n^{\mathrm{th}}$ harmonic level is $n+1$ times
degenerate. We will use the labeling of the states as follows: $n$ is
the principal quantum number and $m=-n,-n+2,...,n$ is the orbital
angular momentum quantum number. For simplicity we will sometimes use
$\kappa$ to label the set of quantum numbers, $\kappa=(n,m)$. Taking
the normalized Harmonic oscillator wave function for a particular
level to be $\Phi_{n,m}(i)$ (where i is the lattice site) we define a
creation operator of a particle in a level as:
\begin{equation}
\Psi_{n,m}^{\dagger}=\frac{1}{\sqrt{N}}\sum_{i}\Phi^*_{n,m}(i)c_{i}^{\dagger},
\end{equation}
which, in the continuum limit, leads to properly anti-commutating
fermionic operators.
 
We calculate the single particle Green function between levels for
each species as follows:
 \begin{equation}
   G_\sigma(\kappa,\kappa') = \langle \Psi_{\kappa,\sigma}^\dagger \Psi_{\kappa',\sigma}^{\phantom\dagger} \rangle
 \end{equation}
 As pairing is our main interest of investigations we also define a
 pair Green function using the creation and annihilation operators of
 a pair of fermions. Similarly to the homogenous case where the pairs
 are formed between particles having different momenta, the pairs here
 can have constituents occupying different harmonic levels:
 \begin{equation}
   G_{pair}(\kappa,\kappa') = \langle \Psi_{\kappa',1}^\dagger \Psi_{\kappa,2}^\dagger \Psi_{\kappa,2}^{\phantom\dagger}
   \Psi_{\kappa',1}^{\phantom\dagger} \rangle
 \end{equation}

\begin{figure}[!htb]
\begin{center}
\includegraphics[width=0.40\textwidth,clip]{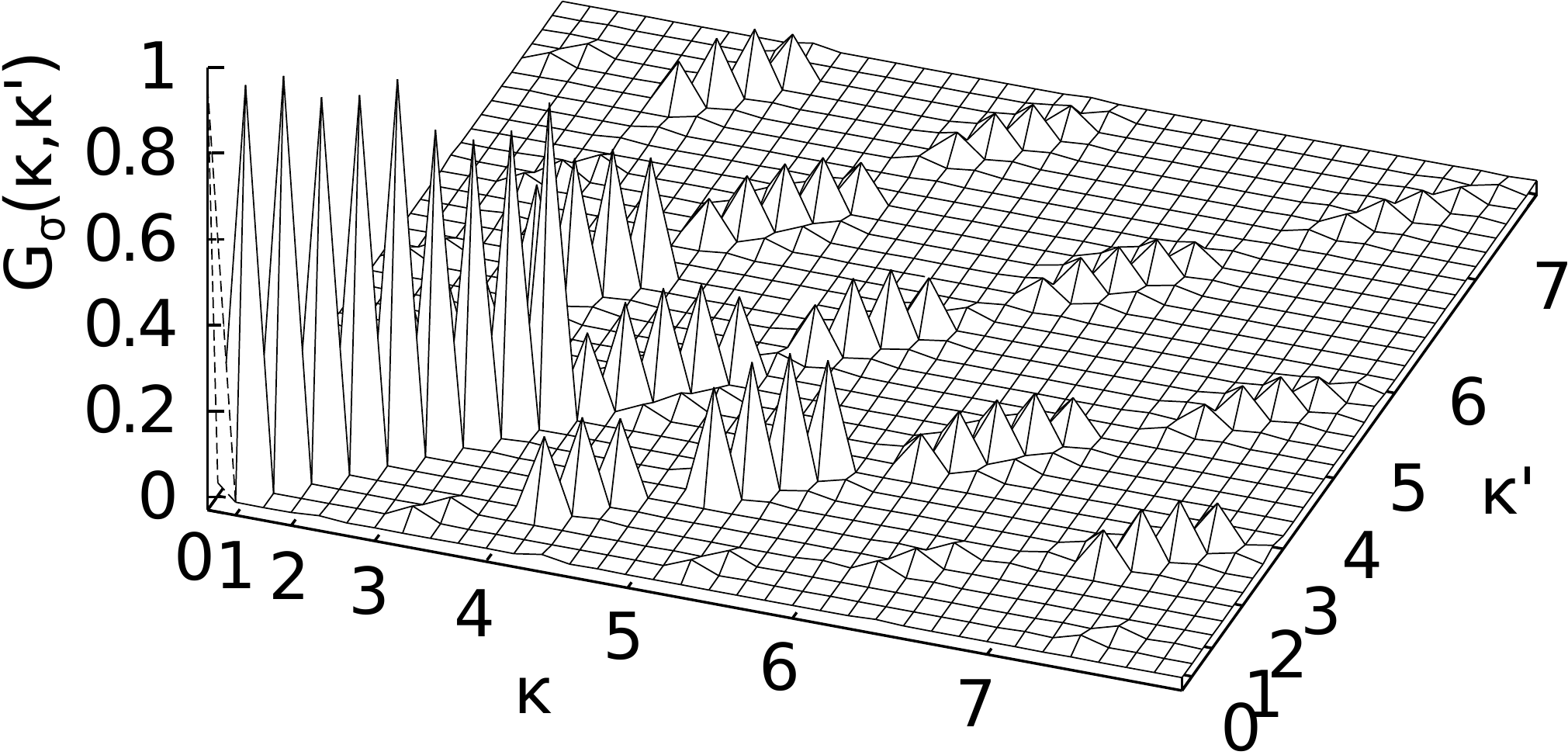}
\includegraphics[width=0.35\textwidth,clip]{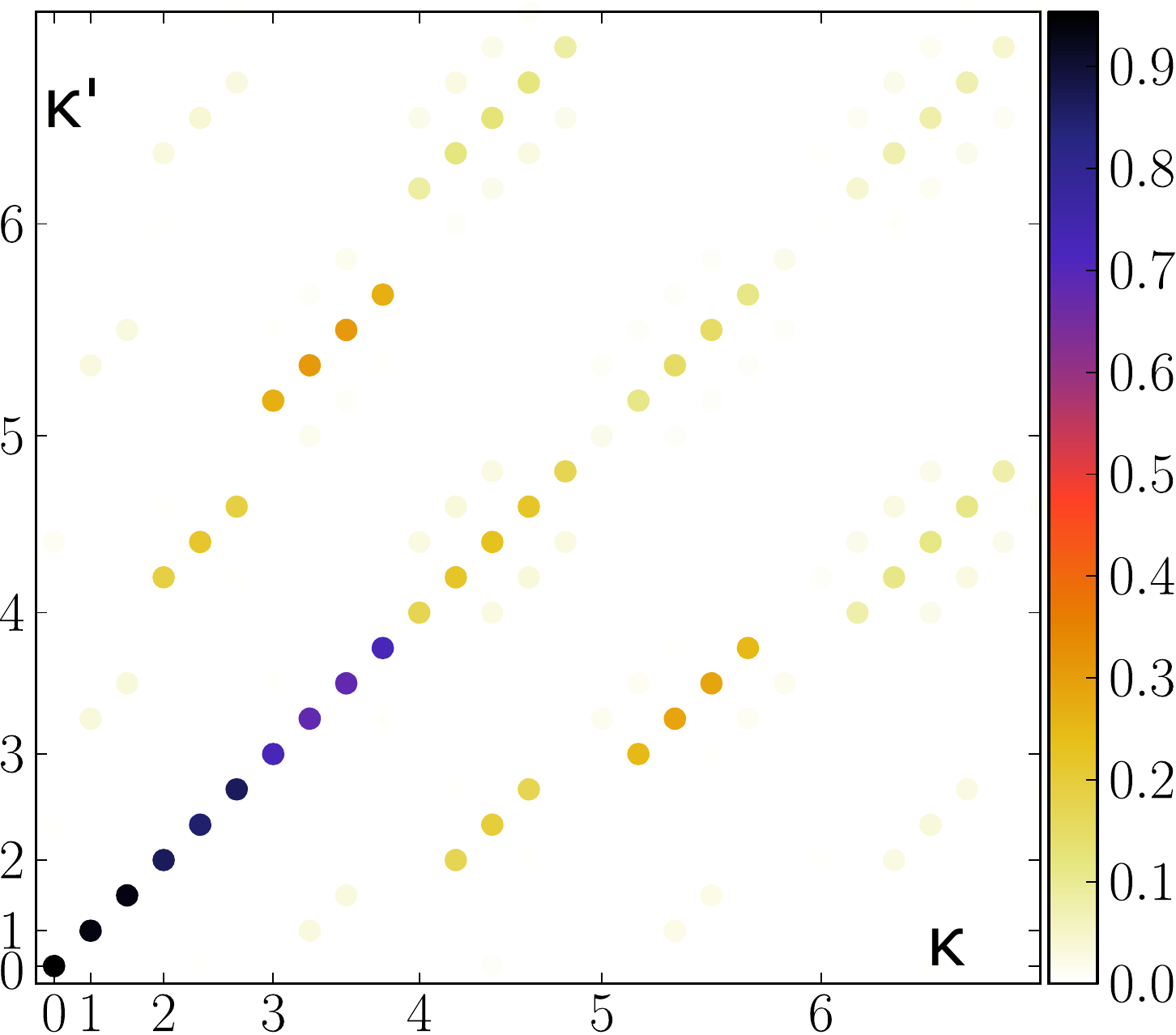}
\end{center}
\caption{\label{fig:HLsingleP=0}(Color online). Single particle Green
  function in the harmonic level basis using QMC, in the balanced
  case.  The total number of particles is $22.3$, i.e $\approx 11$
  particles per spin. As one can see, the single particle Green
  function value on the diagonal sharply drops just before the
  $5^{\mathrm{th}}$ level ($n=4$) corresponding to $10$ harmonic
  states, roughly the number of particles per spin. The off-diagonal
  elements are small compared to the diagonal ones, emphasizing the
  accuracy of the harmonic description of the system. States are
  labelled with $\kappa=(n,m)$ and only the principal quantum number
  $n$ is displayed on the $x$ and $y$ axis.}
\end{figure}
\begin{figure}[!htb]
\begin{center}
\includegraphics[width=0.40\textwidth,clip]{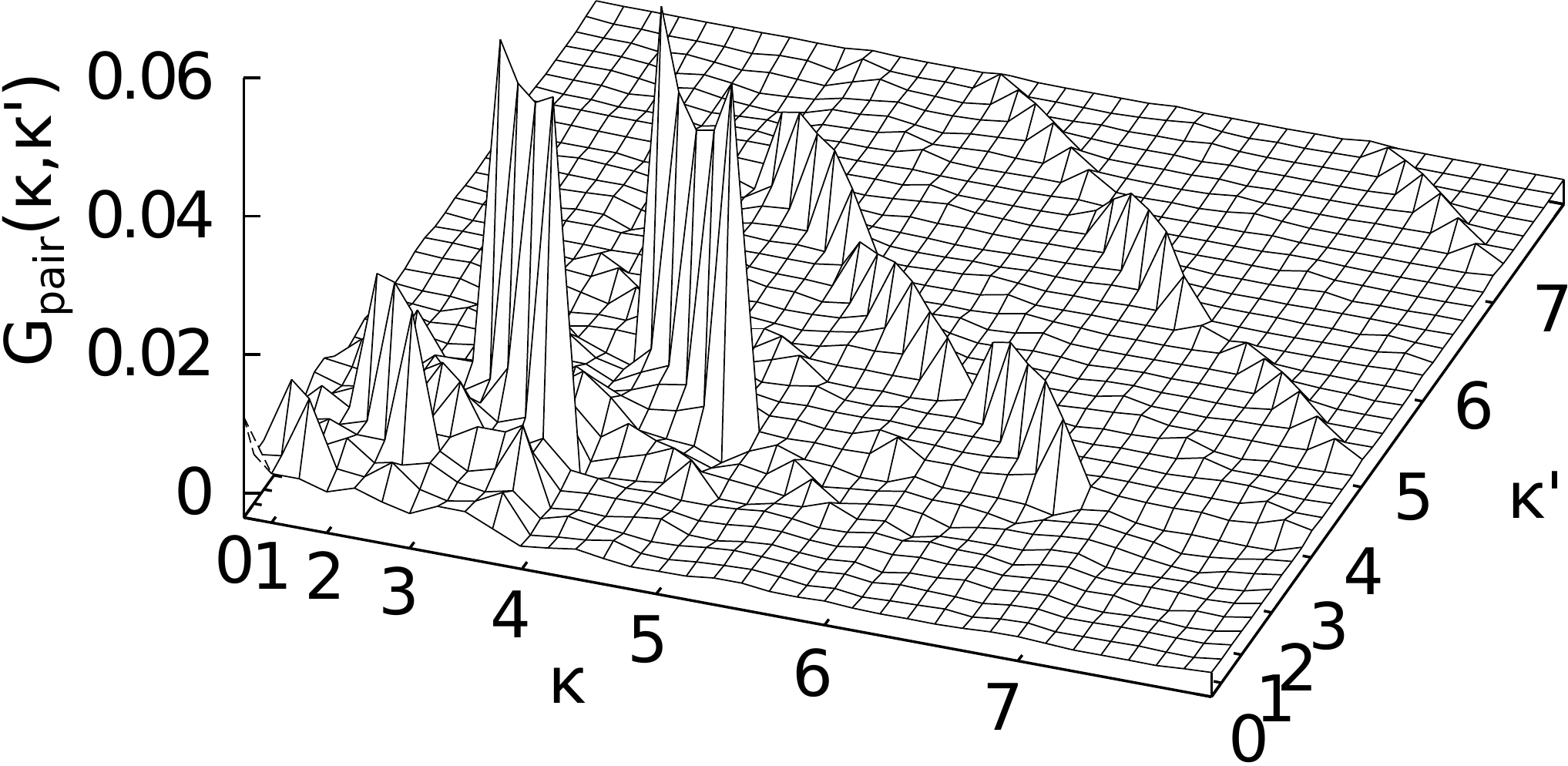}
\includegraphics[width=0.35\textwidth,clip]{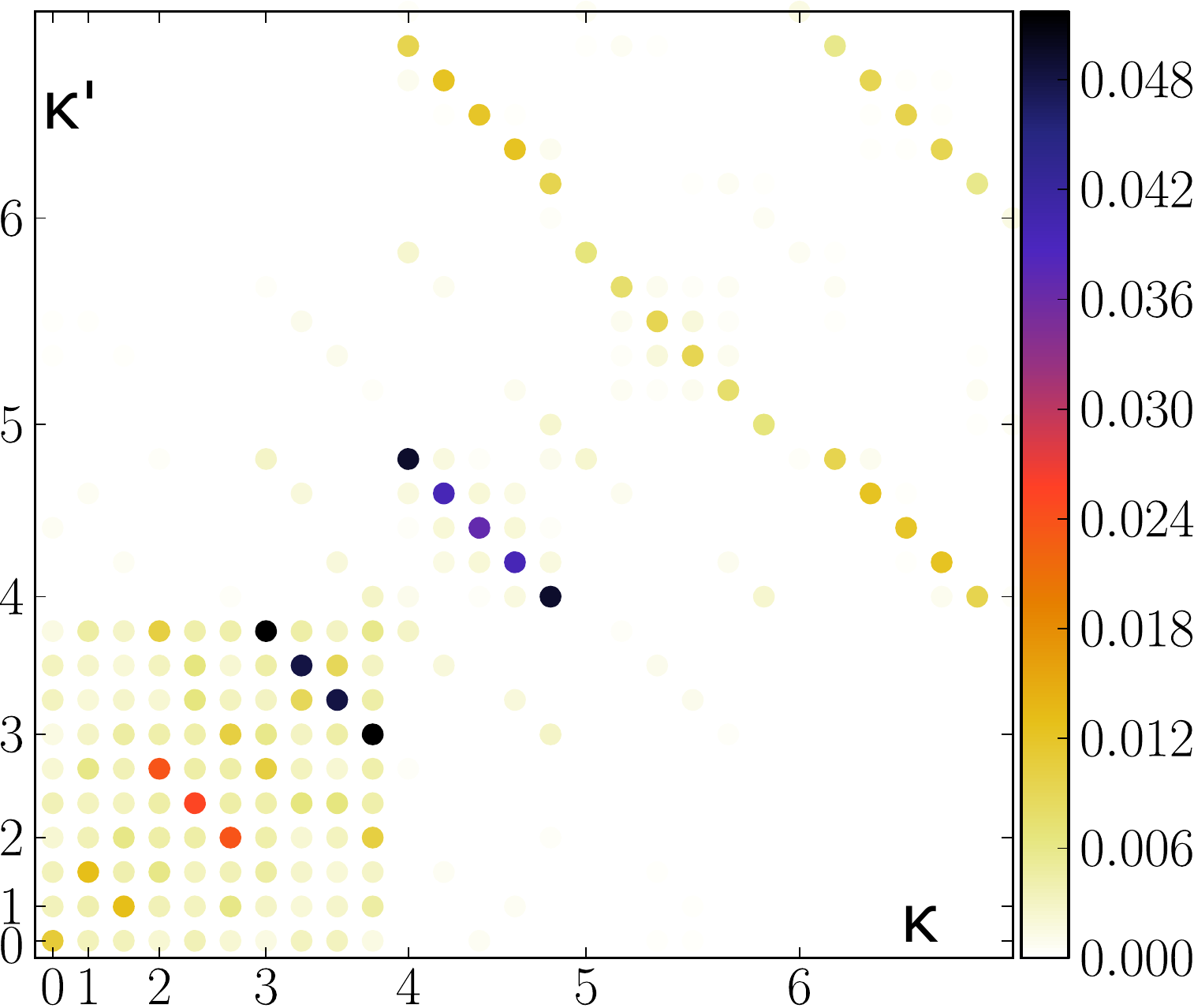}
\end{center}
\caption{\label{fig:HLpairP=0}(Color online). Pair Green function in
  the harmonic level basis using QMC.  The total number of particles
  is $22.3$ and the populations are balanced. One can clearly see that
  the pairing is maximum at the Fermi level $n=4-5$, with opposite
  magnetic quantum numbers $m$. Off-diagonal pairing, \textit{e.g.}
  between $\kappa=(6,m)$ and $\kappa'=(4,m')$, is almost
  negligible. By diagonal pairing we mean pairing between levels with
  equal principal quantum numbers.}
\end{figure}
\begin{figure}[!htb]
\begin{center}
\includegraphics[width=0.3\textwidth,clip]{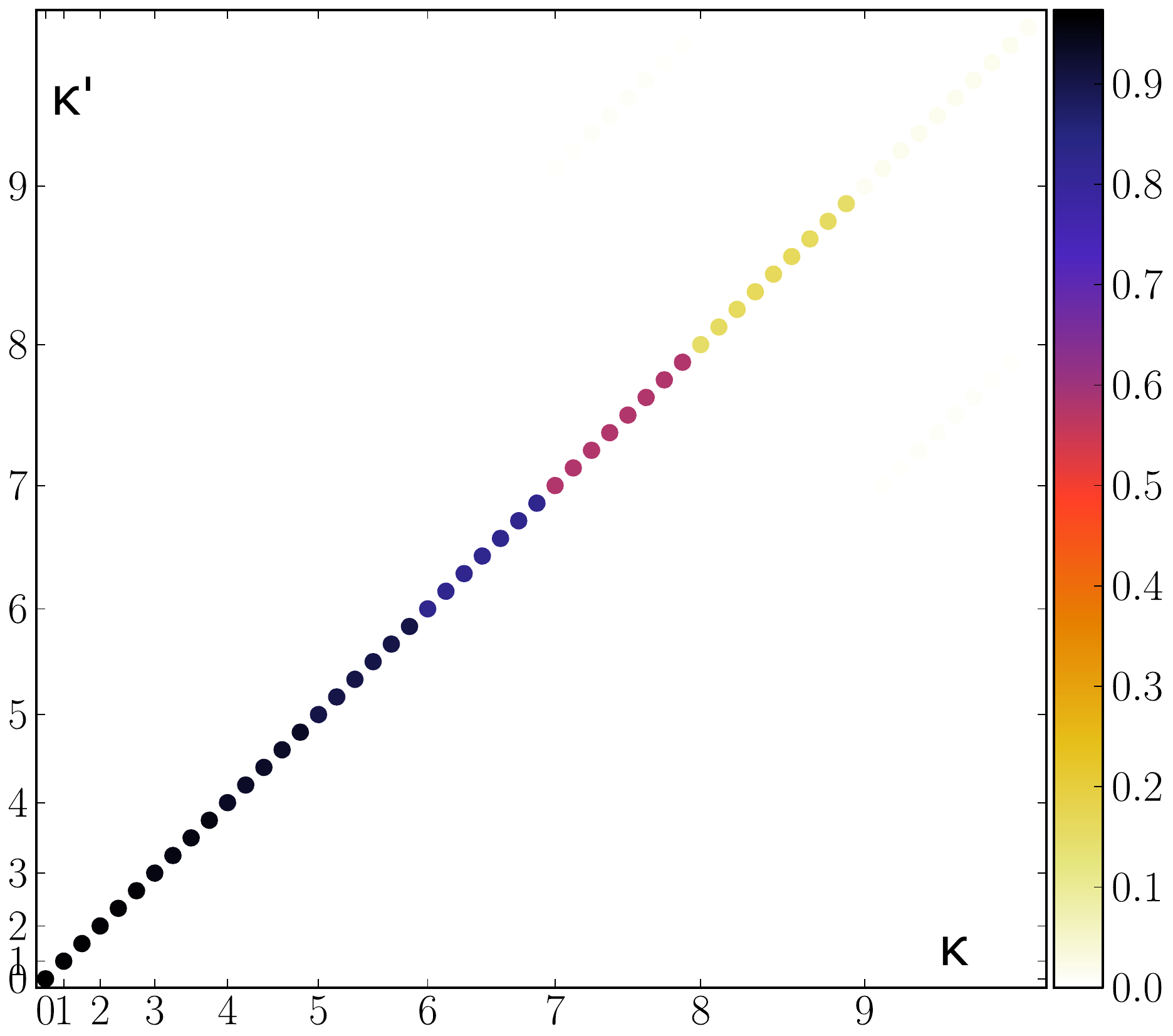}
\includegraphics[width=0.3\textwidth,clip]{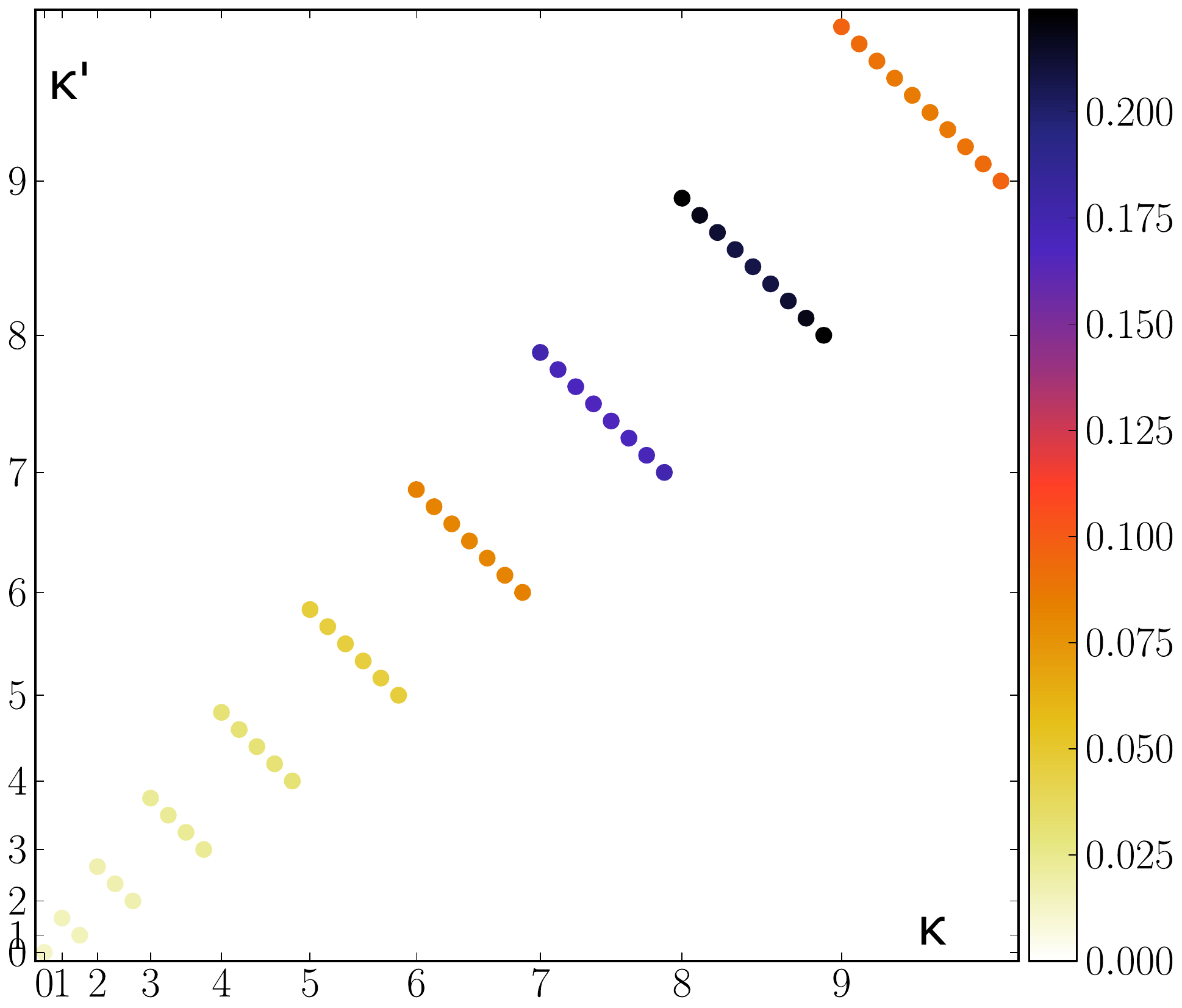}
\end{center}
\caption{\label{fig:MFHLpairP=0}(Color online).  Single particle and
  pair Green function in the harmonic level basis using MF.  Total
  number of particles is $80.4$ and the populations are balanced. As
  in Fig.\ref{fig:HLsingleP=0}, the single particle Green's function
  is diagonal, with a value equal to 1 up to the Fermi level
  ($n\approx8$) dropping to zero after.  The pair Green's function
  emphasizes the diagonal pairing $(n,m)\leftrightarrow (n,-m)$.}
\end{figure}
 
In this section we present results for these correlation functions
obtained using both QMC and MF method of balanced and polarized
systems with low filling of the lattice. All QMC results were done on
a $20 \times 20$ lattice at the inverse temperature $\beta=10$ with
interaction strength $U=-3.5$ and the trap potential $V_t=0.065$ which
translates to an effective harmonic frequency $\omega=0.5$.  The
simulations done using the Mean-Field method were performed on a
bigger lattice of 41 by 41 sites, at the inverse temperature of
$\beta=25$ and taking the interaction strength to be $U=-3.0$ In the
figures only the n values are explicitly written, but correlations are
calculated between all different n and m values. The m levels are
arranged from m=-n to n from left to right (or bottom to top).  In the
balanced case shown in Fig.~\ref{fig:HLsingleP=0} the single particle
green function is mainly diagonal which indicates that in this regime
the harmonic level basis offers a good description of the system. The
diagonal part is the occupation of levels and where it drops to zero
one can define the Fermi level. We compare these results to those
obtained using the Mean-Field method. Both single particle as well as
pair Green function shown in Fig.~\ref{fig:MFHLpairP=0} agree
qualitatively to the QMC results. The small off diagonal values in
QMC, which are not present in the MF results, stem from the exact
treatment of the interactions in QMC, not taken into account in the MF
calculations.  In the regime of much higher fillings of the lattice
(for example around half-filling), the effective mass approach is no
longer valid and the MF results show that the harmonic basis is no
longer a relevant one. We do not have any QMC results in that regime
due to the sign problem.

In the balanced population case, both Fig.~\ref{fig:HLpairP=0} (QMC)
and Fig.\ref{fig:MFHLpairP=0} (MF), emphasize that the pairing is
maximum around the Fermi level and happens between particles from
levels with the same $n$ and for opposite $m$ and $m'$ values such
that the total orbital angular momentum of the pair is 0. This
situation is similar to the free particle case, where the pairing
occurs mostly between the $+\vec{k}_F$ and $-\vec{k}_F$ states.

\begin{figure}[!hhh]
\begin{center}
\includegraphics[width=0.235\textwidth,clip]{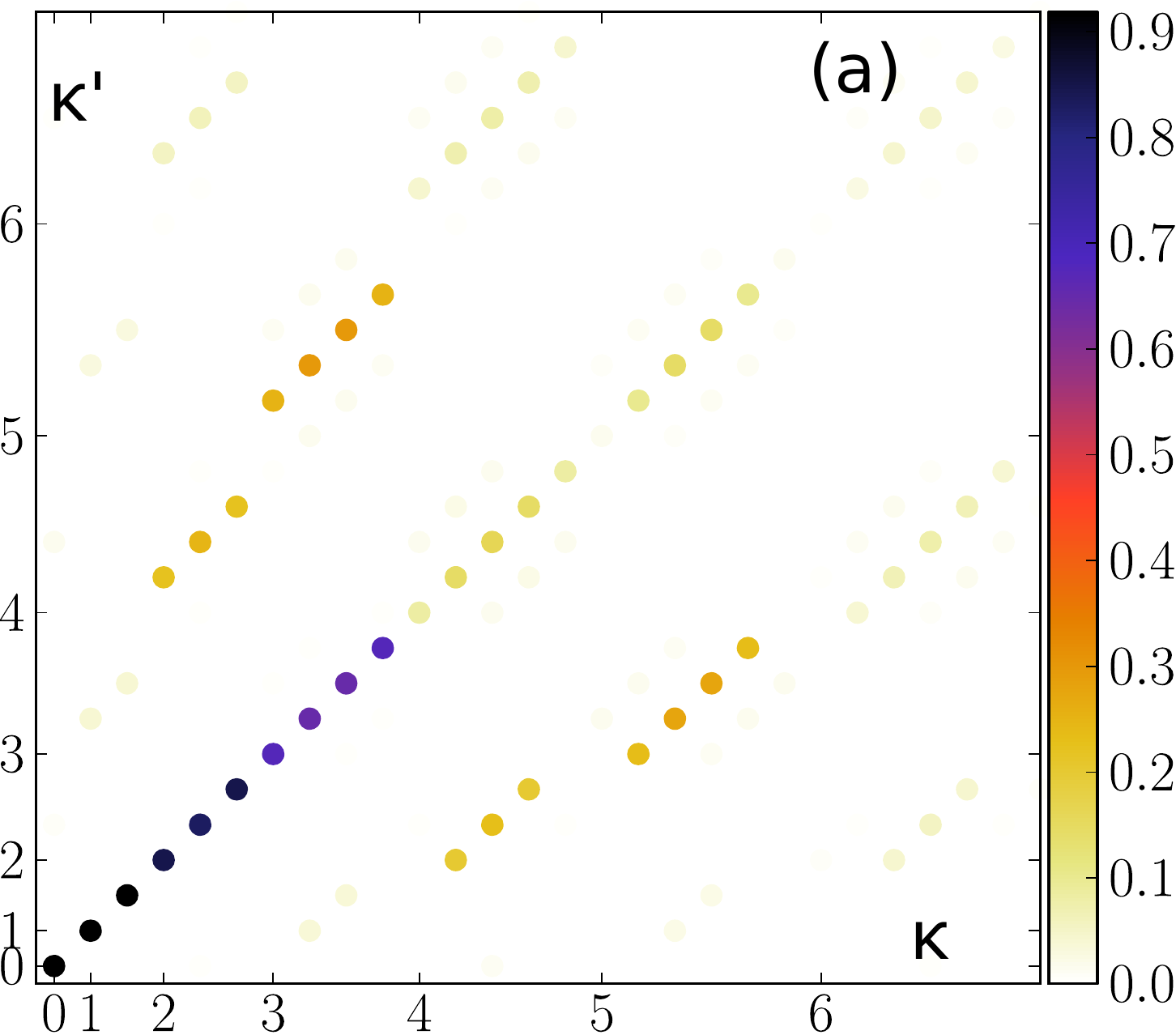}
\includegraphics[width=0.235\textwidth,clip]{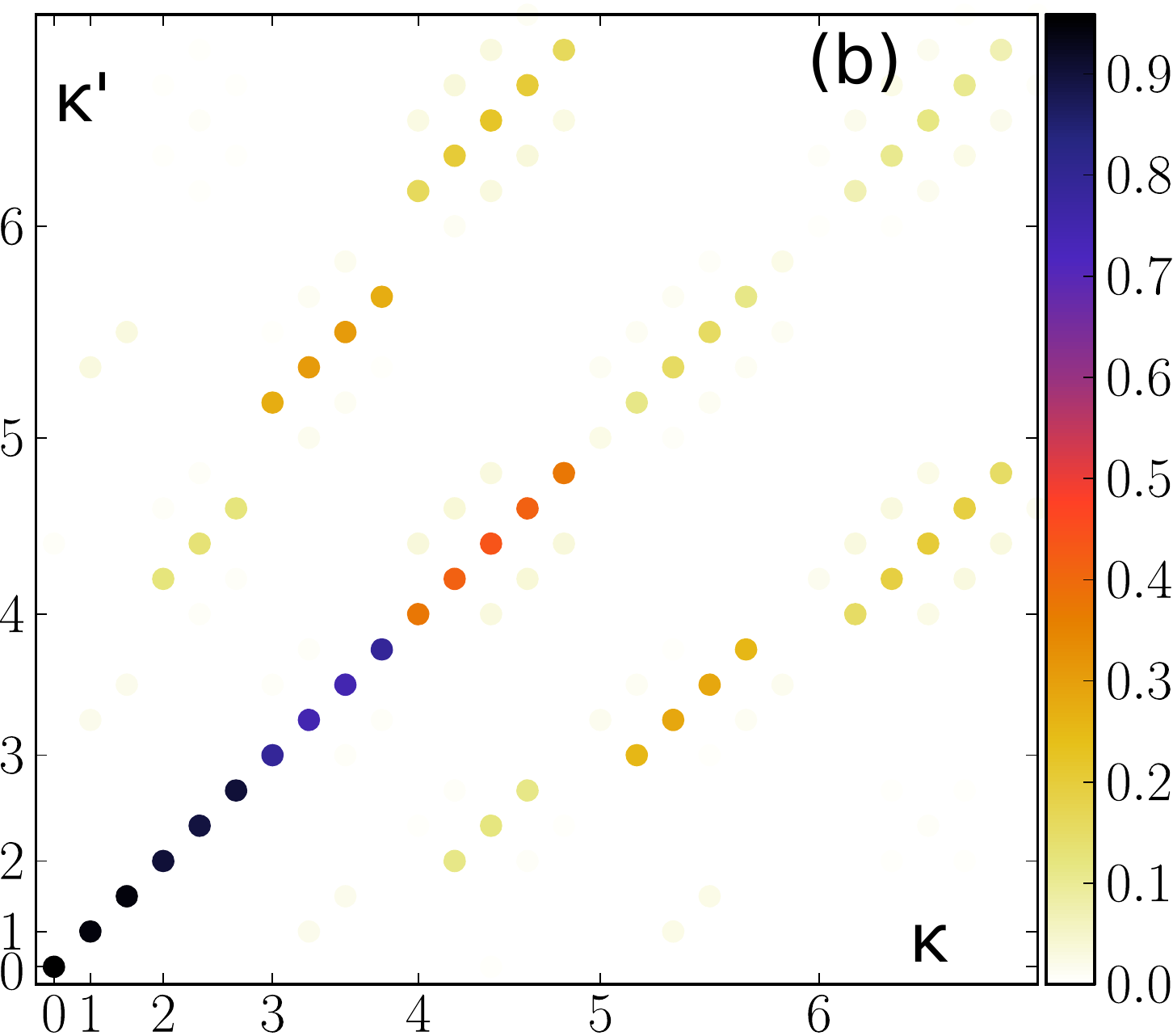}
\includegraphics[width=0.35\textwidth,clip]{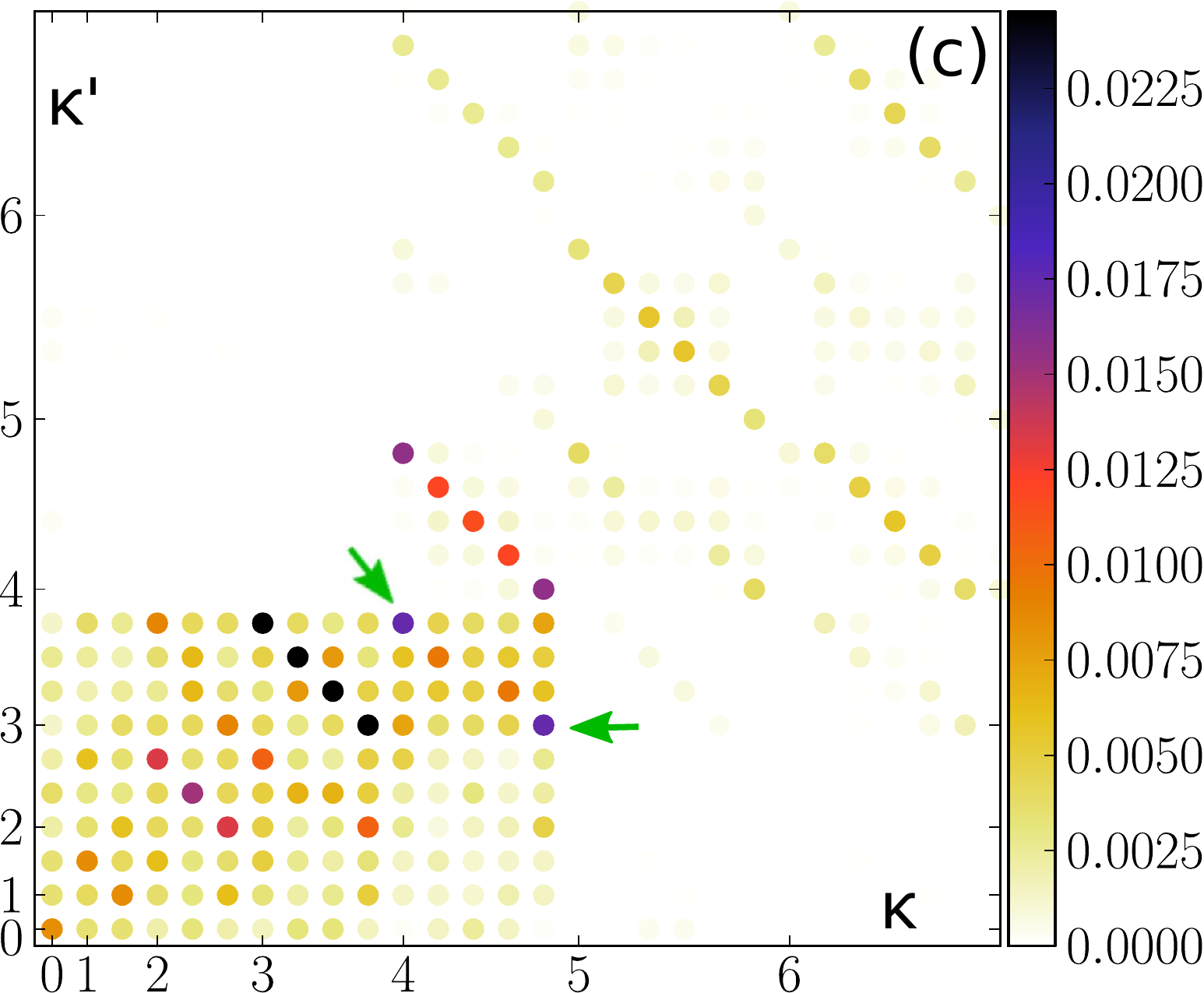}
\end{center}
\caption{\label{fig:HLpairP=0.11}(Color online). Single particle (a)
  and (b) and pair Green functions (c) in the harmonic level basis
  (QMC results) for a low polarization situation (P=0.11). The total
  number of particles is $25.5$. Even though the Fermi-levels between
  the two species no longer match, the pairing is still diagonal for
  $n=3$ and for $n=4$ levels.  However, one observes an off diagonal
  feature appearing that corresponds to pairing between the levels
  $\kappa=(4,-4)$ and $\kappa'=(3,3)$ and respectively $\kappa=(4,4)$
  and $\kappa'=(3,-3)$ as indicated by arrows.}
\end{figure}
\begin{figure}[!hhh]
\begin{center}
\includegraphics[width=0.235\textwidth,clip]{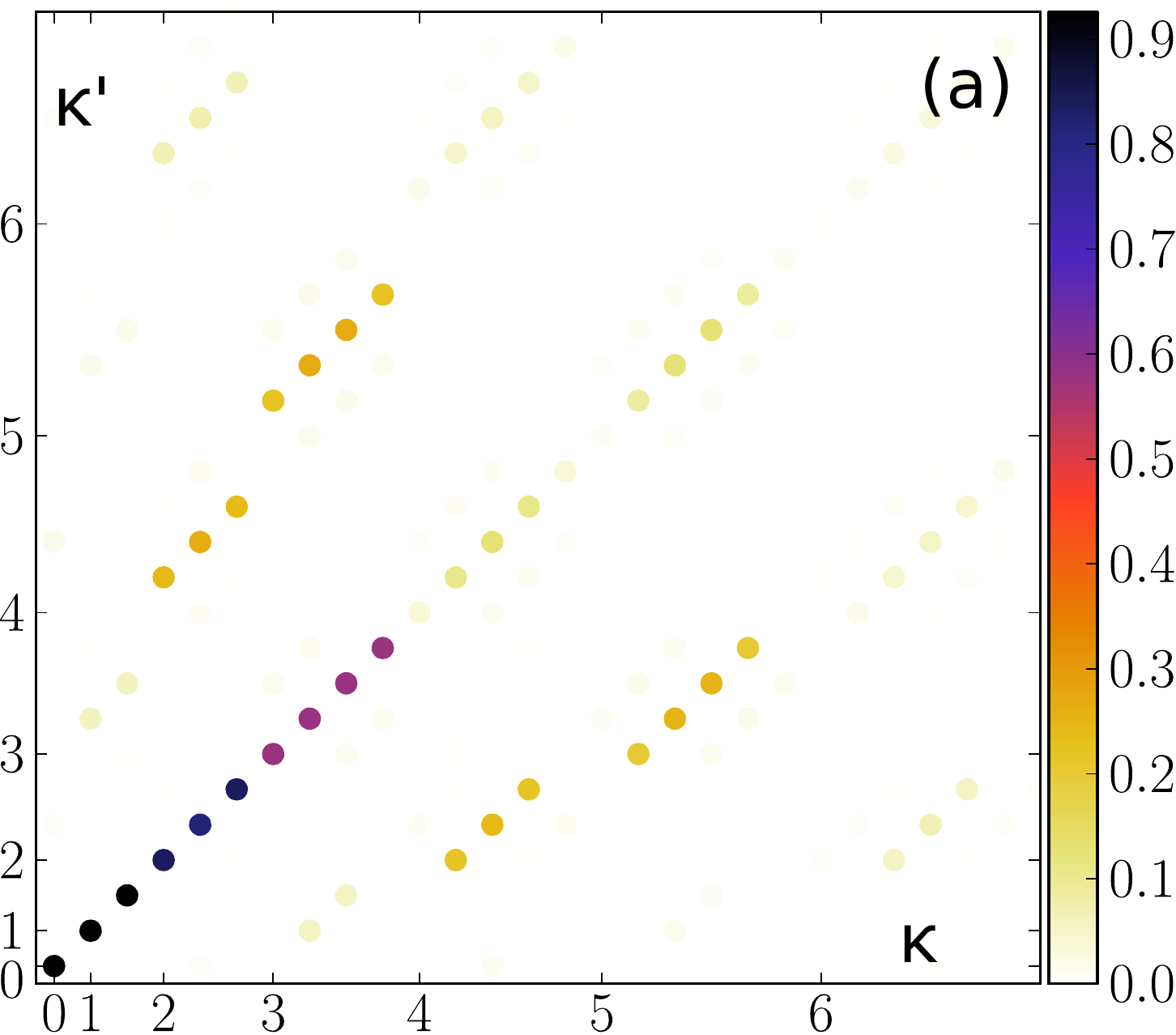}
\includegraphics[width=0.235\textwidth,clip]{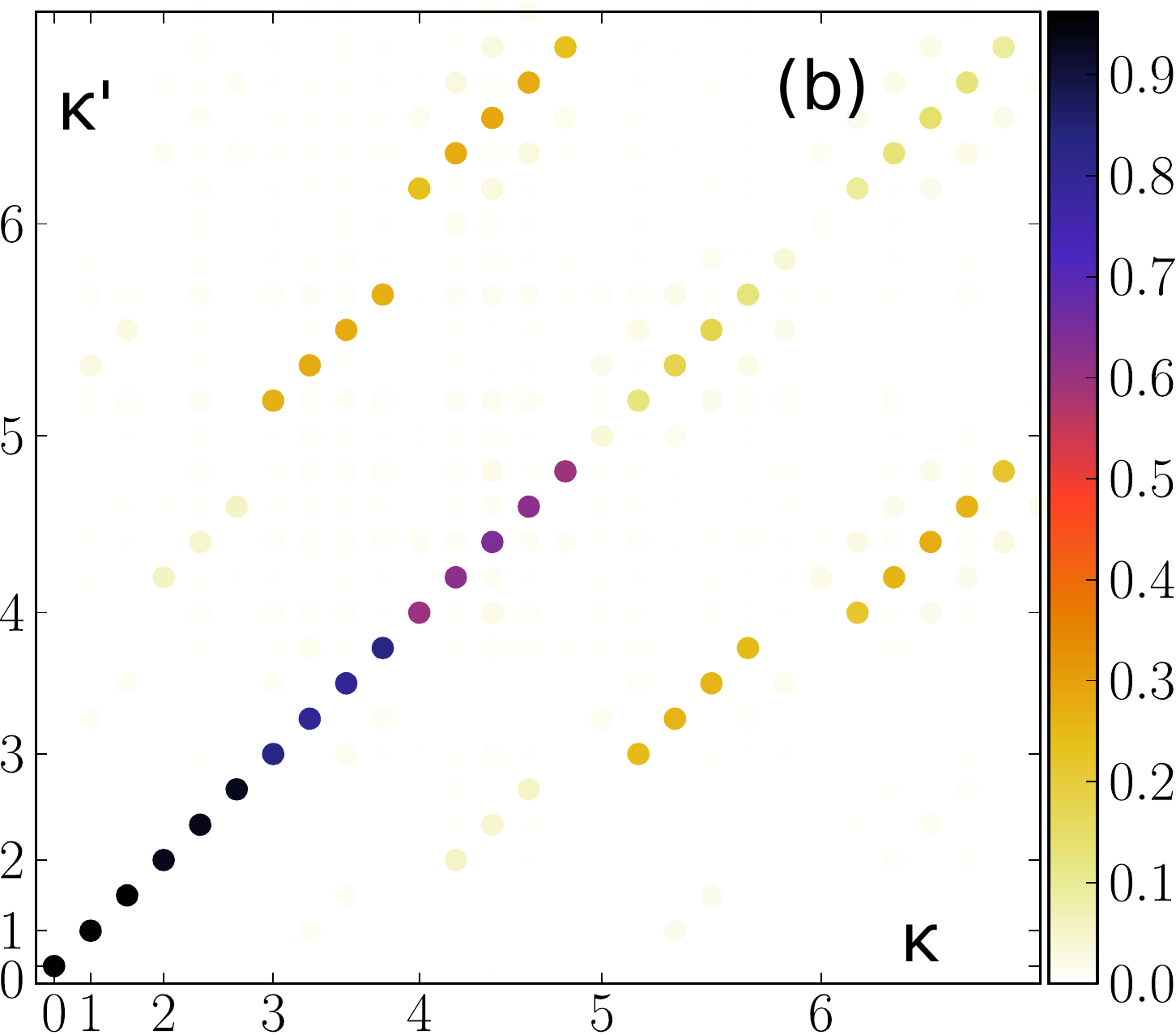}
\includegraphics[width=0.35\textwidth,clip]{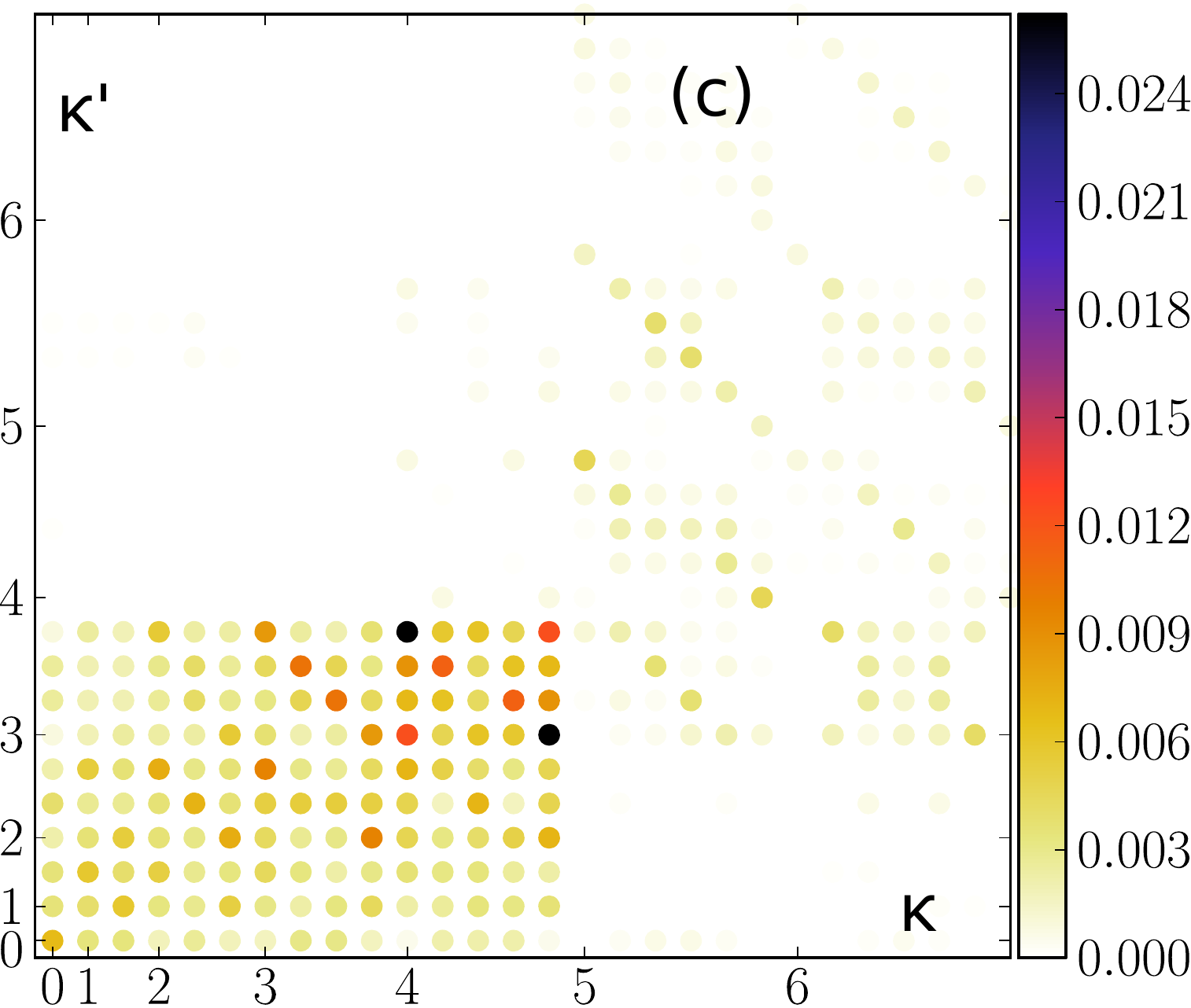}
\end{center}
\caption{\label{fig:HLpairP=0.22}(Color online). Single particle, (a)
  and (b), and pair (c) Green functions in the harmonic level basis
  (QMC results) for a medium polarization (P=0.22). The total number
  of particles is $26.9$.  The diagonal pairing has almost completely
  disappeared and the pairing mostly occurs between the levels $n=3$
  and $n=4$.  More precisely, the strongest pairing occurs between
  $\kappa=(4,-4)$ and $\kappa'=(3,3)$ and analogously between
  $\kappa=(4,4)$ and $\kappa'=(3,-3)$.  There is, in addition, a small
  contribution from the levels $\kappa=(4,-2)$ and $\kappa'=(3,1)$ and
  $\kappa=(4,2)$ and $\kappa'=(3,-1)$. Note that each pairing
  corresponds to a non-vanishing total angular momentum for the pair.}
\end{figure}

\begin{figure}[!hhh]
\begin{center}
\includegraphics[width=0.235\textwidth,clip]{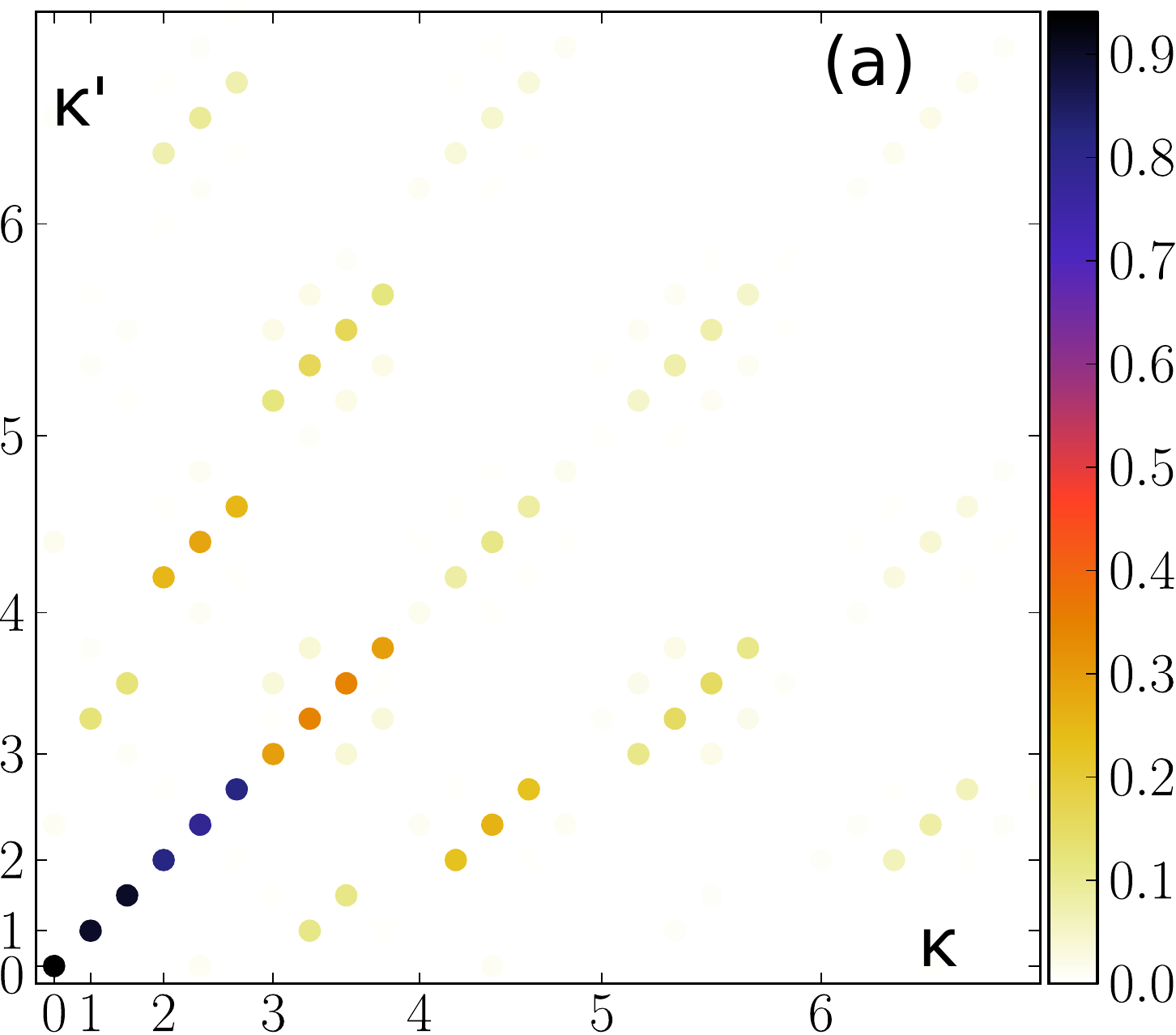}
\includegraphics[width=0.235\textwidth,clip]{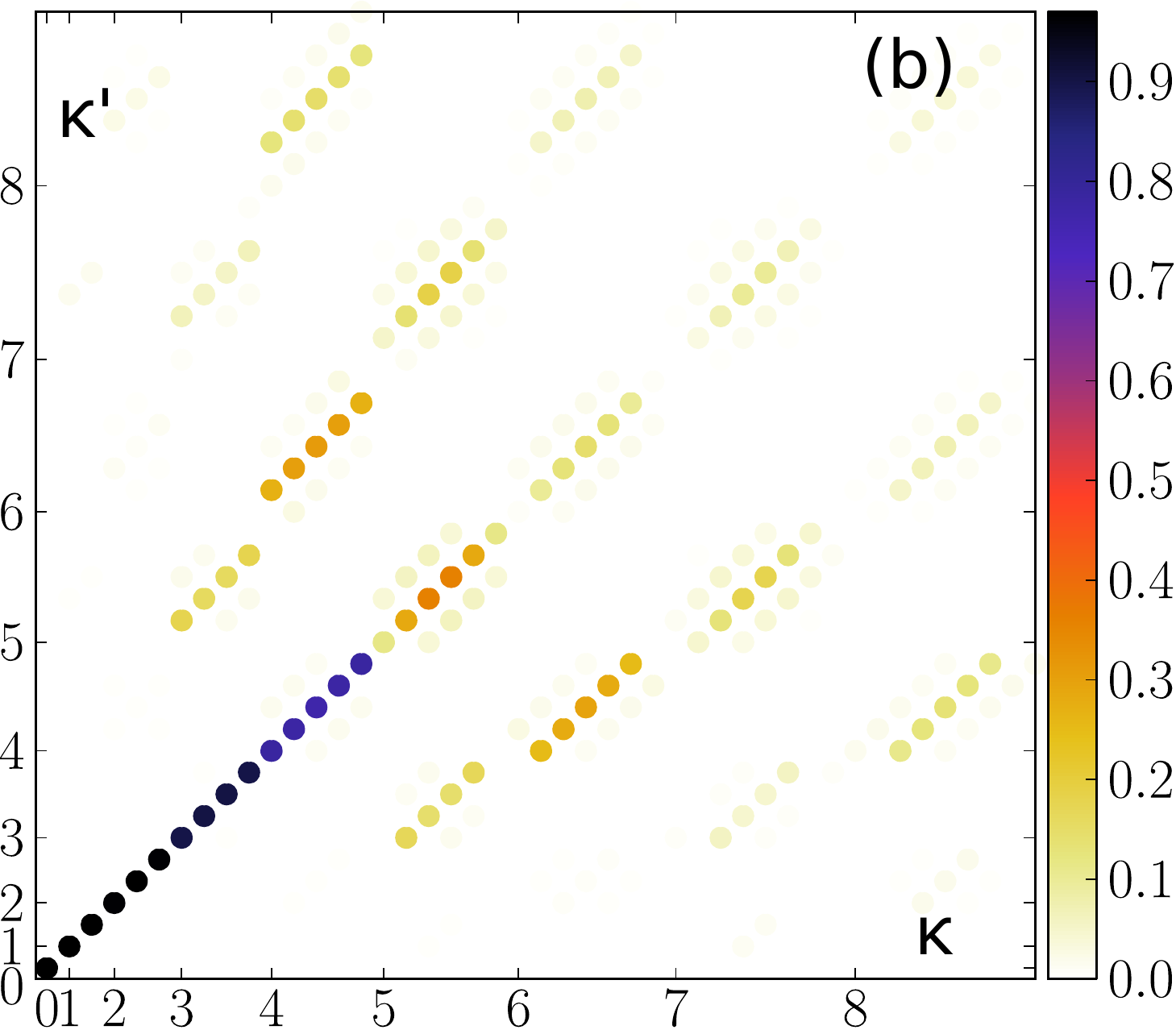}
\includegraphics[width=0.35\textwidth,clip]{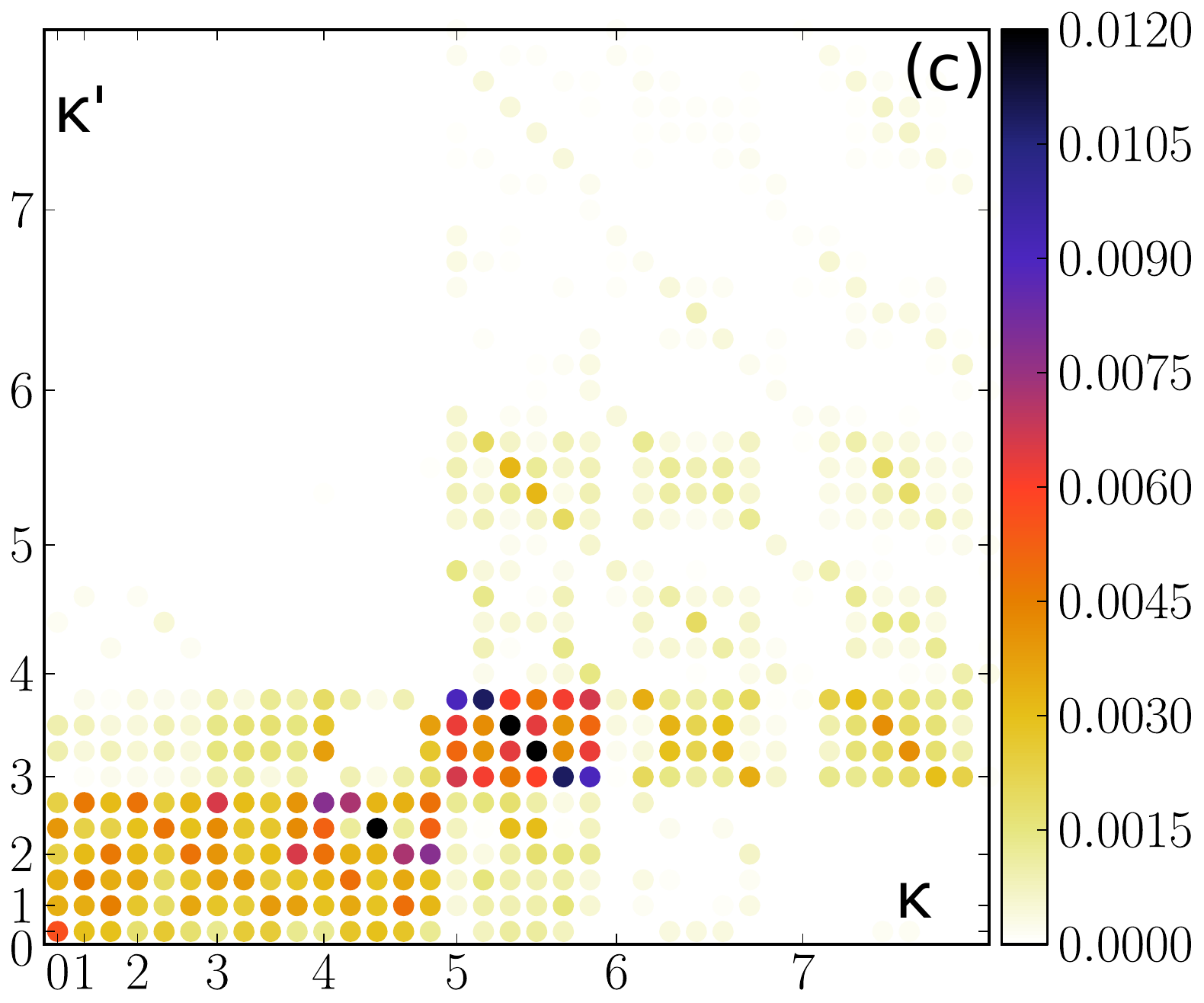}
\end{center}
\caption{\label{fig:HLpairP=0.37}(Color online). Single particle and
  pair Green functions in the harmonic level basis (QMC results) for a
  strong polarization (P=0.37). The total number of particles is
  $27.4$.  The pairing occurs between the levels $n=5$ and $n=3$ and
  also $n=4$ and $n=2$, \textit{i.e.} with total zero orbital angular
  momentum. Still, there is small contribution to pairing between
  $\kappa=(5,-5)$ and $\kappa'=(3,3)$ and $\kappa=(5,5)$ and
  $\kappa'=(3,-3)$.}
\end{figure}

\begin{figure}[!hhh]
\begin{center}
\includegraphics[width=0.37\textwidth,clip]{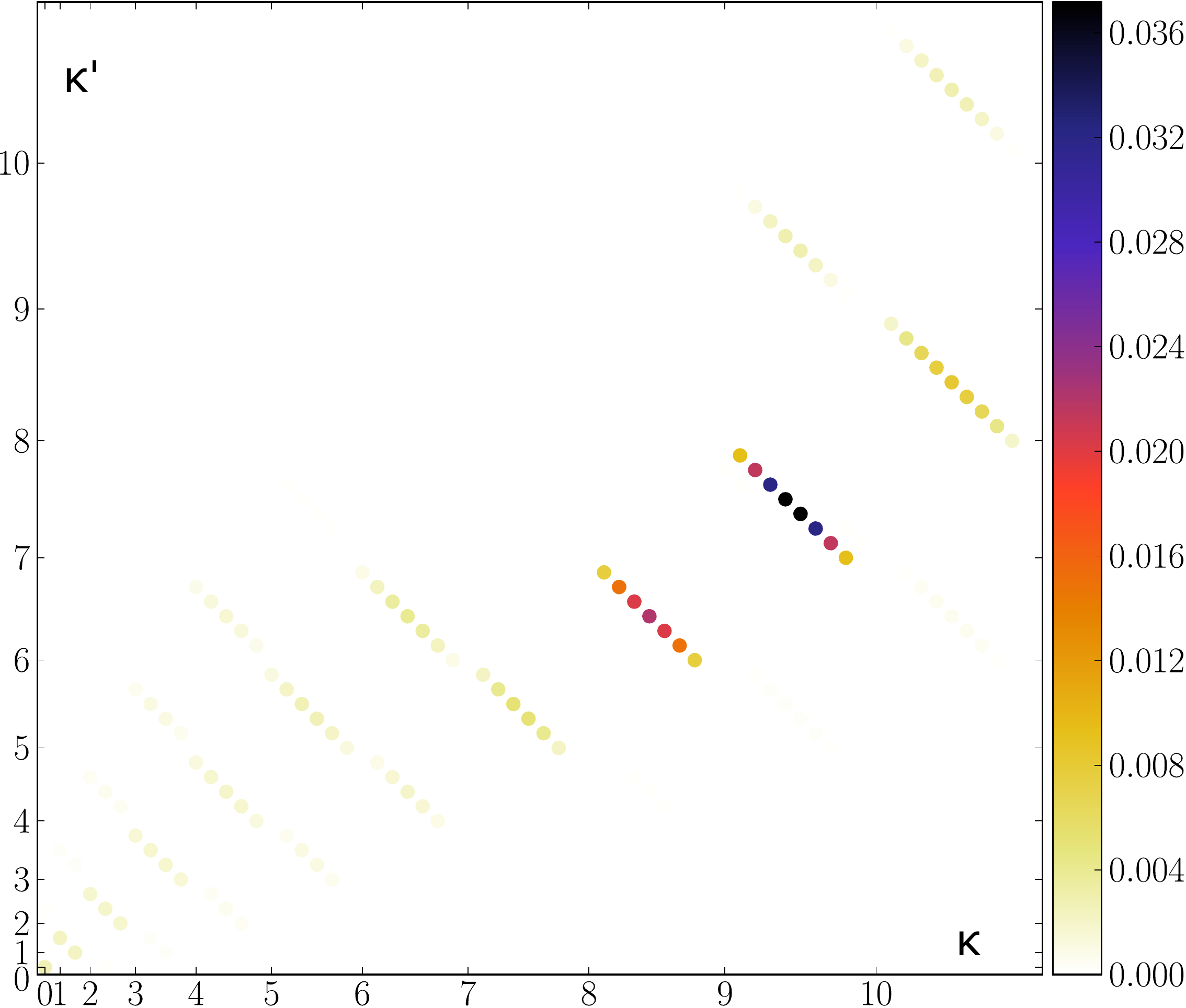}
\end{center}
\caption{\label{fig:MFHLpairP=27}(Color online). Pair Green function
  in the harmonic level basis (MF results) for a polarization
  P=0.27. The results are similar to the QMC results: pairing is
  maximum among the Fermi-levels $n=7$ and $n=9$ and also among the
  two levels below $n=6$ and $n=8$.  The largest $m$ values are almost
  unpaired, for they would have led to non-zero total angular
  momentum. }
\end{figure}

At low imbalance, one observes that the pairing mostly occurs between
the same levels, for instance in Fig.~\ref {fig:HLpairP=0.11}, where
one observes diagonal pairing for $n=3$ and for $n=4$.  However, one
observes an off diagonal feature appearing that corresponds to pairing
between the levels $n=3$ and $n=4$. When the system is imbalanced even
more, the off-diagonal feature becomes the main pairing amplitude. For
instance, as shown in Fig.~\ref{fig:HLpairP=0.22}, corresponding to a
polarization $P= 0.22$, the diagonal pairing has almost completely
disappeared and the pairing mostly occurs between the levels $n=3$ and
$n=4$.  Since it corresponds to a pairing between an odd and even
level, it is impossible to match the $m$ values and get the pairing
with total angular momentum zero. We observed that the strongest
pairing happens, for example, between $\kappa=(4,-4)$ and
$\kappa'=(3,3)$ and analogously between $\kappa=(4,4)$ and
$\kappa'=(3,-3)$.  There is, in addition, a small contribution from
the levels $\kappa=(4,-2)$ and $\kappa'=(3,1)$ and $\kappa=(4,2)$ and
$\kappa'=(3,-1)$. In both cases the sum of the orbital angular
momentum is non-zero.  Imbalancing the system even more we arrive at
the situation where the difference between the Fermi levels of each
species is $n_{F2}-n_{F1}=2$. As illustrated in
Fig.~\ref{fig:HLpairP=0.37} for P=0.37 the pairing occurs between the
levels $n=5$ and $n=3$ and also $n=4$ and $n=2$ which means that the
system can now achieve pairing with zero total orbital angular
momentum. Still, there is small contribution of pairing between
$\kappa=(5,-5)$ and $\kappa'=(3,3)$ and $\kappa=(5,5)$ and
$\kappa'=(3,-3)$, for which $\Delta m=\pm2$.  For a comparison we show
the results from the mean-field simulations, depicting a similar
behavior. In the realization shown in Fig.~\ref{fig:MFHLpairP=27} the
Fermi levels of each species are $n=7$ and $n=9$ and we can see the
pairing occurs between those levels as well as between the two levels
below $n=6$ and $n=8$. The largest $m$ values are almost unpaired, for
they would have lead to non-zero total angular momentum.  We conclude
that in the low filling regime and at intermediate interaction
strength we can understand the FFLO pairing mechanism in a trapped
system as pairing between fermions from different harmonic levels. We
observe that the pairs are formed in such a way so that the total
orbital angular momentum of all pairs is always zero, and the orbital
angular momentum is minimized for each pair.  Finally, similarly to
the untrapped case where the pairs are produced with a finite center
of mass momentum (vanishing for the balanced case), the FFLO state in
the harmonic trap corresponds, in a classical picture, to pairs whose
center of mass is oscillating around the minimum of the trap with an
amplitude increasing with population imbalance.

\textit{Momentum distributions and density profiles at low
    filling of the lattice.}  Fermion systems with imbalanced
populations have been realized experimentally in one- and elongated
three-dimensional harmonic traps.  The density profiles of the
populations were found to be qualitatively different in the two
cases. In three dimensions, one observes the formation of concentric
shells where, for very low polarization, the core is fully paired,
{\it i.e.} zero local magnetization, and the wings are partially
polarized \cite{zwierlein06,partridge06}. On the other hand, it was
observed in one-dimensional systems that, for low polarization, the
unpolarized fully paired populations are located at the edges of the
cloud while the core is partially polarized \cite{hulet}. The role of
dimensionality in this qualitatively different behavior has been one
of the focus of studies on this system.  Consequently, the behavior of
the system in two dimensions is of considerable interest.

We present here results of our DQMC study of the trapped two
dimensional system. The presence of the trap imposes constraints which
make the simulations much harder than the uniform case. The number of
particles should be large enough so that at large $P$ the minority
population will still be appreciable but not so large that the local
density in the core regions is close to half filling. Another
constraint is that the size of the lattice be large enough to ensure
that particles do not leak out. These constraints limit our ability to
do simulations for system sizes beyond $20\times 20$.

As for the uniform system, the most important indicator of the
presence of the FFLO state is the pair momentum distribution. Although
the plane wave basis is not the natural one in the harmonically
confined case, we study the momentum distributions because they are of
experimental interest. We will show that despite the shortcomings of
this language one can still detect the FFLO pairing signal this
way. In addition, since the trap destroys translational invariance it
is very useful to study the density profiles and local magnetization,
$m(x,y)=\rho_2(x,y)-\rho_1(x,y)$ where $\rho_2$ ($\rho_1$) is local
density of the majority (minority).
\begin{figure}[!hhh]
\begin{center}
\includegraphics[width=0.35\textwidth,clip]{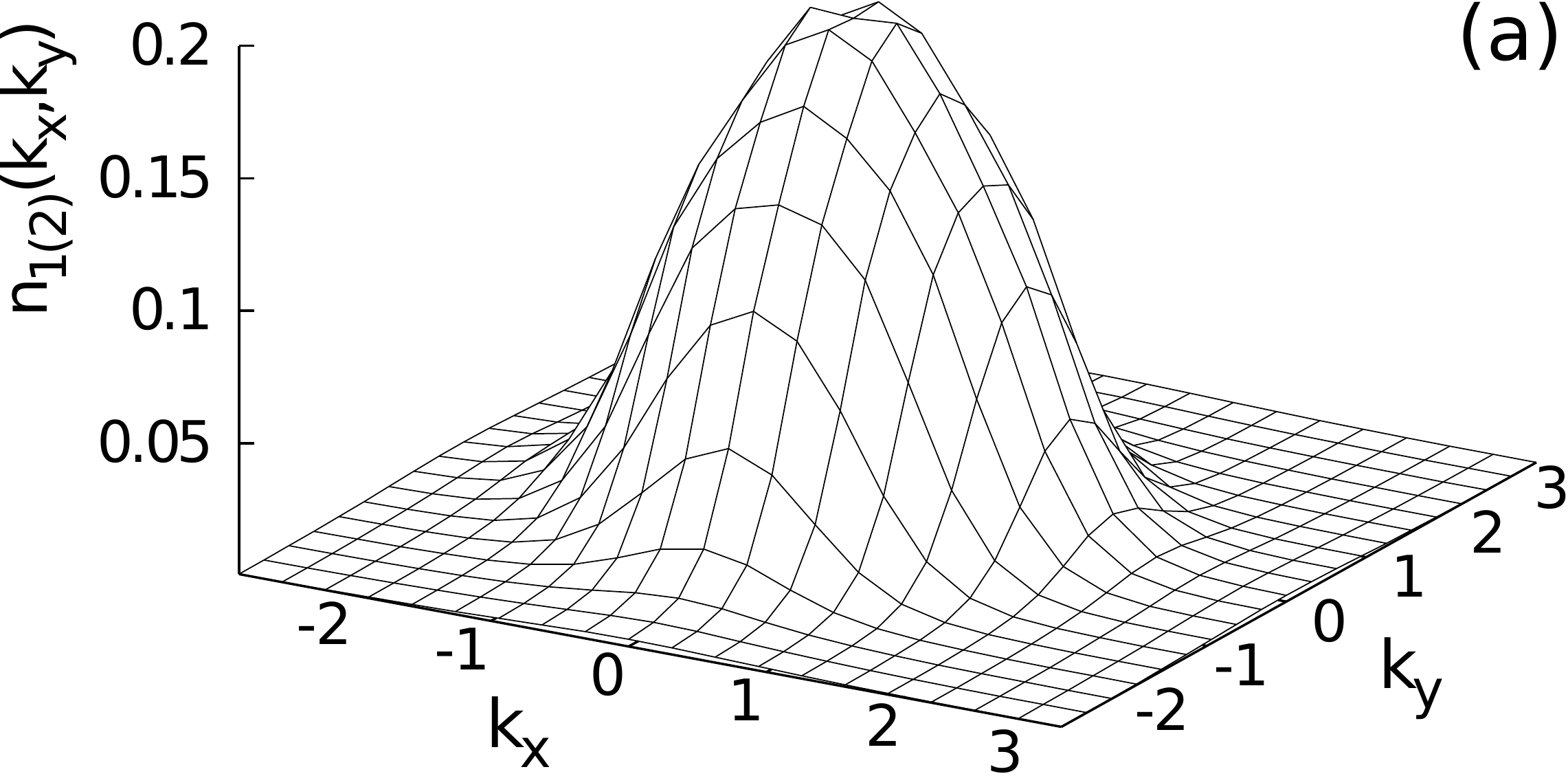}
\includegraphics[width=0.35\textwidth,clip]{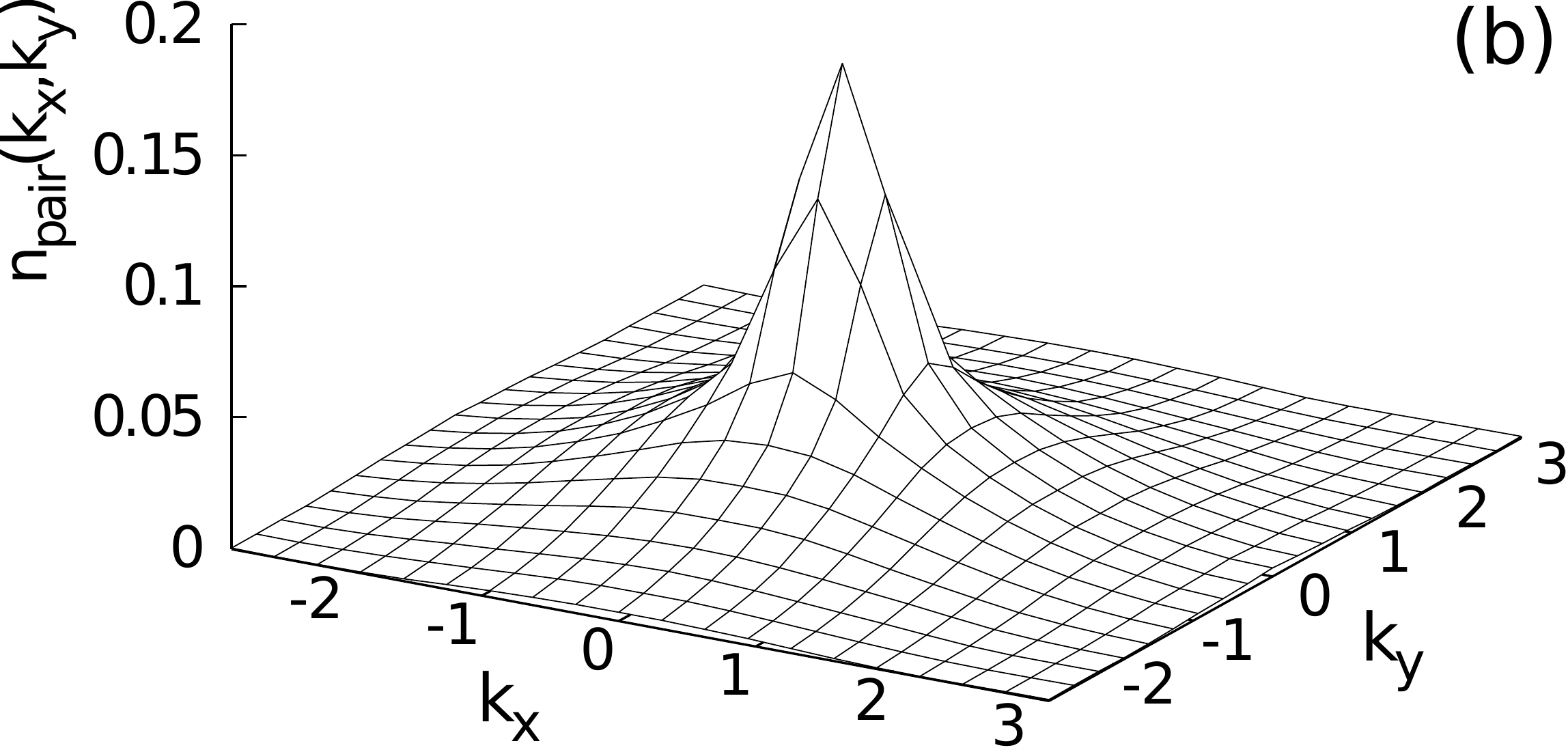}
\end{center}
\caption{\label{fig:balancedmomentumTrap} Momentum distributions of
  (a) the single particles, $n_1(k_x,k_y)=n_2(k_x,k_y)$ and (b) the
  pairs $n_{pair}(k_x,k_y)$. The total number of particles is $22.3$,
  $P=0$, $\beta=10$, $U=-3.5t$, the trap potential is $V_t=0.065$
  and the lattice size $20\times 20$. }
\end{figure}
We start with the unpolarized system. Figure
\ref{fig:balancedmomentumTrap}(a) shows the momentum distribution of
the particles (the two populations are identical) for a system with a
total of $22.3$ particles, $P=0$, $\beta=10$, $U=-3.5t$ and a lattice
size of $20\times 20$. The trap potential is given by
$V_t=0.065$. Figure \ref{fig:balancedmomentumTrap}(b) shows the pair
momentum distribution and exhibits a sharp peak at zero momentum.
\begin{figure}[!htb]
\begin{center}
\includegraphics[width=0.35\textwidth,clip]
{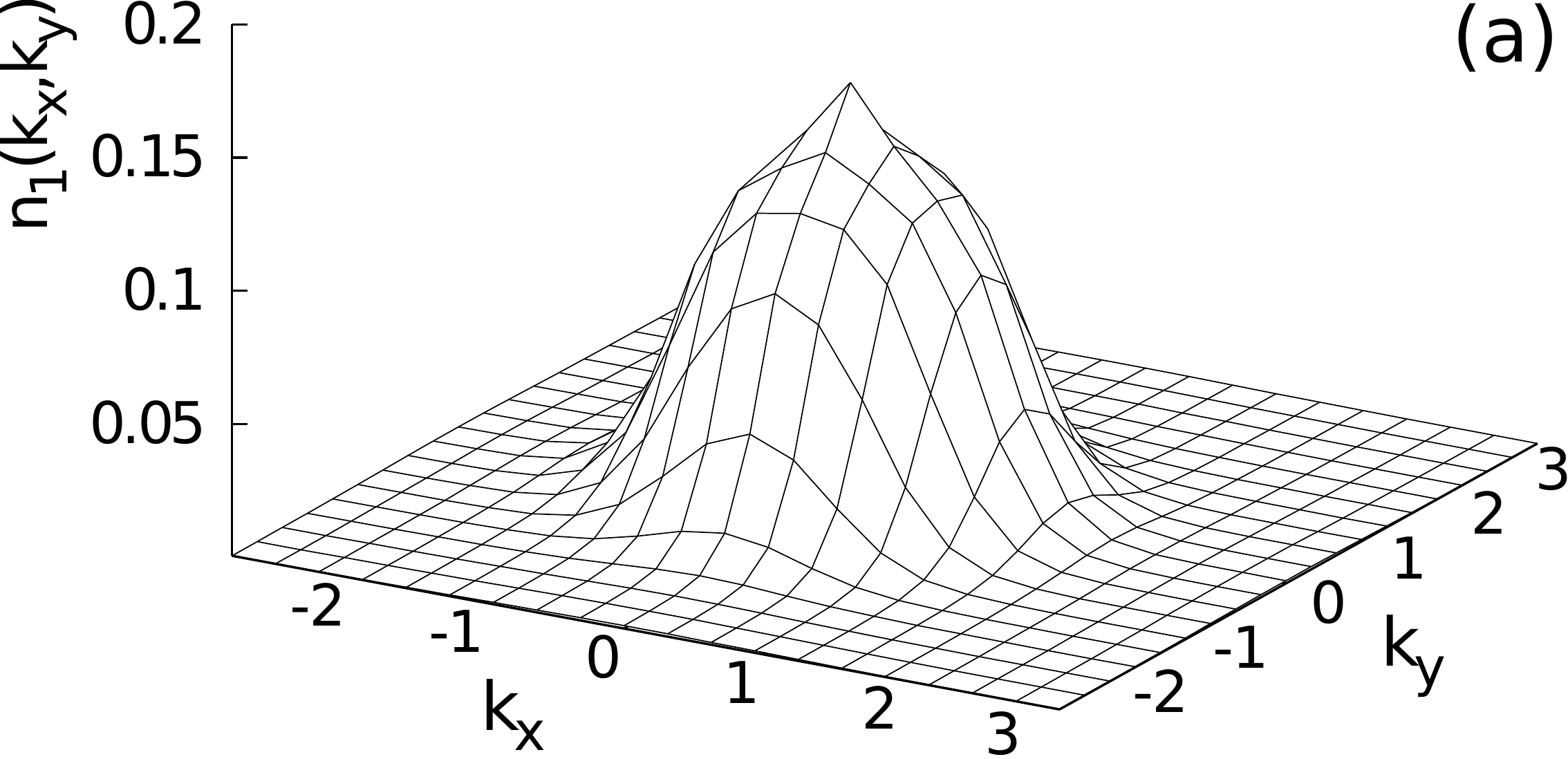}\\
\includegraphics[width=0.35\textwidth,clip]
{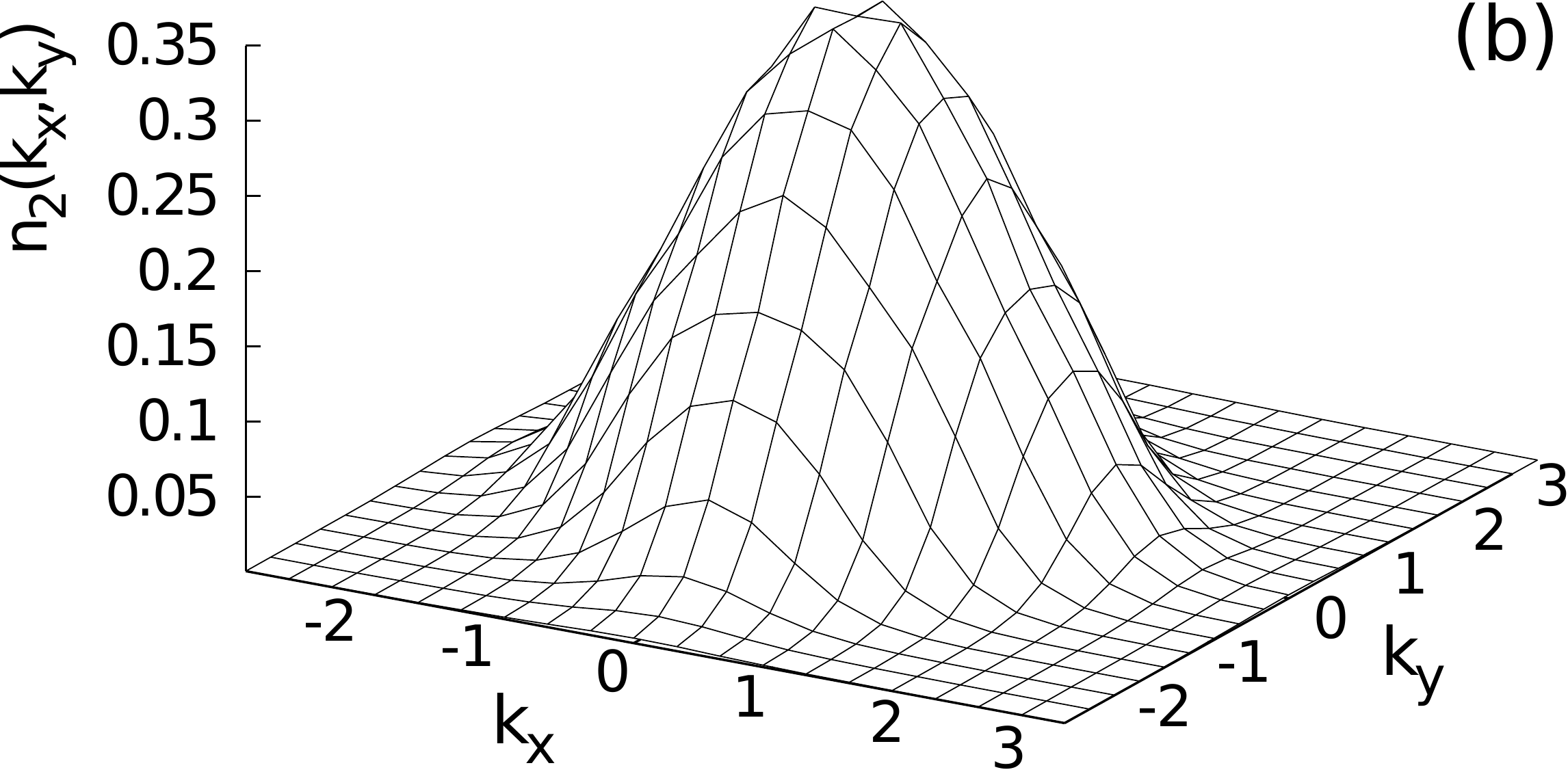}\\
\includegraphics[width=0.35\textwidth,clip]
{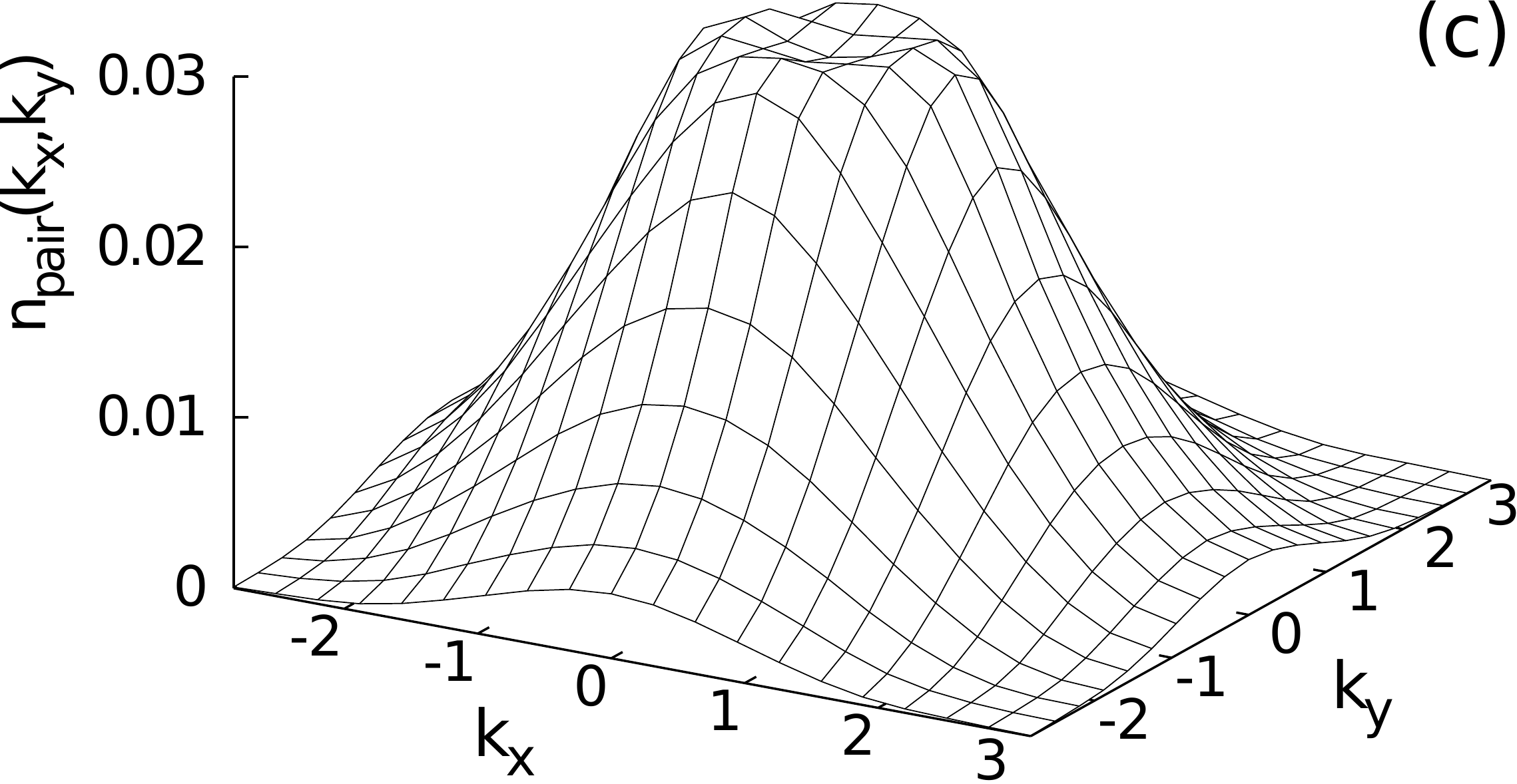}
\end{center}
\caption{\label{fig:imbalancedmomentumTrap} The momentum distributions
  of (a) the minority and (b) majority populations and (c) the pairs.
  The total number of particles is $21.4$, $P=0.55$, $\beta=10$,
  $U=-3.5t$ on a $20\times 20$ lattice. The trap potential is
  $V_t=0.065$ and the Fermi temperature is $T_F=1.86$.}
\end{figure}
Now we polarize the system keeping the total number of particles
constant which corresponds to the experimental situation. Figure
\ref{fig:imbalancedmomentumTrap} shows the momentum distributions of
the (a) minority and (b) majority populations and (c) the pairs. The
system has a total of $21.4$ particles, $P=0.55$, $\beta=10$,
$U=-3.5t$ and a trap potential $V_t=0.065$ on a $20\times 20$
lattice. The Fermi temperature of the system is $T_F=1.86$. Figure
\ref{fig:imbalancedmomentumTrap}(c) is qualitatively different from
Fig.~\ref{fig:balancedmomentumTrap}(b) and shows clearly that when the
confined system is polarized it exhibits FFLO states with pairs
forming with nonzero center of mass momentum. This behavior was
observed for a wide range of polarizations and interaction
strengths. The vertical scale in
Fig.~\ref{fig:imbalancedmomentumTrap}(c) shows that the number of
pairs is very small. This is due to the small total number of
particles in the system. A simulation for a larger system but with the
same characteristic density~\cite{marcos} should give a stronger
signal in the form of higher peaks at nonzero momentum. This effect of
the total number of particles was shown in the one dimensional uniform
case in Ref.~\cite{batrouni08}.

\begin{figure}[!hhh]
\begin{center}
\includegraphics[width=0.35\textwidth,clip]{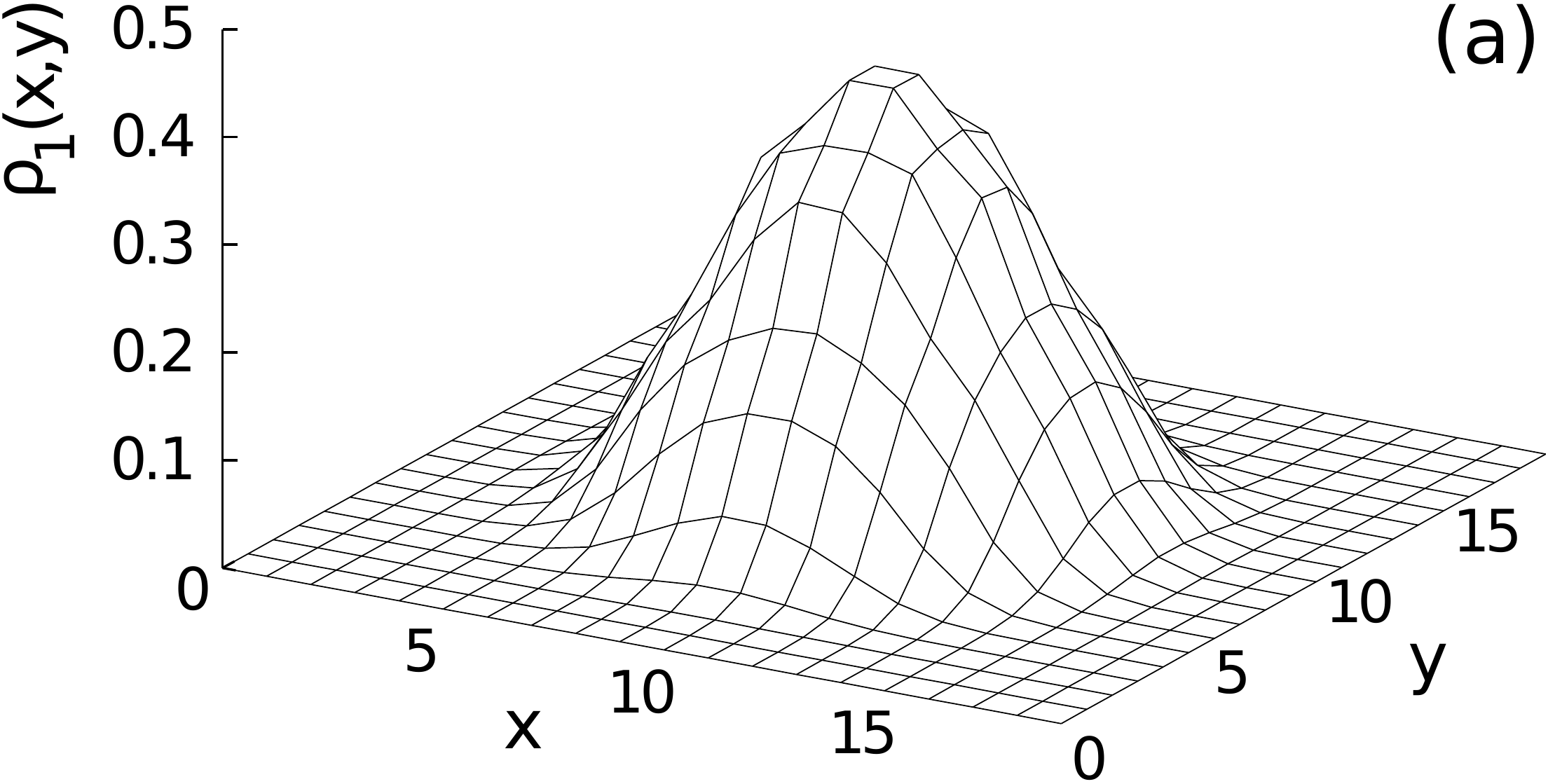}
\includegraphics[width=0.35\textwidth,clip]{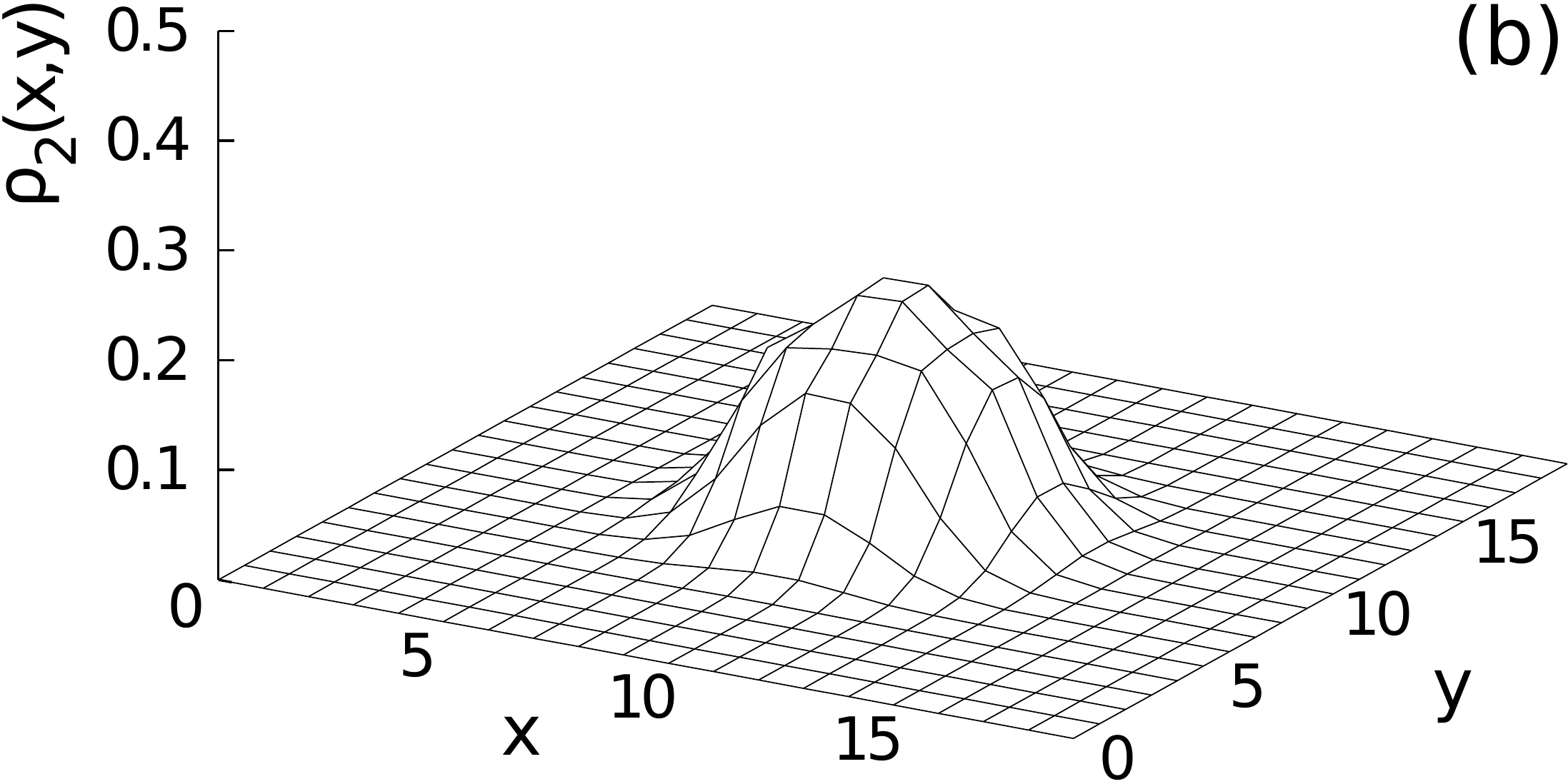}
\includegraphics[width=0.35\textwidth,clip]{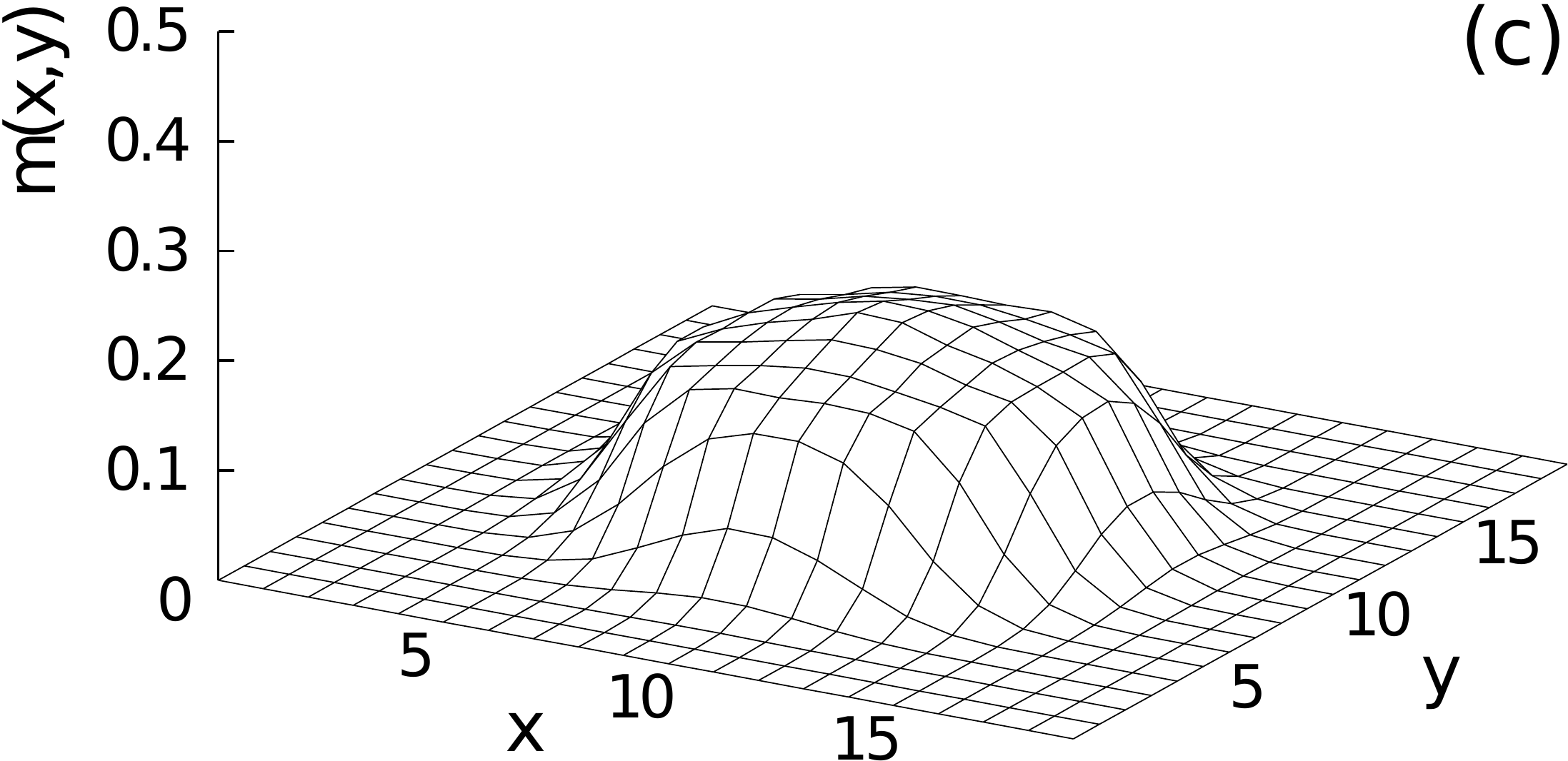} 
\end{center}
\caption{\label{fig:imbalanceddensitiesTrap} Density distributions of
majority ($n_1(x,y)$), minority ($n_2(x,y)$) and the local
magnetization (m(x,y)). Total number of particles is 21.4, P=0.55,
$\beta=10$, Lattice size 20x20, Trap potential $V_t=0.065$.}
\end{figure}

During our simulations we measure the density profiles of each species
and we calculate the local magnetization
$m(x,y)=n_1(x,y)-n_2(x,y)$. The profiles shown in
Fig.~\ref{fig:imbalanceddensitiesTrap} correspond to the situation
when FFLO-type pairing has been observed in the system as in
Fig.~\ref{fig:imbalancedmomentumTrap}. One observes that the system is
partially polarized at the core and fully polarized in the wings
(where we see no minority particles). There is no fully paired phase
where $m(x,y)$ would disappear within the size of the cloud.

Density profiles are the basic quantities that characterize the
trapped system. The first experimental results in a three-dimensional
system show the formation of concentric shells where for very low
polarization the core is fully paired (no local magnetization) and the
wings are partially polarized (see \cite{zwierlein06} and
\cite{partridge06}). On the other hand in the one-dimensional system
it has been observed that there exists a low polarization regime where
the unpolarized superfluid is located at the edge of the cloud, and
the core is partially polarized \cite{hulet}. The issue of this
dimensionally driven transition caused considerable interest. It is
interesting to look at the intermediate two dimensions and study the
behavior of the density profiles to see whether it follows more
closely any of the two limiting scenarios.  During our simulations we
measure the density profiles of each species and calculate the local
magnetization $m(x,y)=n_1(x,y)-n_2(x,y)$. The profiles shown in
Fig.~\ref{fig:imbalanceddensitiesTrap} correspond to the situation
where FFLO-type pairing has been observed in the system as in
Fig.~\ref{fig:imbalancedmomentumTrap}. One observes that the system is
partially polarized at the core and fully polarized in the wings
(where we see no minority particles). There is no fully paired phase
where $m(x,y)$ would disappear within the size of the cloud.

\begin{figure}[!htb]
\begin{center}
\includegraphics[width=0.45\textwidth,clip ]{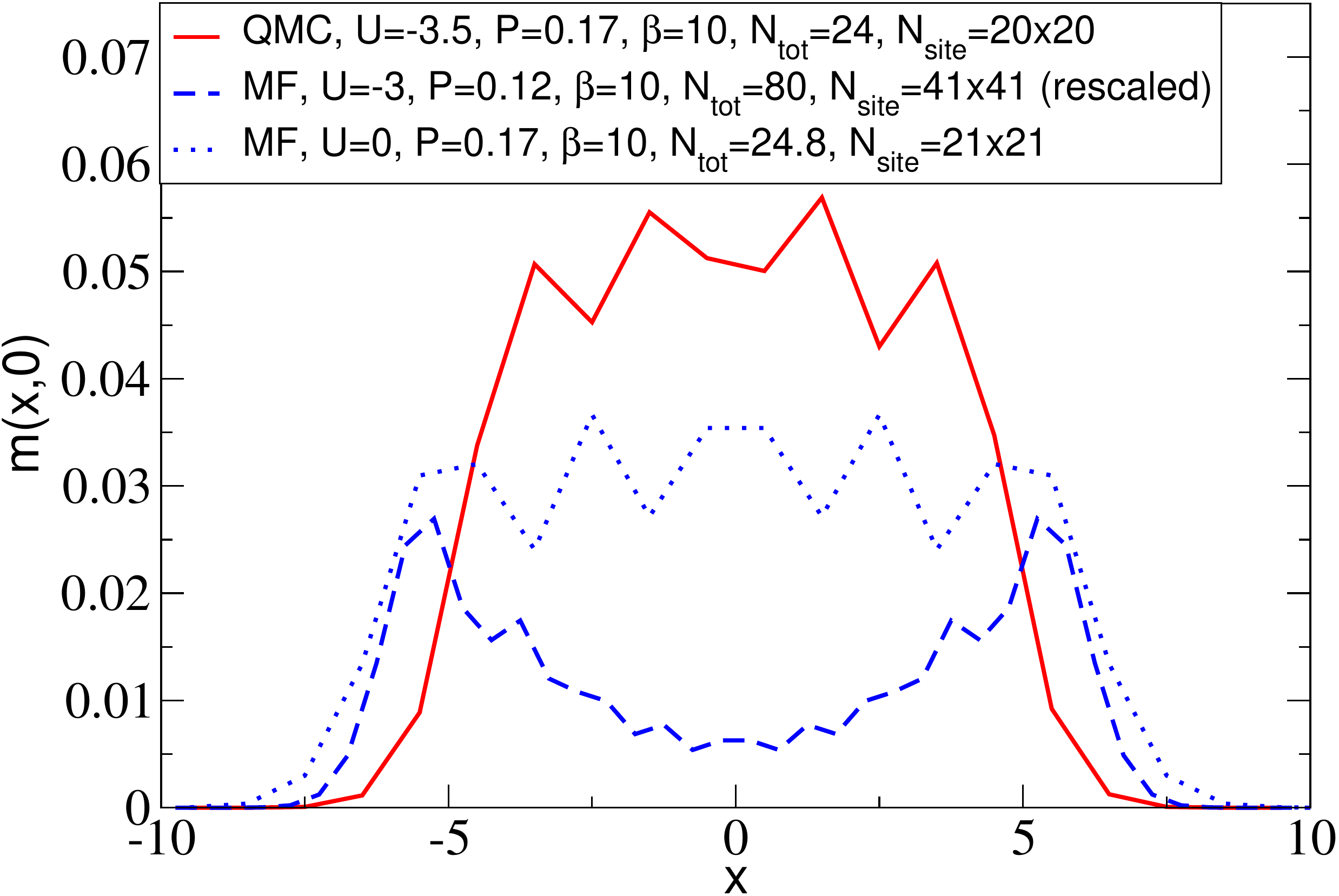}
\end{center}
\caption{\label{fig:localmagU0}(Color Online). Cut through the center of the trap showing the local
magnetization (m(x,y)). Comparison of interacting and non interacting profiles for low polarization using MF and QMC.
The oscillations seen in the magnetization are present even in the non-interacting situation (dashed line). From both
the MF and QMC, one can see that the interaction might change the profile, but does not crucially change the oscillation
pattern. Therefore, we attribute the oscillations to the underlying harmonic levels rather than to the FFLO order.}
\end{figure}

In the very low polarization regime we observe oscillations appearing
in the profile of the local magnetization. We looked in detail into
these results in order to establish whether the oscillations are
linked to the FFLO type pair density wave behavior.  We found however
that the oscillations are present in the system even when there is no
interaction between particles as seen in Fig.~\ref{fig:localmagU0}.
From the discussion in the preceding section, where we have shown the
relevance of the harmonic levels at low fillings, we believe that this
effect stems from the underlying harmonic level structure.  In the
balanced case, it has already been shown that the density of a
fermionic cloud in a trap can exhibit oscillations with minima or
maxima in the center of the trap depending on whether the last filled
state corresponds to an odd or even harmonic level~\cite{burnett}.

\begin{figure}[!hhh]
\begin{center}
\includegraphics[width=0.4\textwidth,clip]{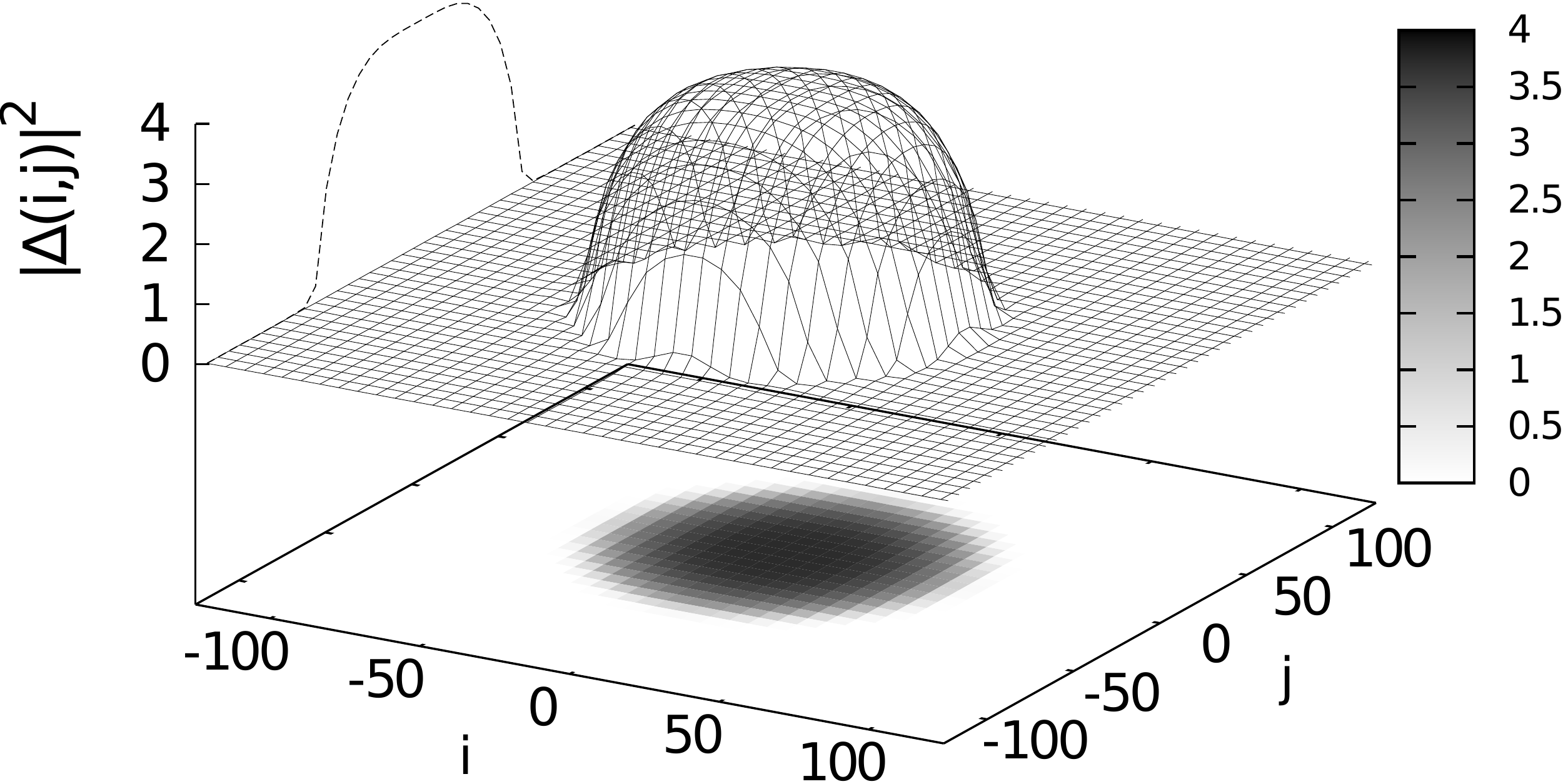}
\includegraphics[width=0.4\textwidth,clip]{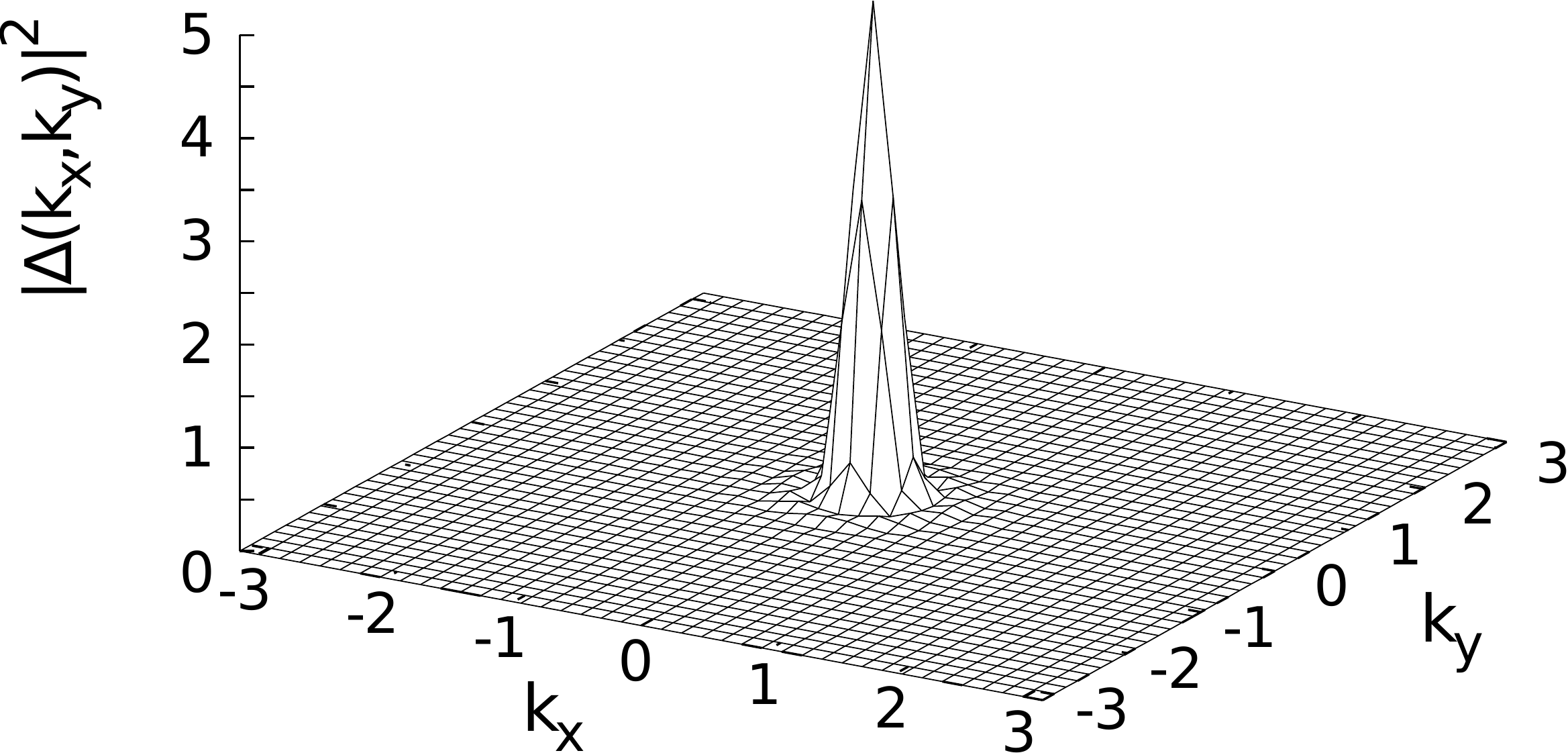}
\end{center}
\caption{\label{fig:TrapMFHF00} Mean Field parameter $\Delta$ as a
  function of the position (top) and in Fourier space (bottom) for a
  low polarization value (P=0.13), around the half-filling situation,
  in the presence of an harmonic trap. The structure is similar to the
  balanced case, \textit{i.e.} a maximum number of pairs at the center
  of the trap, decreasing on the border. The Fourier transform simply
  depicts a peak at $\vec{k}=0$, emphasizing a BCS-like pairing.}
\end{figure}

\begin{figure}[!hhh]
\begin{center}
\includegraphics[width=0.4\textwidth,clip]{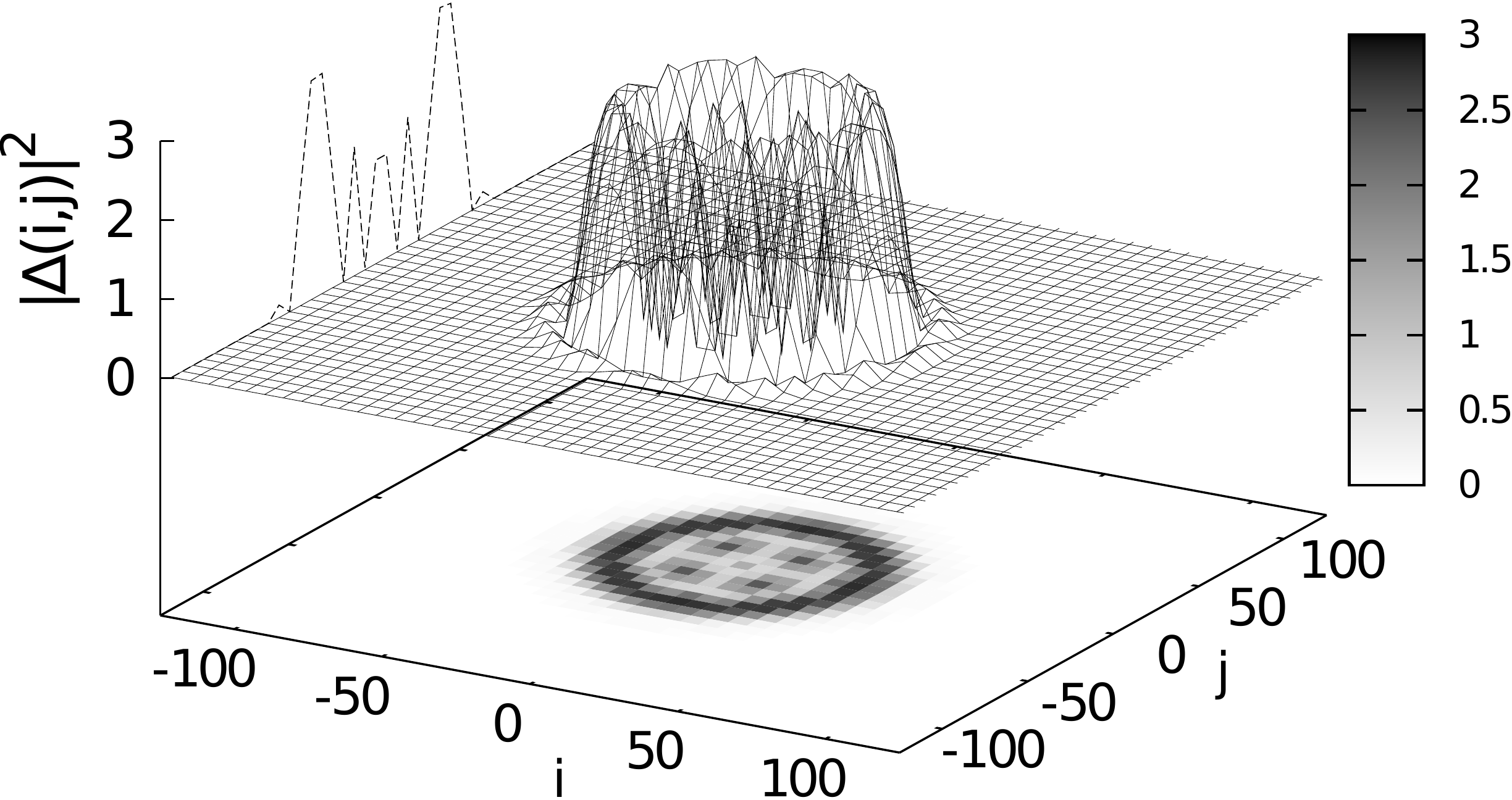}
\includegraphics[width=0.4\textwidth,clip,]{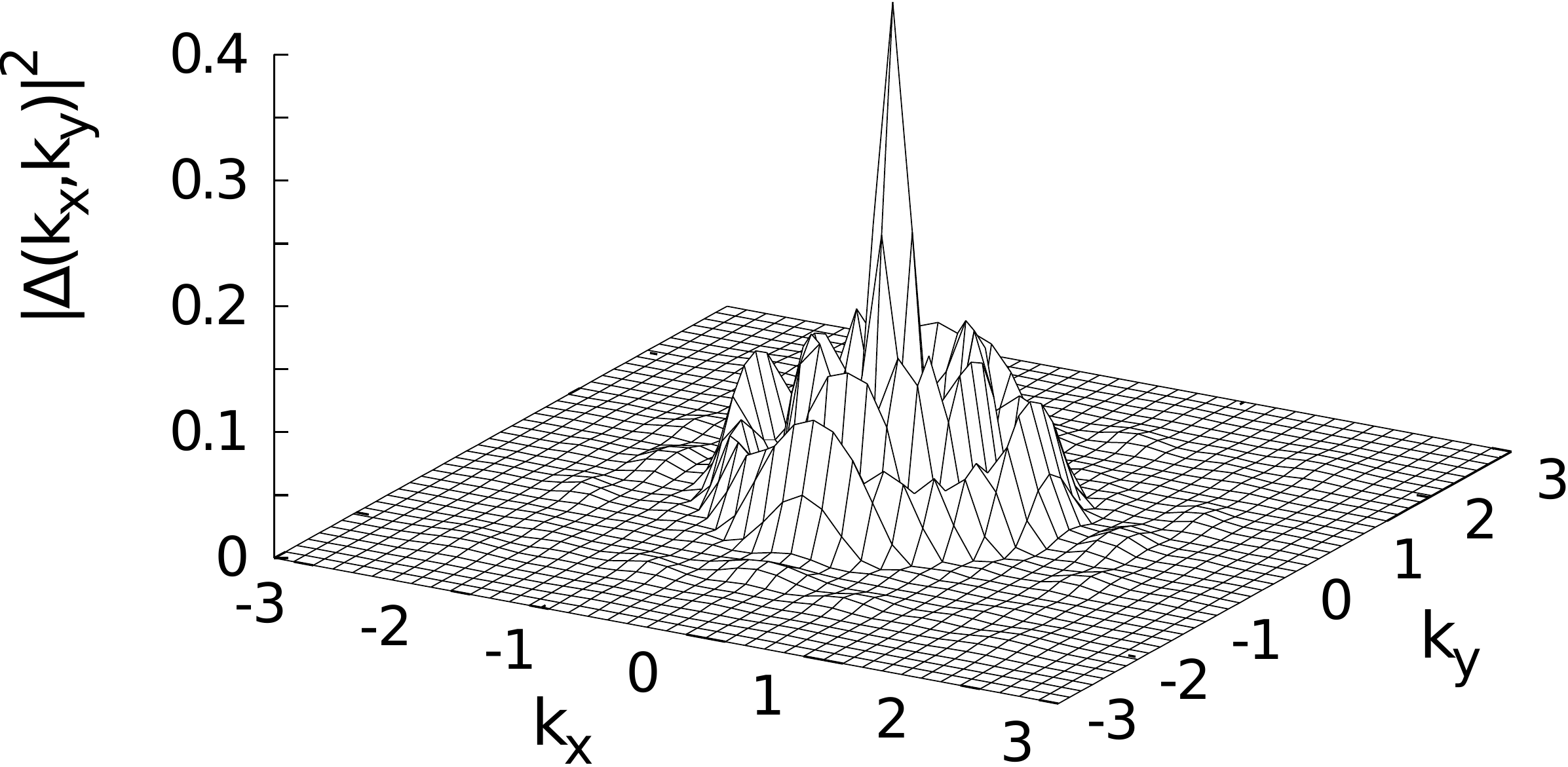}
\end{center}
\caption{\label{fig:TrapMFHF07} Mean Field parameter $\Delta$ as a
  function of the position (top) and in Fourier space (bottom) for a
  polarization value P=0.43. A structure in the center of the trap is
  clearly visible, leading to oscillations in the Fourier transform.}
\end{figure}

\begin{figure}[!hhh]
\begin{center}
\includegraphics[width=0.4\textwidth,clip]{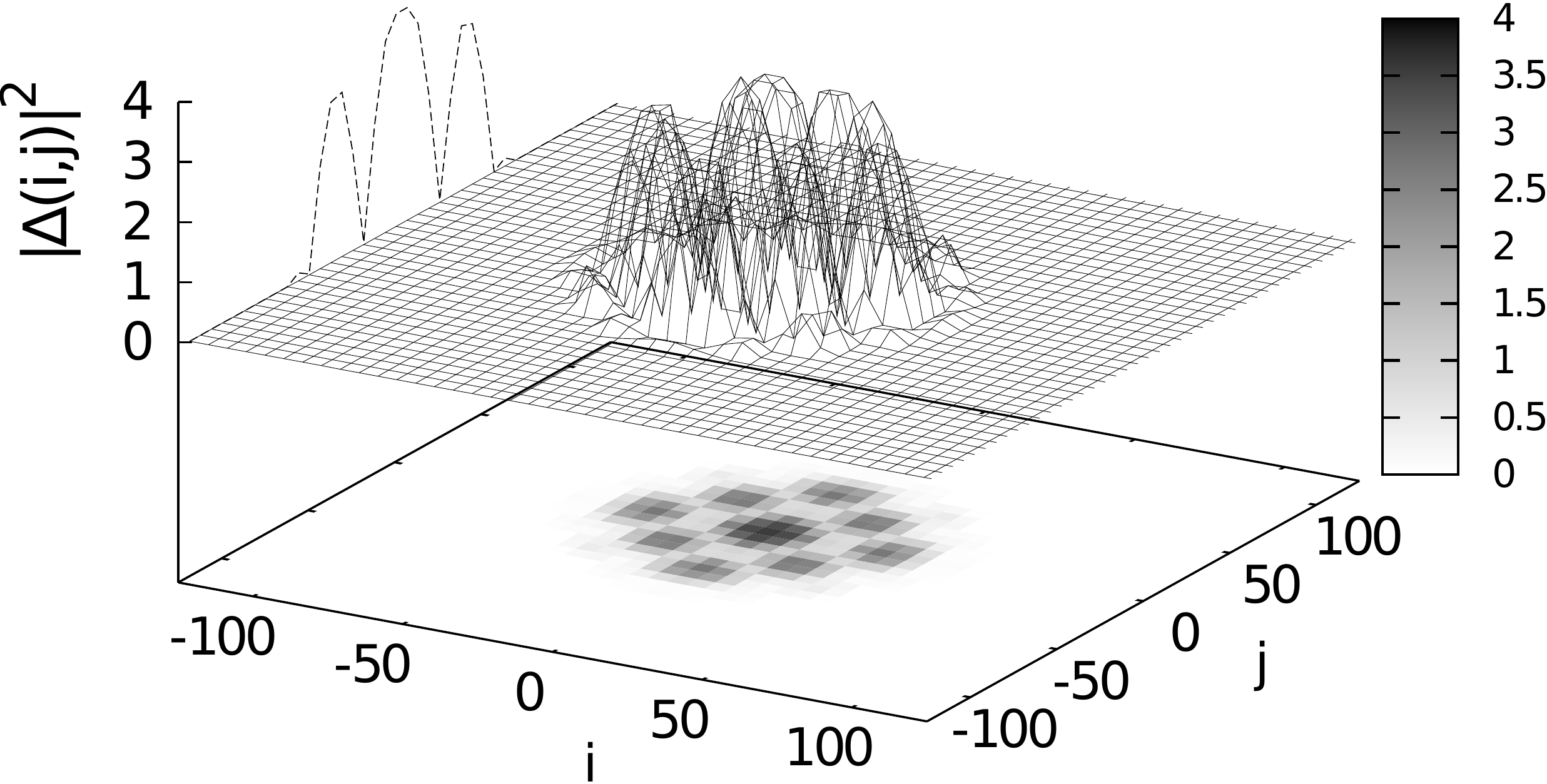}
\includegraphics[width=0.4\textwidth,clip]{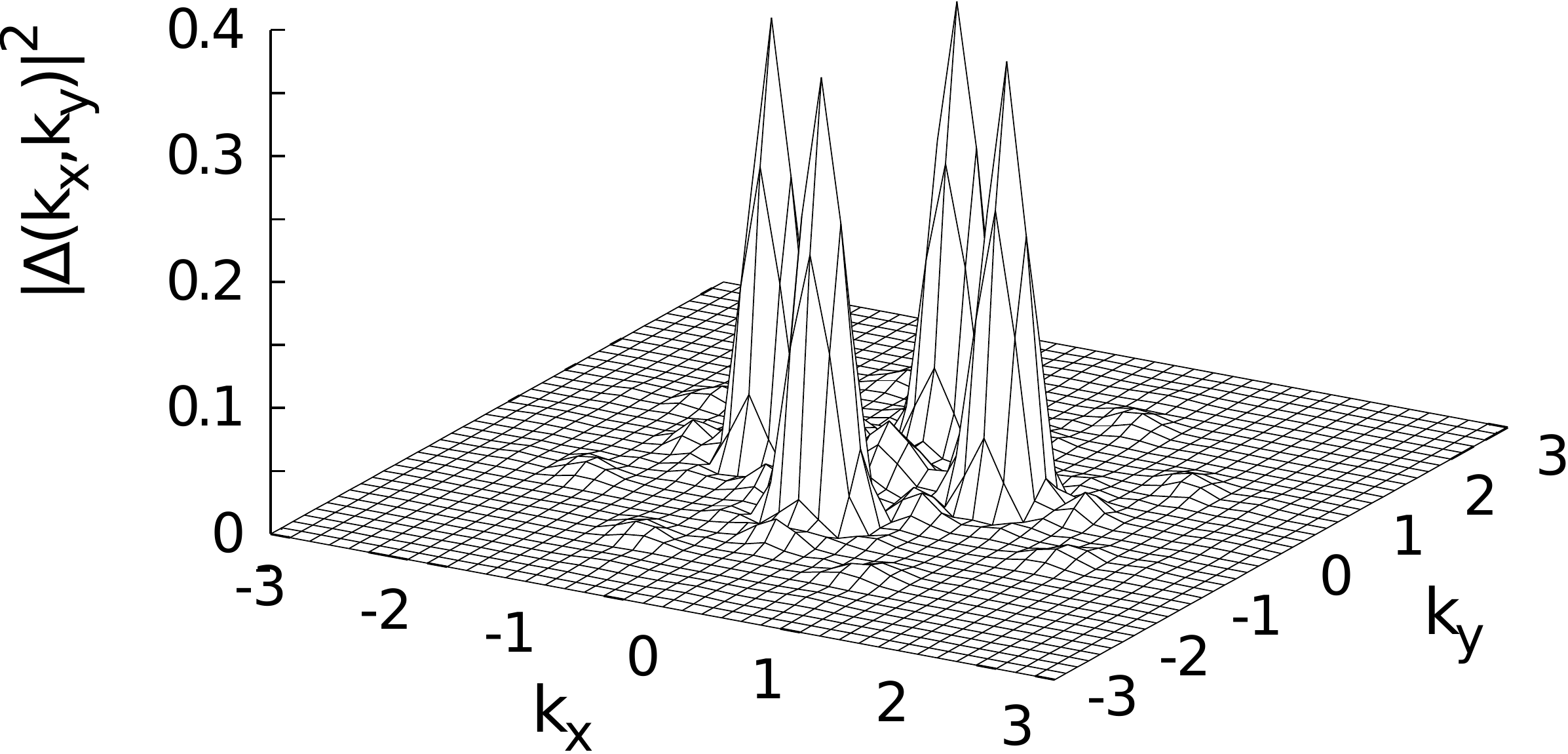}
\end{center}
\caption{\label{fig:TrapMFHF08} Mean Field parameter $\Delta$ as a
  function of the position (top) and in Fourier space (bottom) for a
  polarization value P=0.48. The checkerboard pattern is a clear
  signature of the FFLO state. The Fourier transform depicts four
  peaks at the positions $(kx=0,ky=\pm q)$ and $(kx=\pm q,ky=0)$,
  precisely like in the homogeneous situation at half-filling.}
\end{figure}

\begin{figure}[!hhh]
\begin{center}
\includegraphics[width=0.4\textwidth,clip]{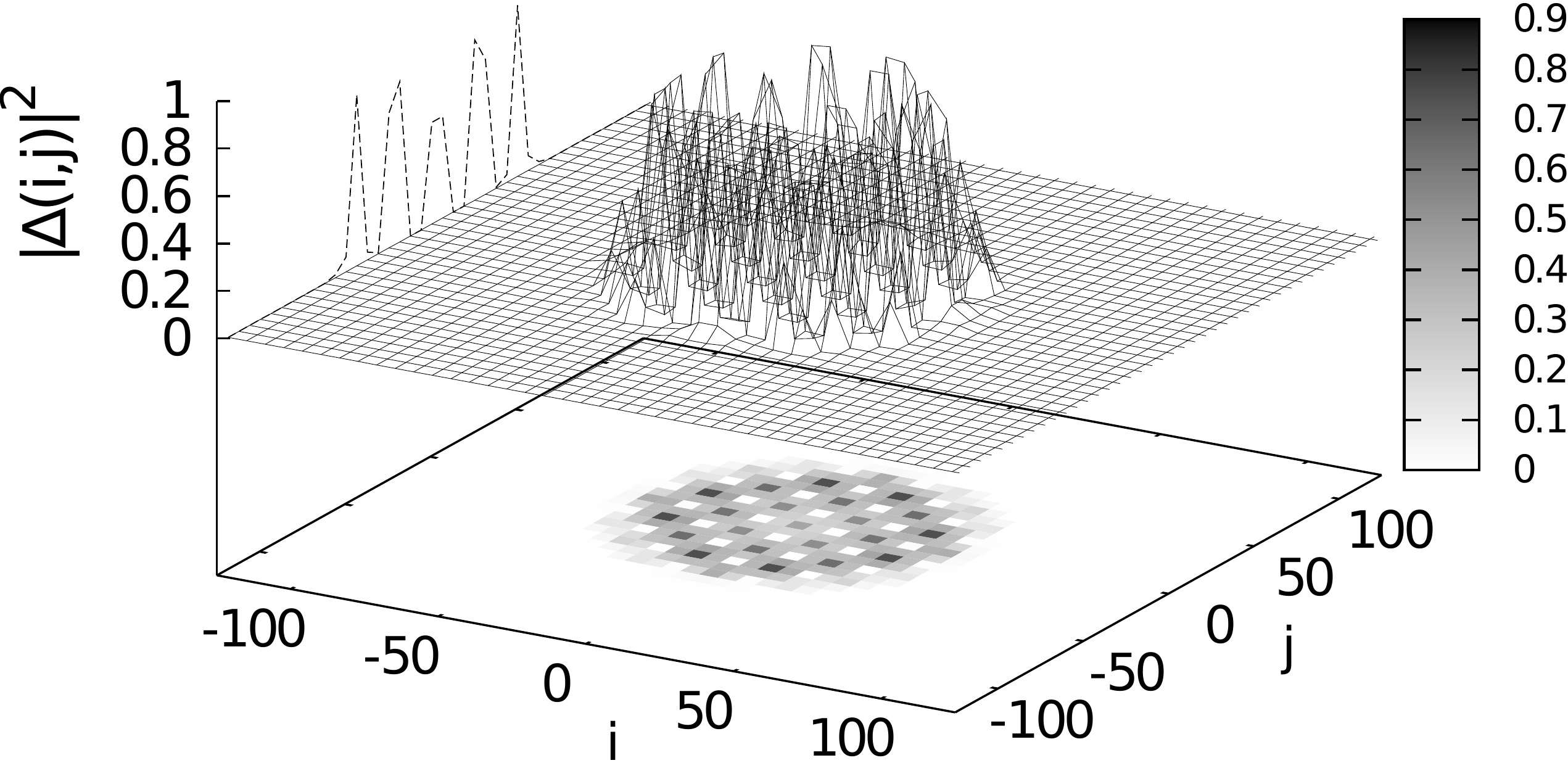}
\includegraphics[width=0.4\textwidth,clip]{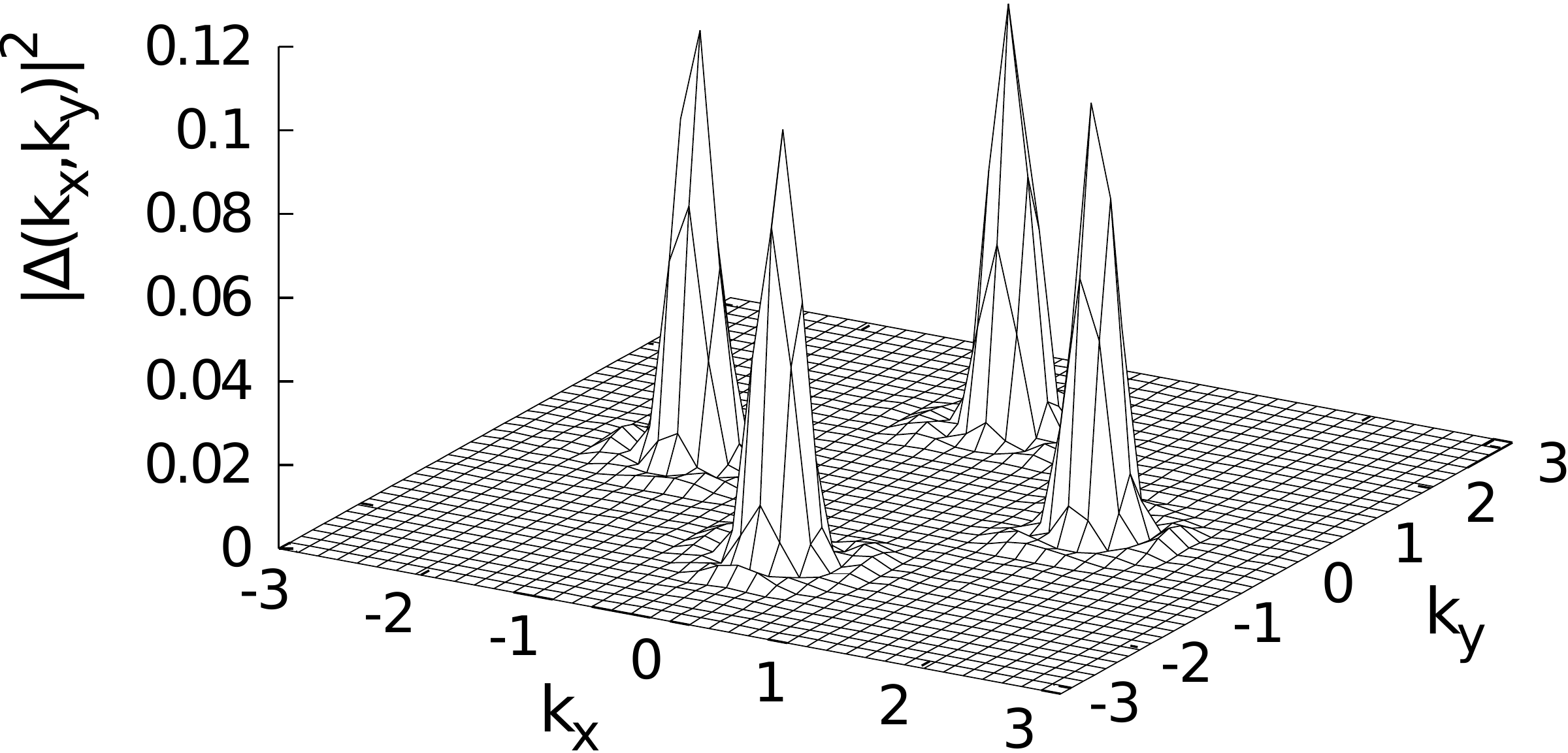}
\end{center}
\caption{\label{fig:TrapMFHF15} Mean Field parameter $\Delta$ as a
  function of the position (top) and in Fourier space (bottom) for a
  polarization value P=0.66. The checkerboard pattern depicts now a
  shorter period in real space, translating into a larger spreading of
  the four peaks in the Fourier space and corresponding to pairs
  having a larger center of mass momentum compared to Fig.~\ref{fig:TrapMFHF08}.}
\end{figure}

\textit{Harmonically confined system around half filling MF study.}
As mentioned earlier, the Quantum Monte-Carlo method suffers from a
stronger sign problem for higher fillings of the lattice with the
trap. However, we successfully studied the system imbalance around
half-filling of the lattice in the trap using the Mean-Field method.
In the figures ~\ref{fig:TrapMFHF00}, ~\ref{fig:TrapMFHF07},
~\ref{fig:TrapMFHF08}, ~\ref{fig:TrapMFHF15} the order parameter is
shown in real space as well as in Fourier space, for increasing value
of the polarization. The numerical results were obtained for a lattice
size $41\times41$, an interaction strength $U=-5$ and chemical
potential at the center of the trap corresponding to half filling.

At low polarization (P=0.13), Fig.~\ref{fig:TrapMFHF00}, the structure
is similar to the balanced case, \textit{i.e.} a maximum number of
pairs at the center of the trap, decreasing on the border. The Fourier
transform simply depicts a peak at $\vec{k}=0$, emphasizing BCS-like
pairing. At higher polarization $P=0.43$, Fig.~\ref{fig:TrapMFHF07}, a
structure in the pairing order $\Delta$ appears at the center of the
trap, leading to clear oscillations in Fourier space.  This pattern
appears first at the center of the trap simply because it corresponds
to half filling which, as explained in a previous section, is strongly
unstable towards the FFLO state. Indeed, this is emphasized by the two
figures \ref{fig:TrapMFHF08} and \ref{fig:TrapMFHF15}, corresponding
respectively to polarization $P=0.48$ and $P=0.66$. The checkerboard
pattern of $|\Delta|^2$ in real space becomes more and more
visible. Note that similar results have been previously shown
in~\cite{Chen}. However we would like to emphasize the link between
this pattern and the nature of the pairing in the homogenous
situation. Indeed, in Fourier space four peaks are clearly observed.
Their positions, $(k_x=0,k_y=\pm q)$ and $(k_x=\pm q,k_y=0)$,
precisely match the ones observed in the homogeneous situation, both
in the QMC results and in the MF ones, around half filling. In
addition, one can see that the oscillation period of the order
parameter becomes shorter with higher polarization, \textit{i.e.}
corresponding to a larger center of mass momentum $q$ of the pair,
which is depicted by the spreading of the four peaks further away from
$\vec{k}=0$.  This also shows that the oscillations in real space are
not related to the underlying harmonic levels, but really to the FFLO
order.  From the experimental point of view, this signature of the
FFLO order could be measured either directly in the density of pairs
or in their velocity distribution. Of course, the present mean-field
calculation does not include the thermal fluctuations which are
crucial to properly describe the condensation of the pairs which, at
large interaction, arises at a temperature $k_B T\approx t^2/U$ lower
than the pair formation temperature $k_B T\approx
U$~\cite{Engelbrecht_97,Iskin_07,Tempere_08,Gremaud_12}.

\section{Conclusions}

Our results, based on QMC and MF calculations, strongly emphasize that
the FFLO state is the ground state of the fermionic Hubbard model on
the square lattice for a large range of parameters, both with or
without harmonic confinement.  At low filling, the FFLO state is
similar to the bulk situation (\textit{i.e.} particles having a
quadratic dispersion relation), where the pairs have a vanishing total
angular momentum, but a finite radial component for the center of mass
momentum. On the contrary, around half-filling, the underlying Fermi
surface due to the lattice structure, leads to a FFLO state having
only discrete value of the center of mass momentum, namely around
$(kx=0,ky=\pm q)$ and $(kx=\pm q,ky=0)$. We have given an explanation
in terms of matching fermionic momentum on the Fermi surfaces. We have
also shown that, in the presence of an harmonic confinement and at low
fillings, the harmonic level basis gives rise to a simple
understanding of the pairing mechanism. In addition, we have shown
that the harmonic levels are at the origin of the oscillations seen in
the local magnetization, which, therefore, are not a signature of the
FFLO state. Finally, still in the presence of an harmonic confinement,
but around half filling, we have shown that the pairing mechanism is
essentially identical to the homogeneous situation, leading to clear
signatures in the pair density, both in real space (checkerboard
pattern) and in Fourier space (four peaks), which allows for a
possible experimental observation with cold atoms.

In the presence of a harmonic trap, it would be interesting to study
the dynamics of the pairs in response to a sudden quench from balanced
to imbalanced populations where our study indicates that one could
expect to observe the oscillations of the center of mass in the
trap. In addition, from a mean field point of view, the following
points would be interesting to consider. By monitoring the wavelength
and the amplitude of the oscillations of the order parameter, one
should be able to determine the nature of the pairing and possible
transitions between paired phases.  One should also take into account
the effects of terms beyond mean field to determine properly the
critical temperature of the transition (BKT-like) and to estimate the
strength of the quantum fluctuations thus allowing for a better
comparison with possible experimental results. Finally, one could
study more exotic situations, like asymmetric tunneling rates, or in
the presence of an effective gauge field.

\acknowledgments We thank C. Miniatura for very helpful
discussions. This work was supported by: an ARO Award W911NF0710576
with funds from the DARPA OLE Program; by the CNRS-UC Davis EPOCAL
joint research grant; by the France-Singapore Merlion program (PHC
Egide and FermiCold 2.01.09) and by the LIA FSQL. Centre for Quantum
Technologies is a Research Centre of Excellence funded by the Ministry
of Education and National Research Foundation of Singapore.

{}

\end{document}